\documentclass[aps,prb,twocolumn, showpacs]{revtex4-2}

\usepackage{epsfig}
\usepackage{graphicx}
\usepackage{amsmath}
\usepackage{mathtools}
\usepackage{amscd}
\usepackage{amssymb}
\usepackage{amsfonts}
\usepackage{amsthm}
\usepackage{enumitem}
\usepackage{stmaryrd}
\usepackage{tikz-cd}

\usepackage[T1]{fontenc}

\usepackage{xcolor}

\def\be{\begin{equation}}
\def\ee{\end{equation}}
\def\bea{\begin{eqnarray}}
\def\eea{\end{eqnarray}}

\def\bfp{\mathbf{p}}
\def\bfA{\mathbf{A}}
\def\bfa{\mathbf{a}}
\def\bfr{\mathbf{r}}

\def\bfv{\mathbf{v}}

\def\nin{\nu_{\rm{in}}}
\def\qin{q_{\rm{in}}}
\def\nout{\nu_{\rm{out}}}
\def\qout{q_{\rm{out}}}

\def\eff{{{\rm{eff}}} }
\def\bfal{\mathbf {\alpha}}

\def\nn{\nonumber\\}

\newcommand{\sbra}[1]{\langle #1 |}
\newcommand{\sket}[1]{|#1\rangle}

\newcommand{\sex}[1]{\langle #1 \rangle}

\begin{document}

\title{Fractional Charge and Fractional Statistics in the Quantum Hall Effects } 
\author{ D. E. Feldman}
\affiliation {Brown Theoretical Physics Center and Department of Physics, Brown University, Providence, RI 02912, USA}    
 \author {Bertrand I. Halperin}
\affiliation{Department of Physics, Harvard 
University, Cambridge, Massachusetts 02138, USA}

\date{\today}

\begin{abstract}
 Quasiparticles with fractional charge and fractional  statistics  are key features of the fractional quantum Hall effect. We discuss in detail the definitions of fractional charge and statistics and the ways in which these properties may be observed. In addition to theoretical foundations, we review the present status of the experiments in the area. We also discuss the notions of non-Abelian statistics and  attempts to find experimental evidence for the  existence of non-Abelian quasiparticles in certain quantum Hall systems. 
\end{abstract}

%\pacs{. }

\maketitle

\narrowtext

\tableofcontents

\section{Introduction}
\label{sec:intro}

{\color{black} 
The experimental discovery in 1982, in a two-dimensional electron system,  of  quantized Hall plateaus with  Hall conductivity $\sigma_{xy} = \nu e^2/h$ showing fractional values $\nu=1/3$ and $\nu=2/3$,  marked the beginning of one of the most surprising and far-reaching developments in condensed matter physics in the second half of the twentieth century.
}
These fractional quantized Hall (FQH)  plateaus, together with plateaus at other rational fractional values of $\nu$, were understood to be manifestations of a new type of correlated electron state, with a number of peculiar properties. Continuing experimental and theoretical efforts have revealed a wide variety of FQH states, as well as other unusual phenomena that can occur in two-dimensional electron systems in a magnetic field at low temperatures, in different materials and under different conditions. Indeed, experiments on these systems continue to produce surprises, and the field of quantum Hall effects remains a vital area of condensed matter research today.\cite{HalperinJainBook} Moreover, 
insights gained from the exploration of FQH states have also inspired predictions of a variety of  other unusual states in other  systems. 

One peculiar feature of the FQH states, which was  understood quite early, is that they must necessarily have well-defined charged excitations (quasiparticles) with  a charge that is a fraction of the electronic charge.  It was also predicted that collections of  these quasiparticles should obey fractional statistics, such that the effective wave function for the quasiparticles would be multiplied by a complex phase factor when two quasiparticles are interchanged, in contrast with  the factor of $\pm 1$ obtained on interchange of the familiar bosons or fermions. 

As we shall describe below, the existence of quasiparticles with fractional charge and statistics is essentially an inescapable logical consequence of the existence of fractional quantized Hall states. 
Thus, in one sense, the observation of an FQH plateau might be considered as a direct demonstration of the existence in principle of quasiparticles with fractional charge and statistics.  However, it is not necessarily true that isolated quasiparticles will form  the lowest energy configurations when  electrons are added to or subtracted from a quantized Hall state, and it is not clear how easy it might be to prepare isolated quasiparticles or to measure directly their charges.

In a similar vein, we may ask whether it is possible to see a direct effect of fractional statistics in an experiment, such as one where there is interference between two possible paths, in which  a pair of  quasiparticles encircle each other a different number of times. We shall see that there are numerous obstacles that need to be overcome to carry out such  an interference experiment in practice. Furthermore, there are many complications, due particularly to subtle effects of  Coulomb interactions and to the possible participation of different species of quasiparticles, which may complicate the interpretation of these experiments. Nevertheless, major progress has been made. 

In addition to quasiparticles with fractional statistics, {\em{certain}}  FQH states have been predicted to have quasiparticles with {\it{non-Abelian statistics}}. In this case, the interchange of two or more quasiparticles can give rise to a unitary transformation between orthogonal quantum states in
a Hilbert space containing  many degenerate ground states. In principle, the existence of such quasiparticles should give rise to some striking experimental manifestations, with possible consequences for technology. However, direct demonstration of the predicted phenomena has, again, proved challenging.

In the following section, we shall introduce precise definitions of fractional charge and fractional statistics, and explain why quasiparticles with these properties  are predicted to occur in fractional quantized Hall states.  In Sections~\ref{charge-expts} and \ref{stat-expts}, 
 we  discuss in greater detail the theory behind experiments designed to demonstrate most directly the effects of fractional charge and fractional statistics, and we  review the current status of these experiments.  In Section~\ref{non-abelian},    
 we discuss in greater detail the concept and implications of non-Abelian statistics,  and we discuss some examples of FQH states where non-Abelian statistics have been proposed to occur.
 The search for a clear manifestation of non-Abelian statistics by means of Fabry-Perot interferometry is discussed in Section~\ref{fpi-2}.  In Section~\ref{MZ}, we discuss the alternate geometry of a Mach-Zehnder interferometer, which has been realized in an integer quantized Hall state but not yet in an FQH state, and we review how the combination of fractional statistics and fractional charge leads to a flux period consistent with the Byers-Yang theorem.  In Section~\ref{other}, we discuss several other experimental techniques, which  reveal  aspects of FQH effect related to fractional charge and statistics, but which would not be considered to be direct observations of these properties. We present  concluding remarks in Section~\ref{conclusions}.

\section{The meaning of fractional charge and fractional statistics} \label{meaning}

\subsection {Fractional charge}

Fractional charge is relatively easy to define in a model where the Hamiltonian $H$ contains only short-range forces.\cite{KivelsonSchrieffer82}   For any state $\sket{\Psi}$ that is an eigenstate of $H$, we can define a charge density $\rho_{\Psi}(\bfr) $, which is the time-independent expectation value of the charge density operator $\rho(\bfr)$.  If there is an energy gap $\Delta E$ separating  $\sket{\Psi}$ from all other states of the Hamiltonian, then the density $\rho_{\Psi}$ may be obtained with arbitrary precision, in principle,  by using an apparatus that measures the density averaged over a time scale large compared to $\hbar / \Delta E$. If properly carried out, such a measurement will not affect the quantum state $\sket{\Psi}$, and the measurement may be repeated many times with the same results. 
In practice, the requirement for an energy gap  of size $\Delta E$ given above may be weakened,  in that one may exclude eigenstates of $H$ of lower energy if they result from excitations, relative to $\sket{\Psi} $ that are localized in space far from the measuring point $\bfr$.

In this paper, we shall consider typically a large system with a Hamiltonian of the form 
\be
H = H_0 + V
\ee
where $H_0$ is at least  approximately translationally invariant in regions far from the system boundaries, and $V$ is a sum of  local perturbations, centered around a set of points $\{\bfr_j \}$, which will be assumed to be far from the boundaries.  Let $\rho_0(\bfr) $ be the charge density in the ground state of $H_0$,  let $\Psi$ be a low-lying eigenstate of the full Hamiltonian $H$, and and let $\delta \rho_{\Psi} = \rho_{\Psi}- \rho_0$. 
We shall say that the state $\sket{\Psi}$ contains one or more  localized quasiparticles if $\delta \rho_{\Psi}  (\bfr)$ differs substantially from zero in the vicinity of at least one of the points $\bfr_j$, but is exponentially close to zero at points $\bfr$ that are far from all $\bfr_j$ and far from the boundaries.  If the point $\bfr_j$ is well separated from other regions where $V$ is non-zero, we may integrate $\delta \rho_{\Psi} (\bfr)$ over the region containing $\bfr_j$ where it is non-zero and thereby obtain the excess charge $q_j$ associated with the quasiparticle or quasiparticles  near point $\bfr_j$.  

For an ordinary insulator,  if one ignores the possible effects of long-range Coulomb interactions,  one finds that the quasiparticle charge $q_j$ will necessarily be an integer multiple of the electron  charge $-e$. For a fractional quantized Hall state, as we shall show below, the quasiparticle charge can have values which are specified rational fractions of $e$.

More generally, we can see that the quasiparticle charge will be a  protected quantity, at least for a system with short range interactions.  Its value must remain constant if the microscopic Hamiltonian is continuously varied, as long as the bulk material retains an energy gap at the Fermi energy and the magnetic field is fixed.  Because the bulk material   remains effectively  an insulator, it cannot carry an electric  current over long distances towards or away from the quasiparticle.  Consequently, the localized charge, as well as the background  charge density in the bulk, must remain constant. 

By an extension of this reasoning, a localized quasiparticle can be moved around if we allow the localizing perturbation $V$  to be time dependent. In particular, if we allow the center $\bfr_j$ of an isolated  localizing well to move sufficiently slowly as a function of time $t$, a state that is initially in eigenstate of the Hamiltonian $H$ at time $t_0$ will be in the corresponding eigenstate of the time-dependent  Hamiltonian $H(t)$  at any later time. If the initial state had a quasiparticle localized at point $\bfr_j (t_0)$, the state at time $t$ will have a quasiparticle at point  $\bfr_j (t)$.  
Clearly, the quasiparticle charge $q_j$ cannot change in this process
if the quasiparticle remains isolated from all other quasiparticles throughout.  

{\color{black} It should be emphasized that the requirements that $V$ varies only slowly and that the measurement of the charge density takes place over a time that is slow on the scale of the ground-state energy gap is essential for these arguments. An instantaneous measurement of the electronic charge in any spatial region will always yield an integer multiple of $e$. }

In the presence of long-range Coulomb interactions, the definition of quasiparticle charge is complicated by the induced polarization  of the dielectric medium.  As a familiar example, for a localized electron embedded in a three-dimensional  insulator with dielectric constant $\kappa$, the total  excess charge in the vicinity of the electron will actually be equal to $ -e / \kappa$, with the remaining  charge  distributed around the boundary of the sample. By convention, we divide the local charge into free charge and bound charge, so that the free charge associated with the electron is said to be $-e$.  Similarly, for a quasiparticle in a fractional quantized Hall state with Coulomb forces embedded in a dielectric medium, we define the quasiparticle charge $q_j$  as the free charge associated with the quasiparticle, which will be equal to the local charge multiplied by  $\kappa$ in this case.  It is this free charge which will be quantized in rational multiples of $e$.

We remark that if an electron is injected at one place on the surface of a three-dimensional insulator and is moved through  the bulk of the sample to another place on the surface, the total charge transferred between the two points will be $-e$, not $-e/ \kappa$. This is because the image charge on the surface of the insulator moves along with the electron so the total charge is transferred.  Thus we may say that the total electric current associated with an electron moving at a velocity $\bfv$ is given by $-e \bfv$, even though the local charge is $-e/ \kappa$. In the case of a quantized Hall system, the current associated with a quasiparticle moving through the bulk is more difficult to define, as the system will necessarily  have conducting states along its edges.

What happens if we turn off the localizing perturbation $V$? As $H_0$ is supposed to be translationally invariant, a localized quasiparticle will not, in general, be an eigenstate of the Hamiltonian.  In an ordinary insulator, in the absence of a magnetic field, the energy eigenstates will be plane-wave-like superpositions of localized states centered at positions throughout the sample. For quantized Hall states, however,  it is possible to create localized states for a charged quasiparticle that are eigenstates of the Hamiltonian.  Of course, these states will be highly degenerate, due to the many possibilities for choosing the center $\bfr_j$, and the localized states can be mixed by an arbitrarily small perturbation. 

As one example, in the presence of a strong magnetic field and a  weak electrostatic potential $V(\bfr)$ that varies slowly in space, energy eigenstates will generally extend all the way along contour of constant potential, while being localized in the perpendicular direction. One finds, in this case, that a quasiparticle {\it{wave packet}}, which  is initially localized at some point in space, will  move along the potential contour line, with a velocity $\bfv_D$, given by the classical  formula, $\bfv_D= \mathbf{E } \times \mathbf{B}/B^2$ , where $\mathbf{E}$ is the local in-plane electric field and $\mathbf{B} $ is the perpendicular magnetic field. 

\subsection {Fractional statistics}

\subsubsection {Definition in terms of effective wave functions and effective Hamiltonian} 
\label{def-fracstat1}

Whereas fractional charge can be easily defined for a single isolated quasiparticle or for a collection of localized quasiparticles, the concept of fractional statistics requires the consideration of two or more quasiparticles that are able to move around each other or to interchange positions. If one is confined to a suitable low-energy subspace, one may hope to describe the quantum mechanical state of  such a system by an effective wave function $\psi_{\eff}$ that depends only on the coordinates of the quasiparticles, rather than of all the electrons in the system. The effective wave function should evolve in time according to a Schr\"{o}dinger equation with some effective Hamiltonian $H_{\eff}$. Fractional statistics will be a characteristic of the combination $\psi_{\eff}$ and $H_{\eff}$.

As was first noted by Leinaas and Myrrheim, in 1977,  in two dimensions it is possible to extend the formulation of quantum mechanics to a situation where the wave function of a set of identical particles is multiplied by a complex phase factor different from $\pm1$, provided we may exclude from consideration points where two quasiparticles coincide precisely in space.\cite{LeinaasM77} Specifically, one may require that if one interchanges the positions of  two identical particles by moving their coordinates in a counterclockwise direction along a closed contour $C$ that encloses $N_C$ other quasiparticles of the same type, the wave function should be multiplied by a phase factor $e^{-i \theta}$, where   
\be
\theta = (1 + 2N_C) \theta_m ,
\ee
where the angle $\theta_m$, defined modulo $2 \pi$, is a characteristic of  of the type of quasiparticle in question.  (We use the index $m$ to distinguish between different species of quasiparticles.)     If the position of a single quasiparticle is moved along a closed loop enclosing $N_C$ other identical quasiparticles, the wave function must be multiplied by $e^{-2iN_C \theta_m}$.  
For cases other than $\theta_m = 0$ or $\pi$, this requires that
the wave function be multivalued, or equivalently that it is defined on a multi-sheeted Riemann surface. Nevertheless, quantum mechanics can be generalized in a straightforward way to deal with this situation. In a case where $\theta_m \neq 0 \mbox{ mod } \pi$, if the effective Hamiltonian $H_{\eff}$ can be written as a local function of the positions $\bfr_j$ and the momenta $\bfp_j = -i \hbar \nabla_j$, with the possible addition of long-range Coulomb forces that depend on position variables only, one says that the quasiparticles obey {\em{fractional statistics}}, with statistical angle $\theta_m$.  Such quasiparticles are often referred to as {\em{anyons}}.\cite{Wilczek-anyons} 

To describe quasiparticles with fractional statistics, however, it is not actually necessary to employ multivalued wave functions. The multiple phase factors can be eliminated by implementation of  a unitary transformation, essentially a singular gauge transformation.\cite{LeinaasM77}  Specifically, if $\psi_{\eff}$ is a multivalued wave function as described above,  let us  define a transformed  wave function 
$\psi'_{\eff}$ by
\be
\psi'_{\eff} \{r_j\}   =  \psi_{\eff} \{ \bfr_j\} \prod_{k<l} \left( \frac {z_k - z_l}{| z_k -z_l | } \right) ^ { \theta_m/2\pi} ,
\ee
where $z_j = x_j + i y_j$ is the position of particle $j$ in complex coordinates. The transformed wave function $\psi'_{\eff}$ will be single valued and will be invariant under interchange of two particles, as would be expected for particles obeying Bose-Einstein  statistics.  The price one has to pay however, is that the transformed Hamiltonian $H'_{\eff}$ will not longer be local in space.  To obtain 
$H'_{\eff}$ from $H_{\eff}$, one must replace the operators $\bfp_j$ by $[\bfp_j - \bfa_j(\bfr_j) ]$, where $\bfa_j$, known as  a {\em{Chern-Simons vector potential}}, depends on the positions of all the other quasiparticles in the system.  Specifically, one has
{\color{black}
\be
\label{csa} 
\bfa_j (\bfr_j) = \frac{\theta_m}{2\pi}  \sum_{k \neq j} \left( \frac {\hat{z}  \times ( \bfr_j - \bfr_k) } { | \bfr_j -\bfr_k |^2} \right) ,
\ee}
where $\hat{z}$ is the unit vector  normal to the plane. Thus, an alternate definition of fractional statistics is that if quasiparticles are described by a single-valued wave function $\psi'_{\eff}$ that is invariant or changes sign under interchange of quasiparticle positions, the effective Hamiltonian must be non-local, containing a Chern-Simons vector potential of the form (\ref{csa}).

 The definitions of fractional statistics must be extended in the case where there may be several kinds of quasiparticles present. Now we must introduce a new set of quantities $\theta_{m m'}$, which will be equal to one-half the phase acquired when a quasiparticle of type $m$ is moved around a quasiparticle of type $m'$, in a representation with multivalued wave functions and no Chern-Simons vector potential. When $m \neq m'$, the quantity $\theta_{m m'}$ is only defined mod $\pi $, but for identical particles, we require  $\theta_{m m} =  \theta_m$ mod $2\pi$.
  
 Again, as an alternative to the above definition, one can make a singular gauge transformation to a representation with single-valued  
wave functions, at the cost of introducing  Chern-Simons vector potentials, analogous to (\ref{csa}), which may couple to the different species in different ways. In particular, to obtain the value of $\bfa_j$ seen by a particle of type $m$, we must include a sum of terms of the form (\ref{csa}), where the coefficient $\theta_m$ is replaced by $\theta_{mm'}$, if particle $k$ is of type $m'$:
{\color{black} 
\be
  \label{csam} 
\bfa_j (\bfr_j) = \sum_{m'} \frac{\theta_{mm'} }{2\pi} \sum_{k \neq j} \left( \frac {\hat{z}  \times ( \bfr_j - \bfr_k) } { | \bfr_j -\bfr_k |^2  } \right) ,
\ee 
}
It follows from the above definitions that in all cases, $\theta_{m'm} =\theta_{mm'}$. Also, if we combine two quasiparticles of type $m$ and $m'$ to form a new quasiparticle, of type $M$, the angle $\theta_{Mm''}$ describing the mutual statistics between the hybrid quasiparticle and a third quasiparticle of type $m''$ will be equal to 
{\color{black} $\theta_{mm''} + \theta_{m'm''} $}.
As a corollary, if we group together $n$ identical quasiparticles of type $m$, the resulting clusters will have a self statistical angle of $\theta_M = n^2 \theta_m$.

\subsubsection {Illustrative  example} 
\label{ilex}
 
 As an example to illustrate a physical consequence of fractional statistics, let us consider a system containing either one or two identical anyons with charge $q_m$ in an external magnetic field $B$. We shall assume that there is at most  a short-range  interaction between the anyons.  We also assume a weak circularly symmetric parabolic electrostatic potential of the form
 \be
 \label{parab}
 \Phi (\bfr) = \frac{K}{2} r^2 ,
 \ee
 with $q _m K > 0$, in addition to a stronger short-range attractive potential that can trap a localized quasiparticle near the origin.      
 The case of a single charged particle in a uniform magnetic field and a weak parabolic potential is exactly solvable.  For a particle  in the lowest Landau level, the energy eigenstates will consist  of  a series of circular orbits with 
 \be
 \label{nr0}
  \sex{r^2} = {\color{black}2}(n+1)\hbar  / |q_m B|  ,   \,\,\,\,\, n=0, 1,2,... ,
  \ee
   and energy given by  
   \be
   \label{En}
   E = E^*_0 + \frac {q_m K}{2} \sex{r^2} ,
   \ee
   where $E^*_0$ is a constant.
The addition of a localized potential well near the origin will have negligible effect on the energies or eigenstates for large values of $n$.

Let us now consider a system with one quasiparticle, say quasiparticle 1,  localized in the well near the origin, and the second quasiparticle sitting in a circular orbit of large radius. According to the Bohr-Sommerfeld rules, we should calculate the allowed radii by requiring that the action for the circular orbit
should be equal to to an integer multiple of $2 \pi$.  Because of the Chern-Simons term due to the presence of particle 1,  the action for quasiparticle 2   will be shifted by an amount 
{\color{black}
\be
\delta S =  \oint  \bfa_2 (\bfr_2) \cdot d \bfr_2 =  2 \theta_m.
\ee
The result is that (\ref{nr0}) will be replaced by
\be
\label{nra} 
 \sex{r^2} = {\color{black}2}(n+1- \sigma \theta_m / \pi) \hbar / |q_m B|  ,
 \ee
 where $\sigma = {\rm{sign}}\,  (q_m B_z)$.}
If $\theta_m \neq 0 \mbox{ mod } \pi$, the set of allowed values for $r^2$, and hence for the energies for quasiparticle 2,  will be different depending on whether quasiparticle 1 is present or not. 

The above arguments can be generalized to the case where one has two indistinguishable quasiparticles in orbits that are not localized near the origin. In this case, one finds that the set of allowed energy levels will be sensitive to $\theta_m$ mod $2\pi$.

\subsubsection {Relation to the microscopic Hamiltonian} 
\label{def-fracstat2}

To make these ideas more concrete, let us return to the microscopic states for a system containing a given number  $N$  of identical quasiparticles.  Let $\sket{\Psi ( \{\bfr_j\}) } $  be the many-electron state  with quasiparticles localized at specified positions $(\bfr_1,..,\bfr_N)$. The set of such states, which we here assume to be unique except for a phase, will form an (over-complete) basis for the set of states we are interested in. The set of allowed positions $\bfr_j$ may include restrictions, such as a minimum separation between two quasiparticles.  We shall  assume that any state in the Hilbert space of interest can be written as a superposition of basis states, in the form
\be
\label{Psimic}
\sket {\Psi} = \int d\bfr_1 ... d \bfr_N  \psi_{\eff} (\{\bfr_j\}) \sket{\Psi (\{\bfr_j\})} .
\ee 

Once we have made a specific  phase choice for the basis states $\sket{\Psi (\{\bfr_j\}) } $, we can define a Berry connection,
\be
\bfal_k (\{\bfr_j \}) = i \sbra{\Psi (\{\bfr_j\}) } \nabla_k \sket{\Psi (\{\bfr_j\}) } ,
\ee
where $\nabla_k$ is the gradient with respect to the position $\bfr_k$. We may  now consider a situation in which the positions of two quasiparticles, labeled $k$ and $l$, are interchanged by moving them around a specified contour $C$ in a counterclockwise fashion, until their final positions are interchanged from their initial positions, while the positions of all other quasiparticles
are held fixed.  The Berry phase for the process is given by 
\be
\theta_{Ckl} = \int d \bfr_k \cdot \bfal_k + \int d \bfr_l \cdot \bfal_l  ,
\ee
where the integral is taken along the contour. Whereas the Berry connections $\bfal_j $ depend on the particular choices made for the phases of the basis states, the Berry phase $\theta_{Ckl}$ may be seen to be independent of those choices, up to an additive multiple of $2 \pi$. Thus the quantity $e^{i \theta_{Ckl}}$ is independent of the phase choice and is therefore gauge invariant. For a system of identical anyons of charge $q_m$ in an external magnetic field, the value of $\theta_{Ckl}$ should be given by  
{\color{black}
\be
\label{tckl}
\theta_{Ckl} \sim (1+ 2N_C)\theta_m + 2 \pi q_m B_z  A_C /\hbar  ,
\ee}
where $A_C$ is the area enclosed by the contour $C$, $\theta_m$ is a constant characteristic of the type of quasiparticle under consideration, and $N_C$, as before, is the number of additional quasiparticles enclosed.  Equation (\ref{tckl}) is supposed to be exact when the contour $C$ is large and quasiparticles $k$ and $l$ stay far from all other 
quasiparticles, but there can be corrections if these conditions are violated. Nevertheless, the implication of (\ref{tckl}) is that with a suitable choice of gauge, $\bfal_j$ may be written in the form 
\be
\bfal_j = q_m  \bfA(\bfr_j ) + \bfa_j  (\bfr_j)   ,
\ee
where $\bfA$ is the vector potential due to the applied magnetic field $\mathbf{B}$, and $\bfa_j$ is just the Chern-Simons vector potential given by (\ref{csa}). The discussion may be readily extended to the case where there are several types of quasiparticle present, in which case the first term in (\ref{tckl}) should be replaced by $(\theta_m  + 2\sum_m' N_{C}^{m'} \theta_{mm'})$, where $N_{C}^{m'}$ is the number of quasiparticles of type $m'$ enclosed by the contour, and the definition of $\bfa_j$ must be  extended, as described in Eq. (\ref{csam}).

Next, we must examine the time evolution of a state $\sket{\Psi}$ of the form (\ref{Psimic}). It is convenient for this purpose to use  path integral approach. Then the state at time $t$ can be related to the state at time 0 by a unitary transformation of the form
\be
\psi_{\eff} (\{\bfr_j \}, t) = \int  d\{ \bfr'_k\} K( \{\bfr_j\}, \{ \bfr'_k \}) \psi_{\eff} ( \{\bfr'_j \}, 0) ,
\ee
where the kernel $K$ is given by the sum of $e^{-iS}$ over all paths connecting the initial and final configurations of positions,  with $S$ being the action associated with the path. 
To a good approximation, we may evaluate  $S$ as
\be
S = \int dt'' U(\{\bfr''_j \})    + \sum_j \int d \bfr''_j  \cdot \bfal_j (\bfr''_j)  ,
\ee
where $U(\{\bfr''_j \})$ is the expectation value of the microscopic Hamiltonian $H$ in the basis state $\sket { \Psi \{ \bfr''_j \} }$, and the integral is taken along the path from the initial to the final configuration. This expression coincides with the formula for the action of a collection of particles with charge  $q_{\color{black}m}$ subject to an applied magnetic field and a Chern-Simons vector potential, in the presence of a position-dependent potential energy $U$, in the limit where the effective mass of the particle is taken to zero, {\it{i.e.}}, in the limit where the particles are all in the lowest Landau level. 

More generally, $U$ should be replaced by an operator that may include terms that are slightly off-diagonal in the position variables, which would lead to additional momentum-dependent 
terms in the Hamiltonian, including, perhaps, short-range momentum-dependent interactions between the quasiparticles. Matrix elements of the microscopic Hamiltonian that mix states in the low-energy subspace we are considering with states outside that subspace may be taken into account via perturbation theory as corrections to the matrix elements of  $U$.   In a similar fashion, the effects of mixing between Landau levels due to interactions in a system of particles with nonzero mass may be included in a model that is projected onto a single Landau level by including suitable corrections to the interactions within the Landau level. As long as corrections to the interaction terms remain short-ranged in space, they can be distinguished from  the Chern-Simons
interaction, and will not affect the behavior of well-separated quasiparticles.  Thus, the values of the statistical angles $\theta_{mm'}$ remain well-defined and unchanged. 

The interplay of Landau-level mixing with long-range forces can change the apparent values of $\theta_{mm}$, as discussed in Refs. \onlinecite{llm-stat1,llm-stat2}. However, the deviations decay as a power of the distance between quasiparticles.

\subsubsection{Non-Abelian statistics}

In our previous discussions, we have assumed that if the locations and types of all quasiparticle are specified, there will be a unique low-energy state of the Hamiltonian corresponding to this specification.  However, a very different situation is believed to occur in some special quantized Hall states.   For these states, in a situation with $N$ localized quasiparticles, there should be a  number of nearly-degenerate low-energy eigenstates which grows exponentially with $N$.  The energy differences between these states should fall off exponentially with the separation between quasiparticles, and they are frequently treated as negligible in theoretical discussions. 

Now, if a set of quasiparticles are slowly moved around each other or interchanged, in such a way that the set of final positions for each quasiparticle type is identical to the initial set,  the final state of the system will be related to the initial state by a unitary transformation in the Hilbert space of low-energy eigenstates. Furthermore, if the braiding process is fast compared to the ``exponentially small"  energy splittings of the Hilbert space, the unitary transformation will  depend on the topology of the braiding, but will be independent of all other details of the paths that are taken. For processes that involve multiple interchanges of quasiparticles, the resulting unitary transformation will generally depend on the order in which the interchanges have been performed. Hence, the quasiparticles are said to obey ``non-Abelian statistics".

Because a full discussion of various types of non-Abelian statistics and the ways in which they may be manifest in quantum Hall systems is complicated, we shall defer that discussion until later sections of this paper, and shall first concentrate on states with Abelian fractional statistics.

\subsection {Application to quantized Hall states} 

\subsubsection {Fractional charge}

In Laughlin's landmark 1983 paper, which described his  trial wave function for the fractional Hall ground state at $\nu=1/3$ and related fractions, he also proposed wave functions for the elementary quasiparticles, often denoted as quasielectrons and quasiholes.\cite{Laughlin83}  The added electric charges associated with these proposed wave functions were, indeed, fractions of an electron charge, {\it{viz.}},  $q_c = \pm e/3$.  Since the trial wave functions are not  exact eigenstates of the Hamiltonian for a realistic model with Coulomb interactions, one might be tempted to  question the exactness of the charge quantization based on them. However, Laughlin presented a more general argument that quasiparticles with fractional charge must be a feature of any fractional quantized Hall state.  

Consider a two-dimensional electron system in a fractional quantized Hall state with filling factor $\nu$ on a large disk of radius $R$. Let us puncture the disk with a hole of diameter $a$ at the center of the disk, and let us thread  an infinite solenoid with radius less than $a$ through  the hole. 
In two dimensions, the scattering cross section of a barrier of radius $a$ will vanish \cite{LL} in the limit $a \to 0$, proportional to $1 / {\color{black}\ln^2} |a|$. Thus, in the limit $a \to 0$, the solenoid will have no effect on electrons in the system when there is no flux through the solenoid. 

Now, start with a situation where the system is initially in its ground state and there is no flux in the solenoid, and gradually increase the flux until the solenoid contains precisely one flux quantum, 
{\color{black} 
pointing in the same direction as the uniform magnetic field. [Note: In our discussions of quantum Hall systems, throughout this paper, we shall assume that the applied magnetic field  $\mathbf{B}$ points along the negative $z$ axis, unless otherwise specified, and $B = |\mathbf{B}| > 0.$]
}
 The time-dependent flux will generate an azimuthal electric field, which will drive electrons in towards the origin, due to the non-vanishing Hall conductance. A simple calculation shows that the total charge accumulated near the origin will be equal to $- \nu e$. 
This extra charge will have come from the edge of the system, where there is necessarily  a reservoir of low-energy conducting states.\cite{Halperin82}   Since there is a finite energy gap in the bulk of the system, we expect, according to the adiabatic theorem, that the final state will again be an energy eigenstate of the system.  (Although, in principle, the adiabatic theorem could break down at an instant where the added energy of the system due to the charge at the origin crosses the energy for adding the charge back to a state at the edge of the system, the matrix element for such a transfer will be exponentially small, if the radius $R$ is very large compared to the magnetic length.
In addition, for a system with circular symmetry, the matrix element will be identically zero by angular momentum conservation.) 

Although the Hamiltonian with the added flux quantum is mathematically different from the original Hamiltonian, we can make a gauge transformation that eliminates the  vector potential due to the solenoid,  multiplying the wave functions by a position-dependent phase factor and restoring the Hamiltonian to its original form. Thus, the original Hamiltonian must have an eigenstate with the same energy and charge distribution as the one we have found for the state with an added flux quantum.

Of course, we can generate a quasiparticle with charge $+ \nu e$  by repeating the above procedure with  solenoid  flux in the opposite direction.  There is no guarantee, however, that quasiparticles with charge $\pm \nu e$ have the lowest energy or the smallest charge of any possible quasiparticle in the system.  In particular if $\nu = p /q$, where $p$ and $q$ are integers with no common divisor,  one can always construct a quasiparticle with charge $q_m  =\pm e/q$. Since $p$ and $q$ have no common divisor, there will necessarily exist integers $n$ and $n'$ such that $nq-n'p = 1$.  Then a combination of $n'$ quasiparticles of charge  $ \nu e$ and $n$ electrons will have total charge $-e / q$. 

These  arguments do not require that $e/q$ is necessarily the smallest charge for a quasiparticle in the system. For example, the various competing models \cite{16-fold} proposed to explain the even-denominator quantized Hall state observed at $\nu =5/2$ have quasiparticles with charge $\pm e/4$. Levin and Stern have  argued\cite{LevinS09}, in fact,  that for any fractional quantized Hall state with even denominator $q$, there must exist quasiparticles with charge $q_m = \pm e /2q$.

\subsubsection {Fractional statistics}

The prediction that quasiparticles in fractional quantized Hall states should obey fractional statistics was made in 1984, by Halperin,\cite{Halperin84}  and slightly later, by  Arovas, Schreiffer and Wilczek.\cite{ArovasSW84}  The analysis of Halperin was based on the behavior of effective wave functions for collections of quasiparticles, similar to the discussion in Subsection (\ref{def-fracstat1}), above. By contrast the analysis of Arovas, {\it{et al.}}, made use of the definition presented in  Subsection (\ref{def-fracstat2}), specifically, by calculating the Berry phase acquired on interchanging the positions of two quasiholes in the $\nu=1/3$ state, using Laughlin's trial wave function for the quasiholes. 

The analysis of Ref. \onlinecite{Halperin84} was motivated by the following set of observations. Laughlin's wave functions for the FQH states at $\nu = 1/m$  involve a factor of
$\prod_{j<k} (z_j - z_k) ^{m}$, in addition to a Gaussian factor which assures that the electrons have the correct density. 
These trial wave functions minimize the kinetic energy, as all particles lie in the lowest Landau level, and they are efficient at minimizing the potential energy, at least in the case of short-range repulsive interactions, as  the wave functions vanish rapidly when two electrons come close together. Moreover, if $m$ is an odd integer, the wave function is antisymmetric under the interchange of two particles, as required by Fermi statistics. If one were to replace the exponent $m$ in this product by a non-integer exponent $\gamma$, and if one multiplied the exponent in the Gaussian factor by a constant $s^{-1}$, one would  have a wave function  that describes a collection of anyons  in the lowest Landau level for particles of charge $\pm e/s$. 
The exponent $\gamma$ would be related to the statistical angle $ \theta_m$ of the anyons by
\be
\label{exponent}
\gamma = 2n \pm \theta_m/ \pi ,
\ee
where the sign in (\ref{exponent}) depends on the sign of the anyon charge and the direction of the applied magnetic field.  With our choice of $B_z<0$,  if one assumes $\theta_m = \pi / 3$ for quasielectrons in  the $\nu=1/3$ state, and one choses $n=1$, one finds that with the negative sign in (\ref{exponent}) the density of $e/3$ quasielectrons is just such as to increase the filling factor to $\nu=2/5$.  If one assumes again $\theta_m = 1/3$ for the quasiholes,   uses the positive sign in (\ref{exponent}), and chooses $n=1$,  the density of quasiholes is such as to decrease the filling factor to  $\nu=2/7$.  FQH states were, indeed,  observed experimentally at both these filling factors.  

Other fractions could be generated using  larger values of $n$.  As noted in Ref. \onlinecite{Halperin84}, this procedure could be repeated, so that starting from a given  FQH state with  $\nu = p/q$, by adding quasiparticles of charge $\pm e/q$ and appropriate statistical angle, one could generate daughter states corresponding to fractions with larger values of $p$ and $q$. An iterative formula was developed for predicting the statistical angle $\theta_m$ at each new fraction, and it was shown that  in this manner one could generate a unique FQH state for any fraction $p/q$ with odd denominator. (Of course, there would be no guarantee that the resulting FQH state would actually be the lowest energy state for a system with any particular form of the electron-electron interaction.)

The form of the effective wave function and the choice of statistical angle were further justified in Ref. \onlinecite{Halperin84} by comparison with microscopic trial wave functions that had been introduced earlier to describe a collection of quasielectrons at $\nu=1/3$ and  the ground state at $\nu= 2/5$.\cite{331}  Of course, these trial wave functions  are only approximate descriptions of the true energy eigenstates for a realistic Hamiltonian, as is also the case for the trial wave function used in Ref. \onlinecite{ArovasSW84} for a pair of quasiholes. However, the statistical angle should be a topologically protected quantity, meaning that its value cannot change under any deformation of the Hamiltonian that does not cause the energy gap to collapse and does not provoke a first-order phase transition.  

In 1989, Jainendra Jain proposed the ``composite fermion'' approach of generating trial wave functions for FQH states, which has proved to give energies and wave-functions that are generally much more accurate than those obtained by previous methods, particularly for states with large denominator.\cite{Jain89}  However, the statistical angles calculated for quasiparticles at a given odd-denominator fraction $\nu$ turn out to be the same as the ones predicted for the same fraction in Ref. \onlinecite{Halperin84}. A general description of all possible FQH states with Abelian statistics, including but not limited to the Jain states, has been given by Wen.\cite{Wen95}   The description includes predictions for the charges and statistical angles of quasiparticles, as well as other topological quantum numbers, such as the ``shift parameter'', for these states. 

As was seen in the case of fractional charge, the necessity that quasiparticles in an FQH state obey fractional statistics  can actually be demonstrated by an argument that does not make any specific assumptions about the form of the ground state at a given fraction $\nu$. Consider a gedanken experiment similar to that described in Subsection (\ref {ilex}), where an FQH system with $\nu=1/3$  is subject to a weak parabolic potential of form  (\ref{parab}). If we add a single quasihole to the system, with positive charge $q_m = e/3$, it will have a series of equally spaced energy levels, with energies given by (\ref{nr0}) and (\ref{En}).  If the quasihole is placed in orbit with radius $r_n$, there will be an electric current around the orbit of magnitude
 $I_n = q_m v_d / 2 \pi r_n$, where $v_d = K r_n / B$ is the classical drift velocity in the perpendicular electric and magnetic fields. 

Now let us place a thin solenoid at the origin and slowly change the flux through the solenoid from zero to one flux quantum  antiparallel to the applied magnetic field. This will produce an additional quasihole at the origin and will modify the orbit of the circulating quasihole. The electromotive force generated by the time-dependent flux will do work given by $2 \pi I / |e| $, which will increase the energy of the circulating quasihole by expanding its orbit in the parabolic confining potential. By equating the work done with the change in radius, we see that the new orbit radius will be related to the old one by replacing (\ref{nr0}) with (\ref{nra}) and choosing $\theta_m = \pi / 3$.  Thus the set of allowed orbits for the circulating quasiparticle is different from the set before the flux was turned on. Since the solenoid flux can be removed from the Hamiltonian by a gauge  transformation,  the change in the set of allowed radii is entirely due to the presence of a new quasihole at the origin, and not to any change in the Hamiltonian itself. Thus we see that the quasihole must have a statistical angle equal to $\pi/3$ mod $\pi$ in this case.  More complicated arguments can be used to demonstrate the necessary occurrence of fractional statistics for other FQH states. 
We shall address one such argument, employing a Mach-Zehnder interferometer, in Section \ref{MZ}. \\

{\color{black} 
The appearance of fractional statistics in FQH states is closely related to the fact that the ground states of these systems are degenerate when studied on a torus or another compact manifold with genus $\geq 1$. It was noted early on that for a translationally invariant system containing   $N_e$ interacting electrons in the lowest Landau level in a finite rectangle with periodic boundary conditions containing $N_\Phi$ quanta of magnetic flux, every eigenstate of the Hamiltonian must be at least $q$-fold degenerate, where $q$ is the denominator of the fraction $N_e/N_\Phi$ reduced to lowest terms.\cite{YoshiokaHL83,HaldaneRtorus85} 
This degeneracy will generally be split in the presence of disorder, but in the case of an FQH state, where  the ground states are separated from all other eigenstates by a finite energy gap, the splittings between the ground states will fall off exponentially with the size of the system, and will therefore be negligible for a sufficiently large system.\cite{NiuTW85} More generally,  it can be shown that any system that supports excitations with fractional statistics must be degenerate on a large torus.\cite{einarsson90}  Predictions for the ground state degeneracies of Abelian and non-Abelian FQH states on a torus and on manifolds of higher genus may be found in various places
in the literature.\cite{WenZ92,t2,top-den,NayakSSFD08}  Although these questions are significant theoretically,  we shall not discuss them further in  the present review, as we are  focused on phenomena that can be studied experimentally.   }

{\color{black} 
\subsubsection{Edge modes}

In our previous discussions, we have focused on the properties of localized quasiparticles, or collections of quasiparticles, that are far from any edges of the sample.  However, fractionally charged quasiparticles can  also  exist in delocalized states along the edges of a sample, or at a boundary between two quantized Hall states with different Hall conductivity. As was originally noted in Ref. \onlinecite{Halperin82}, there must be gapless modes at a boundary between a gapped quantum Hall liquid and a vacuum.  In the case of integer quantized Hall states, these edge modes may be understood as orbits for electrons at the Fermi level, which propagate along the edge in a particular direction. For FQH states, the edge modes may be similarly interpreted as orbits for quasiparticles of various types.   

Though the charge on an edge will be conserved if there are no contacts to the edge and the edge is far from all other edges of the sample, charges can tunnel between two opposite edges of a sample if there is a narrow constriction which brings them close together.  When the tunneling strength is small, transfer of charge from one edge to another can occur in discrete units which may be interpreted as the charge of a tunneling quasiparticle. In a geometry with two or more constrictions, there can be interference features that one can attribute to  the difference in phase accumulated by a quasiparticle in tunneling from {\color{black} one edge to the opposite edge} via the  possible paths involving tunneling at  different constrictions. For quasiparticles with fractional statistics, the accumulated phase will be sensitive to the number of other quasiparticles enclosed by the difference in paths. Therefore, interference experiments can provide a means for observing effects of the fractional statistics.

Although fractional statistics as well as fractional charge may be measured, in principle, with experiments on quasiparticles far from any sample edges, as illustrated by the  gedanken experiments described above, in practice studies of fractional statistics have always employed interferometers with tunneling between edges.  As described below, a few  experiments have succeeded in measuring fractional charge accumulation in localized regions far from any edge of the sample, but even here, the majority of experiments have involved edge modes and have measured the charges  of quasiparticles tunneling from one edge to another  across a constriction. 

{\color{black} Since quantum Hall edges are  interacting gapless systems, it is not possible to define quasiparticle charge in the same way as in the bulk.  In particular, charges propagating in a one-dimensional metal may break up into multiple pieces, 
giving rise to such phenomena as spin-charge separation and charge fractionalization \cite{GiamarchiBook,wire-frac-1,wire-frac-2}.
 These phenomena would be  sensitive to details of the edge.  However,   charges tunneling from one edge to another through a gapped quantum Hall state should be quantized, and can be measured, at least in the dilute limit, by noise experiments.}
 A more detailed discussion of edge modes in FQH states will be given below in the section on non-Abelian statistics. }

\section{Experimental  probes of fractional charge} \label{charge-expts}

Fractional charge was one of the earliest predictions of the FQH theory, but it took more than a decade to directly observe it. 
Three experimental techniques have been implemented: noise  \cite{frac1,frac2,one/5,one/7,two/3,charge-5/2,charge-7/3,josephson,Josephson2},
Aharonov-Bohm interferometry \cite{willett08,AB-heiblum,marcus-inter,manfra19}, and charging spectroscopy 
{\color{black}\cite{goldman-su,alt-goldman-su,GKLZ,local1,local2,marcus-antidot,graph-antidot}}. 
The bulk of our knowledge comes from shot noise experiments, and we start with their review. This includes a discussion of a recent experiment on photo-assisted shot noise \cite{josephson}. We then discuss  two approaches to charging spectroscopy.  Because the interferometry technique uses the same setup to probe fractional statistics and fractional charge, we shall defer discussion of  both applications until  the following section. 

\subsection{Shot noise}

Suppose that particles of charge $q_m$ tunnel through a high barrier between two conductors. The tunneling rate from the higher to lower 
electrochemical potential is $T_q$ so that the average current 
$\langle I(t)\rangle=q_m T_q$. The shot noise technique focuses on the low-frequency fluctuations of the current. The noise is defined as
\be
\label{noise-def}
S=\int_{-\infty}^{\infty} dt \exp(i\omega t)\langle I(t)I(0)+I(0)I(t)\rangle,
\ee
where we are interested in the $\omega\rightarrow 0$ limit. The integral reduces to the mean square fluctuation of the total transmitted charge over a long time $t$,
$S=\lim_{t\rightarrow \infty}2\langle [\Delta Q(t)]^2\rangle/t$.
In the low-transmission limit, this simplifies to 
\be
\label{Schottky}
S=2q_m I_T, 
\ee
where $I_T$ is the average tunneling current. The quasiparticle charge can be extracted, if both noise and current are known.
The derivation does not depend on any details of the Hamiltonian and applies as long as $T_q$ is small and no charges tunnel uphill from the lower to higher electrochemical potential. The latter is true as long as the temperature is low compared to the voltage energy scale $q_mV$. {\color{black}  Measurements 
 of the current noise at a finite frequency $\omega$ can be used to determine the quasiparticle charge provided that  $\omega$ is less than a value  that is necessarily smaller than microscopic frequencies such as  $\hbar^{-1}$ times the energy gap and $I/e$ but in practice is likely to be limited by details such as capacitive lags on the sample or characteristics  of the measuring apparatus.     }

 \begin{figure}[!htb]
\bigskip
\centering\scalebox{0.25}[0.25]
{\includegraphics{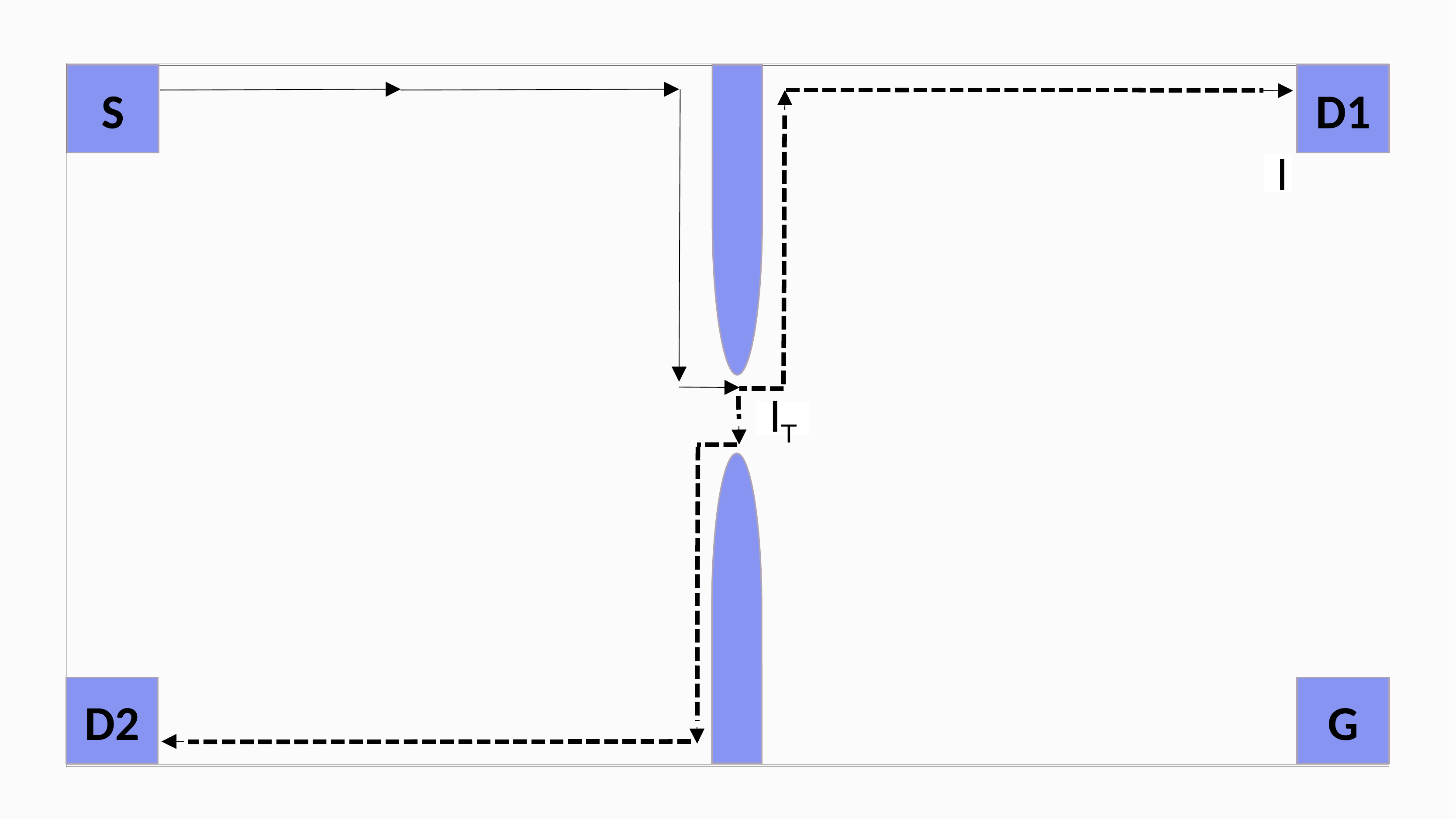}}
\caption{{Shot noise setup with chiral edges. The incoming current from source S splits into the transmitted current $I$ into drain D1 and the  tunneling current $I_T$ into drain D2. Current fluctuations can be measured at D1 or D2.    }}
\label{fig1:review}
\end{figure}

Shot noise  was used with success to measure the electron charge \cite{schottky} as early as in 1918, but almost a century elapsed before it was extended to FQH quasiparticles in Refs. \onlinecite{frac1} and \onlinecite{frac2}. The schematics\cite{noise-theor1,noise-theor2,noise-theor3,martin05} of the experimental setup are shown in Fig.~\ref{fig1:review}. In the quantum Hall effect, the bulk is gapped, 
and charges travel along edges, which are maintained at different voltages in the setting of Fig.~\ref{fig1:review}. 
A narrow constriction allows charge tunneling between the edges. Since tunneling charges cross the bulk of the sample, they are restricted to the allowed quasiparticle charges in the bulk. 
Usually, but not always, the lowest quasiparticle charge dominates the limit of weak tunneling and can be extracted from shot noise.  The noise is detected in the drain at the end of one of the edges. In the absence of the Nyquist noise, at zero temperature, the drain noise is the same as the noise of the tunneling current $I_T$.

In real experiments, the temperature and the frequency $\omega$ remain finite. A finite frequency does not affect the interpretation of the data as long is $1/\omega$ exceeds all other time scales in the problem, such as the thermal scale $\hbar/T$ and the Josephson scale $\hbar/q_mV$. Experiments are typically performed at $\omega\sim 1$~MHz.

The effect of a finite temperature is more complex. Access to fragile FQH states requires simultaneous low temperatures and voltages, and the limit of $T\ll q_m V$ might not be available. Fortunately, a universal relation \cite{noise-formula} exists among the tunneling current $I_T$, the voltage, the temperature, and the noise:

\be
\label{fitting}
S=2q_m I_T{\rm coth}\frac{q_mV}{2k_B T}-4k_BT\frac{\partial I_T}{\partial V}
+4k_BTG,
\ee
where $G=\nu e^2/h$ is the quantized Hall conductance. 
Eq. (\ref{fitting}) contains the non-linear tunneling conductance $\frac{\partial I_T}{\partial V}$ and is a 
consequence of detailed balance \cite{LR} and fluctuation relations \cite{fluct1,fluct2}. It applies irrespectively of microscopic details as long as tunneling is weak and the edges are chiral, that is, all edge modes propagate in the same direction, as is the case at the filling factors $\nu=1/(2n+1)$. 

\begin{figure}[!htb]
\bigskip
\centering\scalebox{0.25}[0.25]
{\includegraphics{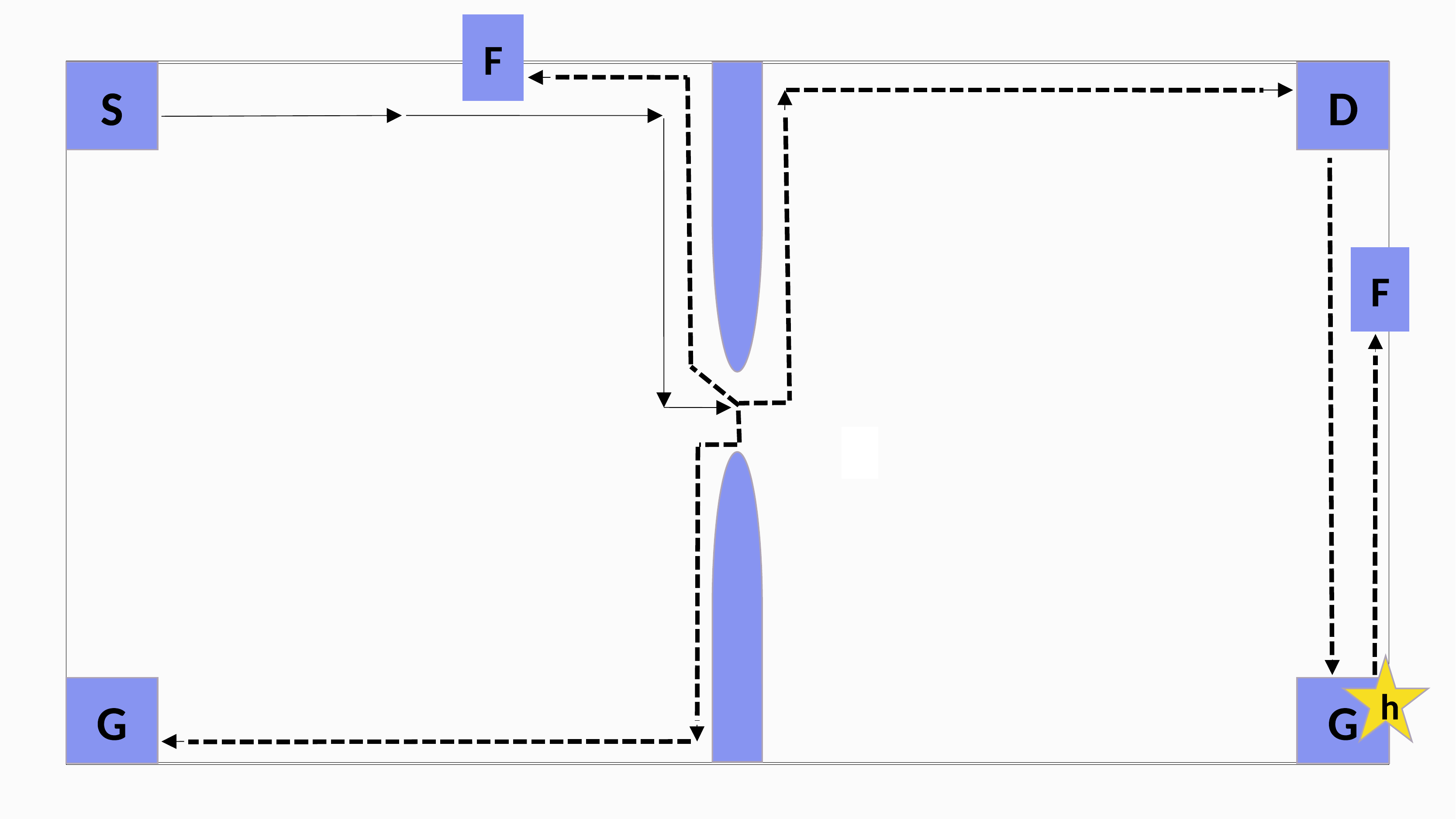}}
\caption{{Shot noise setup useful for edges with upstream neutral modes. Drain D sends the current in a narrow range of frequencies to the amplifier that measures the voltage noise.The rest of the current arrives to the ground, where a hot spot h forms. Floating contacts F absorb excess heat carried by the upstream modes emitted from the hot spot and the tunneling contact.}}
\label{fig2:review}
\end{figure}

The interpretation of experiments on FQH states with non-chiral edges is complicated by hot spot formation \cite{noise-formula}. A non-chiral edge contains a downstream charge mode that carries charge and energy and one or more neutral modes that carry energy in the opposite upstream direction. 
When a biased charged mode arrives to a grounded terminal, Joule heat is dissipated.
Some of it is carried back by neutral modes. This heat arrives to the source and affects the thermal noise of the outgoing current. This, in turn, affects the measured noise in the drain so that Eq. (\ref{fitting})  no longer applies. The problem can be alleviated \cite{noise-formula} with floating contacts along the edge. (See Fig.~\ref{fig2:review}.) 
The contacts absorb the excess heat carried by the neutral modes.

The first experiments \cite{frac1,frac2} revealed charges $e/3$ at the filling factor $\nu=1/3$. Subsequent work \cite{one/5,one/7,two/3,charge-5/2,Josephson2} reported quasiparticle charges 
$e/3$ at $\nu=2/3, ~4/3,~5/3$ and $8/3$, $e/5$ at $\nu=2/5$, and $e/7$ at $\nu=3/7$ in agreement with the lowest quasiparticle charges predicted at those filling factors. 
Charge values, consistent with the lowest theoretical values, $q_m\sim 0.25 e$ at $\nu=5/2$ and $q_m\sim 0.3 - 0.4 e$ at $\nu=7/3$ were also observed \cite{charge-5/2,charge-7/3}, 
but only at intermediate tunneling rates. The observed tunneling charge grows in the weak-tunneling limit.

A challenge for Eq. (\ref{fitting}) is the growth of the observed $q_m$ as the temperature goes down at several filling factors. This does not  happen at $\nu=1/3$, where $q_m$ stays at $e/3$ (see, however, Ref. \onlinecite{charge-7/3}). 
On the other hand \cite{one/7}, $q_m$ reaches $2e/5$ at the lowest temperatures at $\nu=2/5$, the low-temperature $q_m$ reaches $2e/3$ at $\nu=2/3$, and $q_m$ 
reaches $2.4e/7$ at the lowest available temperatures at $\nu=3/7$. A possible explanation consists in the competition of the tunneling of quasiparticles 
with the charge $e/q$ and composite quasiparticles of the charge $pe/q$ at $\nu=p/q$. 
The bulk energy cost of a quasiparticle grows with its charge. Hence, the bare tunneling amplitude of a quasiparticle through 
a constriction between two edges is higher for a lower charge. This is not necessarily the case for the renormalized tunneling amplitude 
that controls low-temperature transport. It was indeed observed in Refs. \onlinecite{sassetti_1} and \onlinecite{sassetti_2} that the tunneling of the charges $pe/q$ is 
more relevant in the renormalization group sense than the tunneling of the charges $e/q$ at low energies at the filling factors $p/q=p/(2p+1)$. Thus, high-temperature and low-temperature tunneling may be dominated by different charges, and both charges compete at intermediate temperatures. 
Similar physics was also proposed at $\nu=5/2$ in the presence of $1/f$ noise \cite{sassetti_5/2}. Recent data raise a question about this interpretation. 
The tunneling charge is usually extracted from the autocorrelation of the drain current. It is also possible to extract it from the cross-correlation of the currents in two drains 
(Fig. ~\ref{fig1:review}). It was observed \cite{glattli-pc} that the autocorrelation gives the effective charge that grows at low $T$, 
yet the charge extracted from the cross-correlation in the same sample remains at its high-temperature value.

\begin{figure}[!htb]
\bigskip
\centering\scalebox{0.25}[0.25]{\includegraphics{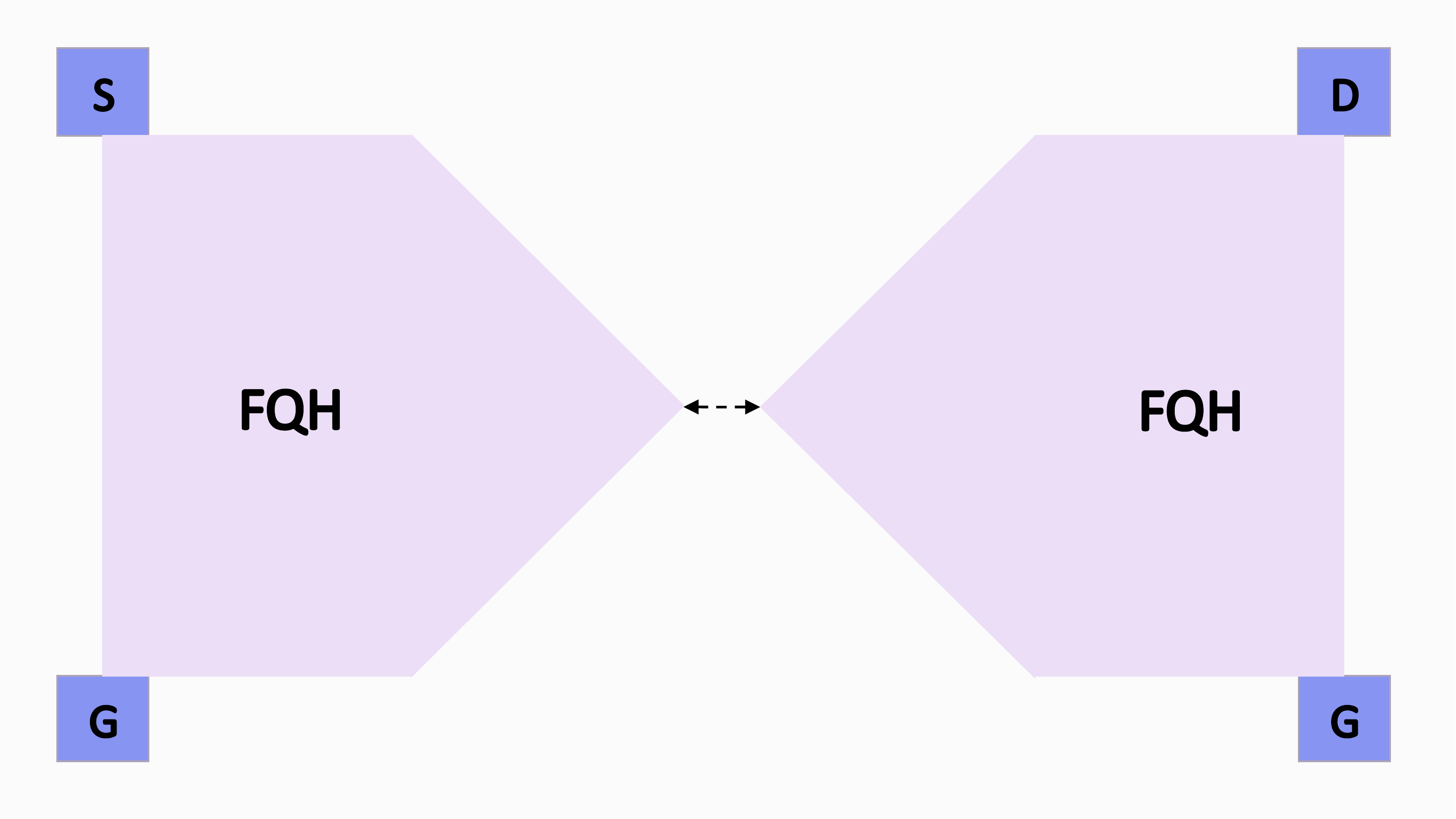}}
\caption{{The geometry with two FQH regions is dual to the  geometry from Fig.~\ref{fig1:review}.}}
\label{fig3:review}
\end{figure}

The above discussion focuses on weak tunneling since Eq. (\ref{Schottky}) holds only in that limit. Much interesting physics is observed at intermediate transmissions (see, e.g., Ref. \onlinecite{noise-formula}), but the Fano factor of the noise cannot be interpreted in terms of the tunneling charge in that case, which is thus beyond the scope of the review.
Theory predicts a rather boring picture at strong quasiparticle tunneling. This case is best understood in terms of the dual geometry (Fig.~\ref{fig3:review}), where electrons tunnel between two separate FQH liquids. The Schottky noise follows Eq. (\ref{Schottky}) with the electron charge in place of $q_m$. However, a surprise was found when a dilute beam of fractionally charged quasiparticles 
impinged on a weak link between two FQH liquids. The observed tunneling charge equaled a fractional quasiparticle charge \cite{dilute}. 
The standard theoretical toolbox sheds no light on this puzzling phenomenon \cite{KF-dilute}.

It should be noted that charge density in an electrostatically defined constriction is lower than than in the FQH bulk. This may affect the filling factor in the constriction and the nature of the quasiparticles whose tunneling is allowed. In particular, fractional tunneling charges were observed in the integer quantum Hall effect \cite{depleted-hashisaka}. 
Charge fractionalization on integer edges is also possible \cite{frac-int} due to purely edge physics that  is beyond the scope of this review.

\subsubsection{Photo-assisted shot noise}

This technique combines a dc bias $V$ with an ac bias $V_{\rm ac}\sim\cos(\omega t)$ in the geometry of Fig. ~\ref{fig1:review}.  A charge $q$, emitted from the source, acquires a time-dependent phase $\phi(t)=q\int dt V_{\rm ac}(t)/\hbar$.
This can be interpreted as a shift in the energy of tunneling quasiparticles. Without an ac bias, the available energy is $q_mV$. The absorption of $n$ quanta of the ac field shifts the energy to $q_m V+n\hbar\omega$. The observed shot noise \cite{PhysRevB.69.205302,safi2014,safi2019} is then a weighted sum 
of the noises at dc voltages $V+n\hbar\omega/q_m$,

\be
S=\sum_{n=-\infty}^{\infty} w_n S_{\rm dc} \left(V+\frac{n\hbar\omega}{q_m}\right),
\ee
where the weight $w_n$ reflects the probability to absorb $n$ quanta. The low-temperature
noise is singular at $V=\hbar\omega/q_m$. This was used \cite{josephson} to verify $q_m=e/3$ at $\nu=1/3$ and
$q_m=e/5$ at $\nu=2/5$.

In a related experiment \cite{Josephson2}, high frequency noise measurements at $f\approx 7$~GHz  with dc bias show $q=0.34e$ at $\nu=4/3$ and $q=0.38e$ at $\nu=2/3$.
{\color{black} See Ref. \onlinecite{PhysRevLett.118.076801} and references therein for related theoretical ideas.}

\subsection{Charging spectroscopy}

{\color{black}

In principle, the most direct way of measuring the charge of a quasiparticle is to form a weak potential well, using a local external gate, which is just strong enough to bind a single quasiparticle, and to measure the change in the electric charge in the region about the well when a quasiparticle is induced to enter or leave the trap. This could be done by varying the depth of the potential well,  by applying voltage to a contact which varies the electrochemical potential of the two-dimensional electron system, or by changing the magnetic field to vary the chemical potential of the surrounding FQH state. Alternatively, one could employ a potential well big enough to accommodate many quasiparticles,  and one could measure the jump in charge each time a new quasiparticle enters the well.  

In practice, however, an absolute measurement of the  local electric charge is difficult.  It is much easier  to measure  values of the varying gate voltage or other parameters where the quasiparticle enters,  and to calculate the quasiparticle charge based on the spacing between successive charge jumps.  Even if discontinuities  in the accumulated charge are largely smoothed out due to finite temperature effects or external noise, weakened sinusoidal oscillations in the accumulated charge may persist, and one could measure their period. Under proper conditions, the dominant factor determining the number of quasiparticles in a well will be a charging energy, which would be minimized when the accumulated charge $Q$ is as close as possible to a value $Q^*$ that varies continuously with parameters such as the gate voltage. If charges can only enter in units of the quasiparticle charge, $q_m$, then the spacing $\Delta Q^*$  between charge jumps will be equal to $q_m$.  Furthermore, if one can carry out  the same  experiment  in  the FQH state and an integer quantized Hall state, and one can be confident that the geometry of the well is the same in both cases,  then the value of $q_m/e$ is given by the ratio between  the periods of oscillation as a function of gate voltage in the  FQH and integer cases. 
}

\subsubsection{Measurements of tunneling through an antidot}

\begin{figure}[!htb]
\bigskip
\centering\scalebox{0.17}[0.2]{\includegraphics{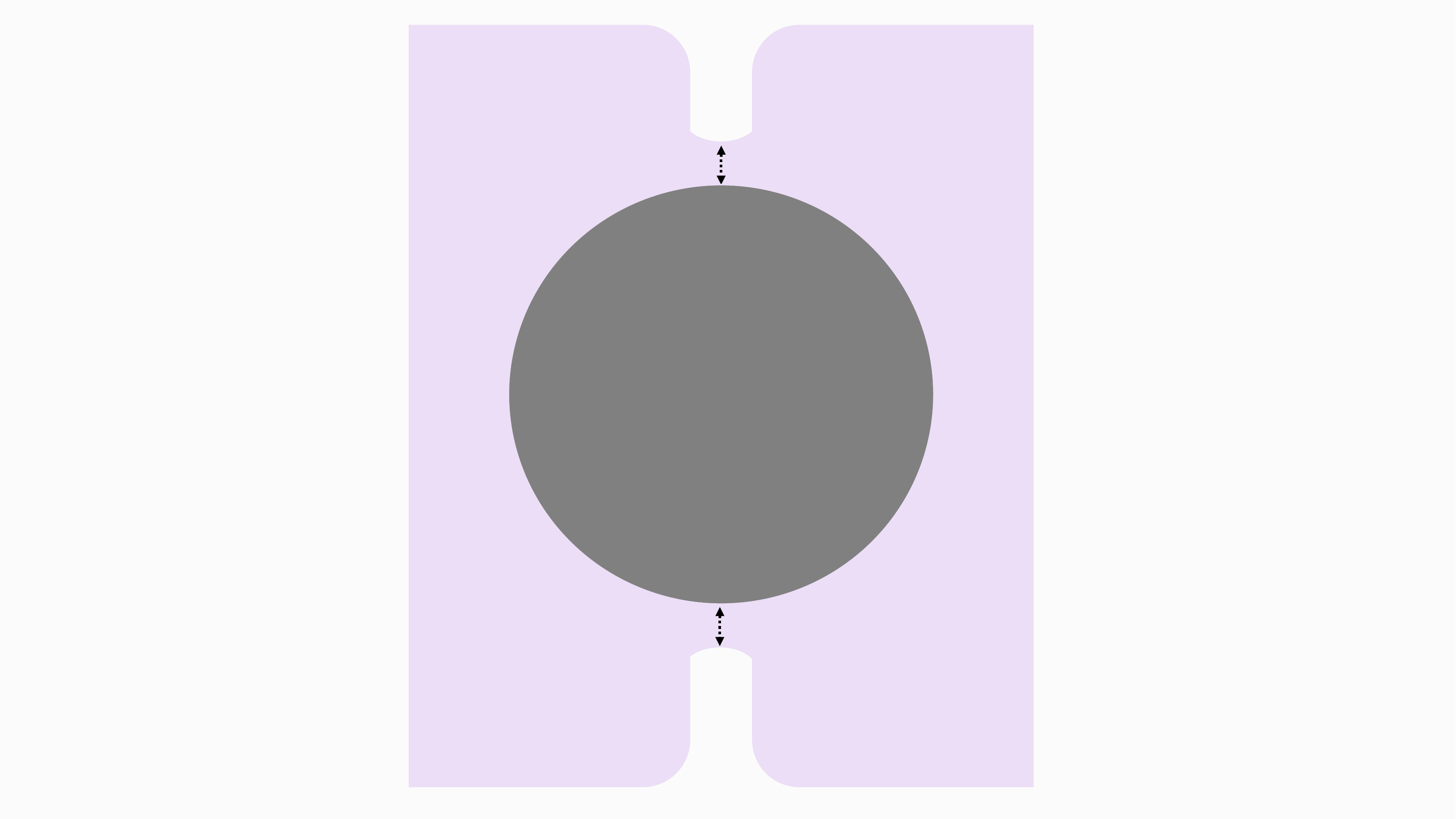}}
\caption{{Antidot geometry. Charge tunnels between the upper and lower edges through the edges of an antidot (grey disk) where the charge is depleted. Purple area is occupied by electrons in an FQH state. }}
\label{fig4:review}
\end{figure}

The earliest version \cite{goldman-su,GKLZ} of charging spectroscopy involved tunneling through an antidot inside a constriction between two FQH edges (Fig.~\ref{fig4:review}). 
This setting is related to that of  interferometry, addressed in the next section.
The electron gas is depleted inside the antidot and hence an FQH edge forms around it. 
Quasiparticles travel through the constriction by tunneling in and out of that edge. The size of the antidot is controlled by  the depleting gate voltage. The technique probes how the conductance through the antidot depends on the gate voltage and the magnetic field.

The dependence can be understood from the picture of quasiparticle orbits, introduced in Section \ref{meaning}.B.2. Changes of the magnetic field or the gate voltage result in an orbit periodically crossing the chemical potential. When this happens, a resonance is seen in
 the transmission through the constriction. The period in voltage corresponds to adding or subtracting a quasiparticle from the antidot. The quasiparticle charge can be found from the observed period in the voltage and the geometric capacitance. 
 The latter can be approximately extracted from the nominal area  of the antidot, and can be checked  with the magnetic field periodicity of the conductance. The calibration may be further checked by comparing with measurements in an integer quantized state.

{\color{black}To understand the magnetic field periodicity, 
it is necessary to take into account fractional statistics as well as fractional charge.}  
The allowed orbits are determined by the Bohr-Sommerfeld quantization rule in terms of the phase, accumulated by an anyon on a closed orbit. The phase has two contributions, 
First, there is an Aharonov-Bohm phase for a particle making a circle around the antidot. For an $e/3$ particle in the Laughlin state at $\nu=1/3$, the Aharonov-Bohm phase is $\phi_{\rm AB}= 2\pi \Phi /3\Phi_0 $, 
where $\Phi$ is the magnetic flux through the antidot and $\Phi_0=hc/e$ is the magnetic flux quantum.  In addition, each localized quasihole in the dot contributes the statistical phase $-2\pi/3$. 
Since a new quasihole is created or destroyed every time the flux changes by $\Phi_0$ at a fixed charge density, the total phase accumulated on an orbit is periodic with the period of a flux quantum. Thus, in relating the magnetic field period to the area of the antidot, one must take into account the fractional statistics as well as the fractional charge. 

The experimental results \cite{goldman-su} are consistent with the theoretically predicted charge $e/3$ at $\nu=1/3$. 
Yet, it was argued that the fractional charge is not the only way to understand the data \cite{alt-goldman-su}. 
One could instead start with a picture of single-electron orbits around the antidot and assume that electron correlations ensure that only $1/3$ of them are populated. 
Resonant transmission would still be observed when an orbit crosses the chemical potential. This predicts the same periodicity as the quasiparticle picture.
Besides, electrostatic effects may not be captured by the above single-particle picture (see the next section). 

An additional drawback of this technique is that in order for tunneling to occur, the antidot must be physically close to an edge of the sample.  Thus one might question whether the results are necessarily reflective of the properties of excitations in the bulk. One might also question whether the gate period obtained from a transport measurement necessarily reflects the period for charge occupancy of the antidot.

It should be noted that a similar technique \cite{marcus-antidot} showed excitations of charge $2e/3$ at $\nu=2/3$.
{\color{black} Very recently, charge $e/3$ was reported in an antidot tunneling experiment\cite{graph-antidot} in graphene at $\nu=\pm 1/3$.}

\subsubsection{Single-electron transistor technique}

Difficulties in the interpretation of the antidot data necessitated a different strategy.
Thus, later experiments used a different approach \cite{local1,local2}: A single electron transistor (SET),   which is sensitive to variations in the  local electrostatic {\color{black} potential,
was} placed on the surface of the heterostructure that embeds the FQH {\color{black}   liquid, or on }
a scanning tip just above the surface. The SET was used to detect potential jumps when a quasiparticle or quasihole enters or leaves a local potential well created by fluctuations in the doping density, which were found to have a physical scale on the order of 200 nm, large compared to the magnetic length.

The SET technique was first developed \cite{set1} for 2D heterostructures outside the quantum Hall regime and allowed the spatial resolution of 100 nm. It was extended to studies  of the integer quantum Hall effect in Refs. 
\onlinecite{set2,set3}.

In a subsequent development, fractional charges in FQH liquids were reported in Refs. \onlinecite{local1,local2}. 
As was explained above, the entry of new quasiparticles into a well can be controlled with the gate voltage, and the quasiparticle charge, relative to that of an electron in an integer quantized state,  can be extracted from the spacing of the jumps as a function of the voltage,  The experimental results \cite{local1} are consistent
with charge-$e/3$ excitations at $\nu=1/3$ and $\nu=2/3$. The absolute value of the quasiparticle charge could also be extracted, with lesser accuracy, from the SET measurements and were consistent with the value $e/3$. In the second Landau level, comparison between measurements at $\nu=5/2$ and $\nu=7/3$ obtained the ratio
  \cite{local2} $q_{m,5/2}/q_{m,7/3}=3/4$ in agreement with the theoretical expectation that the charges should be $e/4$ and $e/3$ in the two cases.
 For a detailed theoretical discussion at $\nu=5/2$, see Ref. \onlinecite{charge-spec-theor}.

\section{Experimental probes of fractional statistics} \label{stat-expts} 

Fractional statistics were defined in terms of phases accumulated by anyons exchanging their positions or running around other anyons. This makes interferometry \cite{interferometry} 
the most direct probe of statistics, since 
that technique is directly sensitive to phase differences accumulated by particles on different possible paths between the same endpoints, which could depend on whether  the difference in paths encloses some other quasiparticles. 

\begin{figure}[!htb]
\bigskip
\centering\scalebox{0.25}[0.25]{\includegraphics{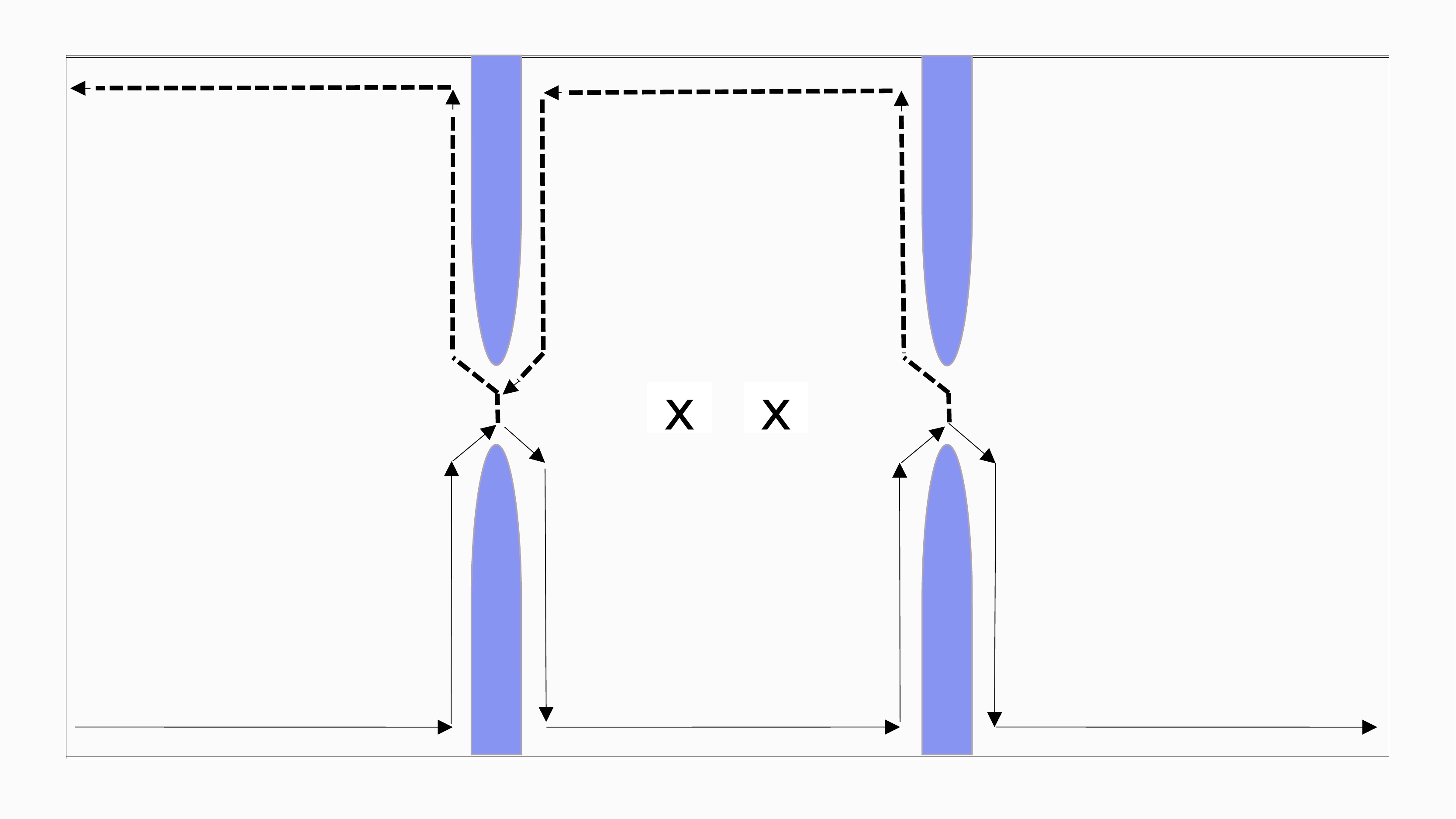}}
\caption{{Ideal Fabry-Perot interferometer. Current tunnels between the lower and upper edges at two tunneling contacts. Two crosses show localized quasiparticles inside the interference loop.}}
\label{fig5:review}
\end{figure}

The simplest Fabry-Perot geometry \cite{interferometry} is illustrated in Fig.~\ref{fig5:review}. In the illustrated ideal  case, the bulk of the system is in an almost perfect  quantum Hall state, where the Fermi level falls inside an energy gap of the pure system, but there are  a small number of localized states  inside the gap, due to impurities, which may become occupied or empty as the Fermi level is varied inside the gap. We have also assumed that there is only one chiral mode at the sample boundaries, carrying quasiparticles in the direction shown by the arrows, and that tunneling between opposite edges can take place only in the two constrictions.
In the weak tunneling limit, two paths connect the source in the lower left corner and the drain in the upper left corner. Their phase difference combines an Aharonov-Bohm phase in the external magnetic field with a
statistical phase. Thus, the technique allows probing both fractional charge and statistics. Several other geometries have been proposed,  with Mach-Zehnder 
interferometry \cite{MZ-heiblum}
attracting particular interest. 
Experimental implementation proved difficult for all geometries, but recent years have brought promising results \cite{willett08,willett10,manfra19,manfra20}
in the Fabry-Perot  approach, and we restrict ourselves to  that geometry in the current  Section.

Very recently, a somewhat less direct observation of fractional statistics was accomplished with an anyon collider \cite{collider,collider-exp}, which will be discussed in Section \ref{stat-expts}.B., below. 
Mach-Zehnder interferometry will be discussed in Section \ref{MZ}.
Several other techniques yield information about statistics, which we  briefly address in Section~\ref{other}, with the emphasis on thermal transport and tunneling experiments. 
Interferometer experiments designed to reveal effects of non-Abelian statistics near filling factors $\nu = 5/2$ and 7/2 will be discussed in Section \ref{fpi-2}.

As we shall see below, there are   several major challenges to  the interpretation of interferometry data.  Some major complications arise due to effects of Coulomb interaction. The situation  also  becomes more complicated in states with more than one propagating edge mode.  Most significantly, in most  experiments, the region inside the interferometer is not in the ideal 
quantum Hall state described above, which we {\color{black} refer to as} an {\it {incompressible}} state.  Rather, the bulk is typically in a {\it{compressible}} state, where the Fermi level does not fall inside an energy gap of the pure system.\cite{CSG,CSG-erratum}   In this case, there will be a large density of {\color{black} localized } quasiparticles or quasiholes present in equilibrium, and it costs relatively little energy to add or subtract one quasiparticle. The result, after thermal fluctuations are taken into account, is that interference patterns tend to fall into one of two categories, which are generally described as {\it {Aharonov-Bohm}} (AB) and {\it{Coulomb-Dominated}} (CD), as, at least for integer quantum Hall states, the difference between the two states is determined by the importance of  Coulomb interactions between charges in the bulk and charges on the interferometer edge, relative to an energy scale set by characteristics of the edge.  \cite{coulomb07,coulomb11}.  For FQH states, these labels may be somewhat of a misnomer, as one predicts in some cases that behavior of the Coulomb-Dominated type may be found in the compressible regime, even when the interaction between bulk and edge is very weak. {\color{black}To emphasize this point we will sometimes use ``CD-like'' in place of ``CD''}.  It should also be emphasized that the   behavior of an FQH state in an incompressible regime is different than  either the Aharonov-Bohm or Coulomb-Dominated  behavior in the compressible regime, {\color{black} as will be discussed below. }

\subsection{Fabry-Perot Interferometry}

\subsubsection{The ideal case}

We shall first consider the ideal case described above. We assume that the electron density drops rather sharply to zero  at the boundary, and there is just a single edge mode propagating along the boundary, as indicated in Fig.~\ref{fig5:review}.  A quasiparticle propagating on the interferometer boundary may then 
 be described by  a Hamiltonian of the form
\begin{eqnarray}
\label{edge-H}
{\hat H}=\hat H_{\rm edge}\nonumber\\
+ [\Gamma_1\exp(i\phi_1)\hat T_1+\Gamma_2\exp(i\phi_2)\hat T_2 +h.c.],
\end{eqnarray}
where  $\hat H_{\rm edge}$ describes charge propagation on the upper and lower edges,
the operators $\hat T_{1,2}$ move a quasiparticle of charge $q_m$ from the lower edge to the upper edge at the two constrictions, and  $\Gamma_{1,2}\exp(i\phi_{1,2})$ are the associated tunneling amplitudes.  We shall focus on the limit of weak tunneling, where the current between the lower and upper edges is much less than the incoming current $\nu e^2 V / h$, where $V$ is the voltage difference between the lower and upper edges. The quasiparticle tunneling rate can then be extracted from Fermi's golden rule,
\be 
\label{int-p}
p=[\Gamma_1^2+\Gamma_2^2]r_0(V,T)+2\Gamma_1\Gamma_2\cos(\phi_1-\phi_2)r_1(V,T),
\ee
 where $r_0$ and $r_1$ depend on  $V$ and the temperature $T$.   Hence the tunneling current between the lower and upper edges of the interferometer will be given by 
\begin{eqnarray}
\label{int-I}
I_t =q_m[\Gamma_1^2+\Gamma_2^2]r_0(V,T)\nonumber\\
+2q_m\Gamma_1\Gamma_2\cos(\phi_1-\phi_2)r_1(V,T).
\end{eqnarray}
To realize  the experimental configuration  illustrated in Figure 5, one can connect a current source at voltage $V$ to the lower left corner, and attach grounded contacts to the lower right and  and upper left corners.  The back-scattered current $I_t$ will then be equal to the  current flowing into  the upper left contact.

The phase difference $\theta=\phi_1-\phi_2=\alpha-\phi_{AB}-\phi_s$  combines two key pieces of information: the quasiparticle charge $q_m$ through the Aharonov-Bohm phase $\phi_{AB} = -2\pi q_m\Phi/e\Phi_0$ and the statistical phase $\phi_s$ accumulated by a quasiparticle on the trajectory around the anyons, trapped inside the interferometer.
Here $\Phi = BA$ is the total magnetic flux through the area $A$ enclosed by the paths of the edge states between the constrictions.    We ignore any additional slow dependence of the matrix elements $r_{0,1}$ on the magnetic field. The constant $\alpha$ will be set to zero without the loss of generality.

{\color{black}Although the above equations can be justified relatively easily in the case of a single edge mode in an ideal system with a sharp edge, there are subtleties involved in applications to a real  system, with a continuously varying electron density and at least some disorder near the edge.  It has been argued that in this case there will still be a discrete set of propagating modes at the boundary, which will be embedded in a region of weakly localized states but will still have 
a well-define phase rotation $e^{i \theta}$ along the edge \cite{coulomb11}. 
Since the edge state will have a finite width, at least as large as the magnetic length, there will clearly be some ambiguity as to the physical area $A$ enclosed by the state.  However, it is assumed that this ambiguity can be resolved in such a way that $\phi_{AB}$ is precisely given by the expression above.}

To get access to the information encoded in $\theta$, an experimentalist needs to look for
 oscillations in the current as one varies the magnetic field and/or the area of the interferometer.  The area may be varied by applying a voltage $V_g$ to external gates along the sides of the device.  We assume that the area does not depend on the magnetic field. 
 This assumption is not crucial for the interpretation of the data in the incompressible regime. We will lift it in the discussion of the non-ideal compressible case.

 If  no quasiparticles enter or leave the interference region,  contours of constant $ \theta $ should lie along lines in the plane of ${\color{black}|}B{\color{black}|}$ and $V_g$  with slope 
\be
\frac {dV} {dB } = - \frac {A} {  B    \,  (\partial A/ \partial V_g )} .
\ee
In the case of $\nu=1/3$, the spacing $\Delta B$  between successive conductivity maxima at fixed $V_g$ should equal $3 \Phi_0 / A$, while the spacing $\Delta {V_g}$ at fixed $B$ should correspond to  an area change that contains one electron. 
At certain values of the parameters, however, it may be favorable for a quasiparticle to enter or leave a localized impurity state in the interferometer, at which point we would expect a jump in  phase by an amount equal to $\pm 2 \theta_m$, caused by a change in the value of $\phi_s$.  For a quasiparticle in the Laughlin state at $\nu=1/3$, it is predicted that  $2 \theta_m =2 \pi /3$.

 \begin{figure}[!htb]
\bigskip
\centering\scalebox{0.25}[0.25]
{\includegraphics{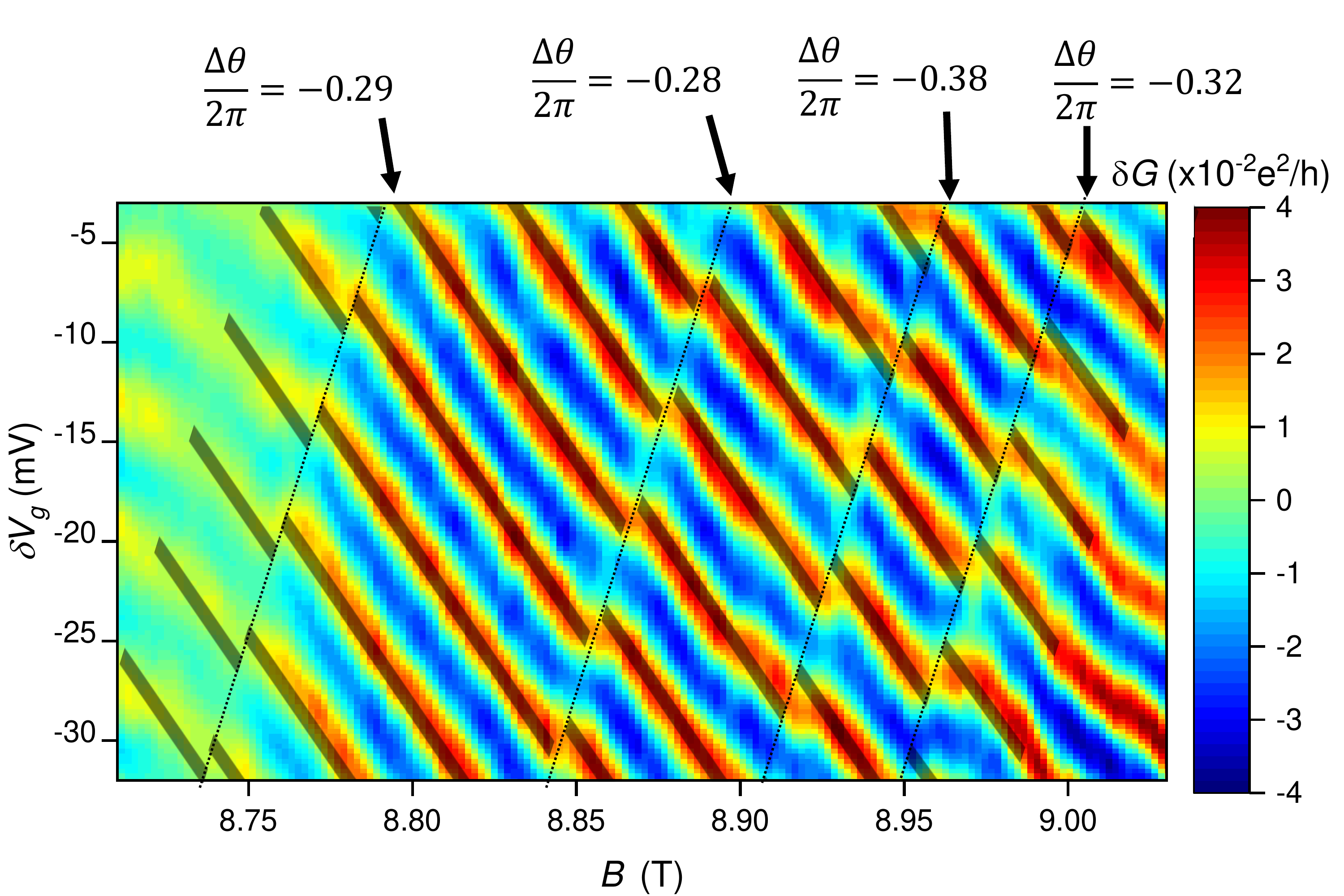}}
\caption{{From Ref. \onlinecite{manfra20}. Conductance through the Fabry-Perot interferometer oscillates when the magnetic field or the side-gate voltage changes.
The Aharonov-Bohm-type behavior is combined with phase jumps $\frac{\Delta\theta}{2\pi}$ when anyons enter the interferometer. Gray lines and dashed lines are guides to the eye. }}
\label{fig-manfra:review}
\end{figure}

Behavior of this type  was observed in recent experiments\cite{manfra20}  by Nakamura, et al.,
 as shown in Fig. 6.
It is possible, of course,  to ask whether the reported  phase jumps could have been caused by some effect other than fractional statistics, such as Coulomb interactions between localized quasiparticles and the conducting  states at the interferometer edges, which could cause a jump in the area enclosed by the interfering trajectories. (See discussion in Subsection \ref{compressible}.) However, the sample in these experiments had nearby conducting planes designed to screen  Coulomb interactions as much as possible. Moreover, it would be peculiar if  phase jumps caused by residual Coulomb interactions would all have the same size for  quasiparticle states localized at  different  impurity positions in the sample, and that these phase jumps just happened to be  close to the value predicted  by theory.   The alternate possibilities should be further checked and hopefully ruled out by additional experiments, but assuming that the interpretation is correct, the results of Ref.~\onlinecite{manfra20} provide as direct a demonstration as one could imagine of fractional statistics and a measurement of  the statistical phase of quasiparticles in the $\nu=1/3$ FQH state.

{\color{black}
 In order to prepare a sample where one could enter the ideal incompressible regime and still see  Aharonov-Bohm oscillations, the authors of Ref.  \onlinecite{manfra20} had to overcome major difficulties. The challenge
comes from conflicting demands on the interferometer size imposed by weak  Coulomb interaction and strong phase coherence. The interaction can be suppressed in a large interferometer, but 
phase coherence is favored by a small device size.  A key improvement, described  in Refs. \onlinecite{manfra19} and \onlinecite{manfra20} 
came from introducing  
ancillary wells that screen Coulomb forces in the heterostructure.

It is important to note that the simple results shown in Fig.~\ref{fig-manfra:review} were only observed  over a limited range of magnetic field. This is to be expected because outside a certain range, the Fermi level will no longer be inside the energy gap of the ideal FQH state.  In that case, we can expect that the sample would fall into the compressible regime described above, where there will be a large number of quasiparticles or quasiholes inside the interferometer,  with only a small energy barrier to add or subtract an additional quasiparticle\cite{SR-manfra}.

}

\subsubsection {Interferometer with a compressible bulk} \label{compressible}

We present here a brief summary of our current theoretical expectations for the behavior of a Fabry-Perot quantum Hall interferometer in  the compressible situation, which will apply to the integer quantum Hall regime  as well as to  FQH systems.  Although these theoretical predictions have been confirmed in a variety of experiments in the integer regime, the reader should be warned that there has been little success so far in {\color{black} observing oscillations of the predicted type in FQH states.}

 \begin{figure}[!htb]
\bigskip
\centering\scalebox{0.25}[0.25]{\includegraphics{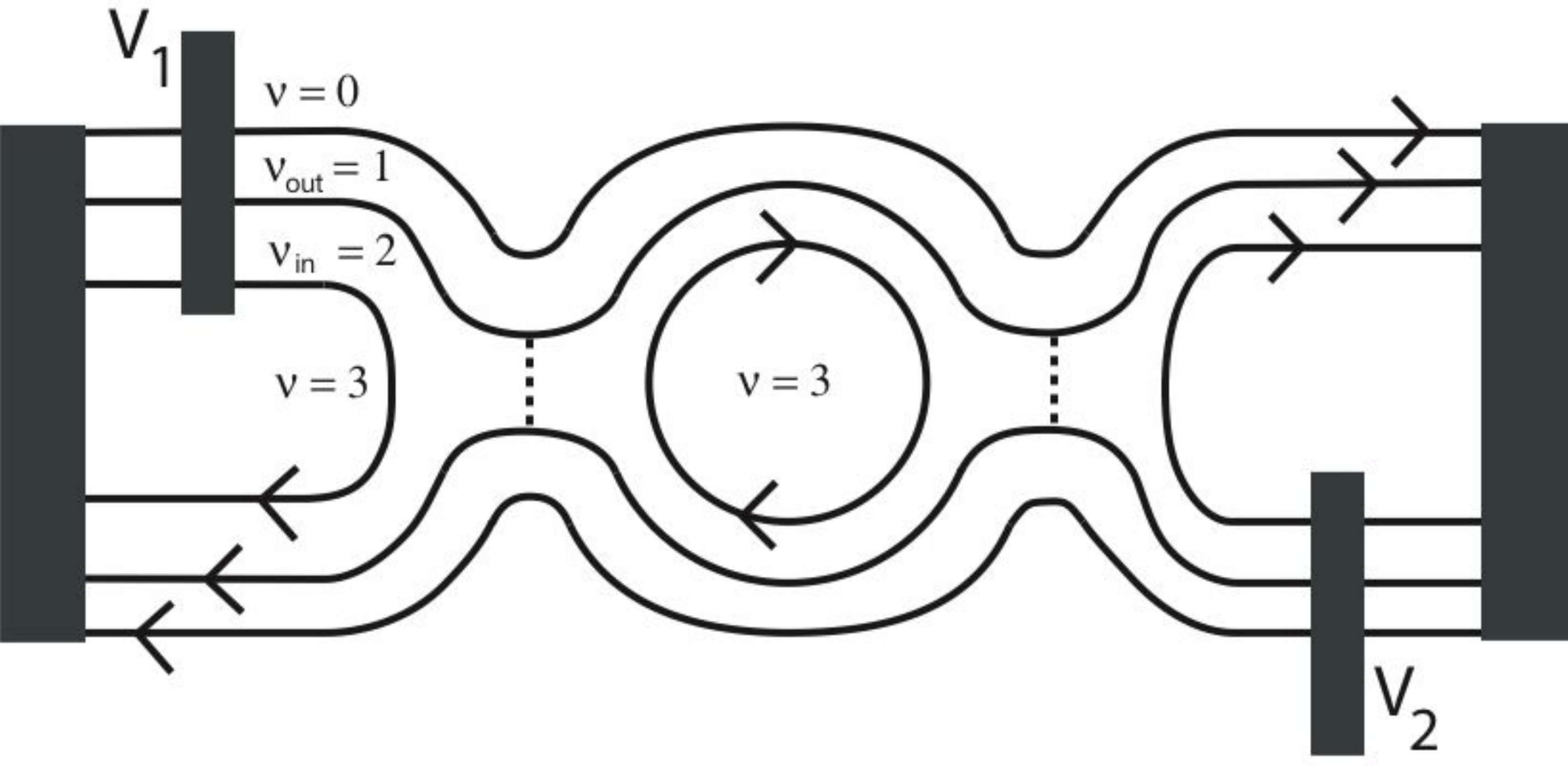}} 
\caption{{From Ref. \onlinecite{coulomb11}. Fabry-Perot interferometer in the integer quantum Hall regime. The channel that separates $\nu=0$ from $\nu=1$ is fully transmitted.
The channel that separates $\nu=2$ from $\nu=3$ is fully reflected. The channel that separates $\nu=1$ from $\nu=2$ is partially reflected in the constrictions, where the charge density is lower than in the center of the device. {\color{black}  Dotted lines show tunneling across the constrictions, applicable to the case of weak backscattering.   Arrow heads, which indicate the  directions of particle propagation, are shown here for the case of a magnetic field pointing {\color{black} towards} the viewer. } }}
\label{fig-interferometer:review}
\end{figure}

Consider the example from the integer quantized Hall regime, illustrated  in Fig. 7.  The bulk of the system is in a  state with a  quantized Hall conductance given by $\nu=3$. The LL filling factor $f$ near  the center of the sample could be anywhere in the range $2.5 < f < 3.5$; deviations from the ideal value of $f=3$ are accommodated by a finite density of localized electrons in regions with $f>3$ and localized  holes  for $f< 3$.  Near edges of the sample, the electron density drops to zero, and  $f$ drops off accordingly. In our somewhat simplified model, we assume that there are quantized Hall strips with quantum numbers $\nu = 2, 1$ and 0 in the edge region, with a single propagating edge mode separating each region.  The locations of the propagating modes should fall roughly where the electron density is such that the local Landau-level filling factor is 2.5, 1.5, or 0.5. 
{\color{black}For situation illustrated in the figure,  the density in the constriction is supposed to be slightly less than $f=2$.} 
The edge state separating $\nu=2$ and $\nu=3$ is totally reflected outside the constriction, the edge state separating $\nu=0$ and $\nu=1$ is totally transmitted, while the edge state separating $\nu = 1$ and $\nu = 2$ is partially transmitted.  It is this last mode that is  relevant in an interference experiment. 

The situation in Fig. 7 can readily be extended to FQH states. For example, if we interpret the labeled filling factors as  effective fillings for composite fermions, with two flux quanta attached {\color{black} to each electron,} the Hall states become quantized  states with $\nu = 1/3, 2/5$ and $3/7$.   More generally, we shall assume that there is a single partially-transmitted edge state, which separates inner and outer regions with quantum numbers $\nin$ and $\nout$, with $\nin > \nout$
We shall assume that the  tunneling  processes occur at  one well-defined point within each constriction, and we shall define the interference area $A_I$  as the area enclosed by the interfering edge state between these points. We also define $\qin$ and $\qout$ as the charges of the fundamental  quasiparticles in quantized Hall states with $\nin$ and $\nout$.   For the fractional case, we shall confine our discussion to the situation where $\nin$ corresponds to a Jain state in the bottom half of the lowest LL, so we may write
\be
\label{jainstates}
\nin = \frac {p} {2ps+1} , \,\,\,\,\,\,\qin= \frac {e} {2ps+1} , 
\ee
where $p$ and $s$ are positive integers. The values of $\nout$ and $\qout$ are obtained by replacing $p$ by $p-1$ in the above formulas. The situation in Fig. 7 corresponds to $p=2.$

Next we define $N_L$ as the net number of  quasiparticles of charge ${\color{black}\qin}$ inside the area  $A_I$. The number $N_L$   takes into account the total excess charge in area, including charges  in regions with $\nu > \nin$ as well as positive or negative quasiparticles localized at density inhomogeneities within the $\nin$ region.  Specifically, it is related to the  total electric charge $Q$  inside the interference area $A_I$  by 
\be  
Q = N_L \qin {\color{black} - } A_I \nin e B / \Phi_0 
\ee
If we assume that the dominant tunneling processes at the constrictions involve quasiparticles with charge $\qin$, then the interference phase seen by the tunneling particles will be given by 
\be
\theta = -2 N_L \theta_{\rm{in}} {\color{black} +} 2 \pi B A_I \qin / e \Phi_0 ,
\ee
where
{\color{black}
 $\theta_{\rm{in}}  $ is the statistical phase associated with the quasiparticles of charge $\qin$, given by \cite{Wen95}  
 \be
 \theta_{\rm{in}}  = \pi \left[ 1 - \frac {2s}{2ps+1} \right]  . 
 \ee 
}
We remind the reader that our sign conventions assume that the field points along the $-z$ direction, and $B=|{\bf B}|$ is the magnitude of the field. 
The phase  $\theta_{\rm{in}}$  would have had the opposite sign  if the magnetic field had been chosen to be in the positive $z$ direction.
Note  that the statistical phase is the same for a particle and its antiparticle.

A key assumption is that $N_L$ is restricted to take on integer values (positive or negative), because the localized states inside the interfering edge are isolated from the states outside and from the edge itself. \cite{coulomb11}   This does not mean that $N_L $ is frozen in time, only  that it is constant on the time scale for a quasiparticle in the edge state to move along the length of the interferometer. We assume that on the longer laboratory time scale, charges can hop readily from one localized state to another and that occupations will take on  an equilibrium distribution determined 
by the temperature, the magnetic field, and any voltages applied to the gates and the current contacts. From this point of view, the entire region inside interfering edge state, as well as the region surrounding the edge state, should be considered as compressible in most cases.\cite{CSG,CSG-erratum}  

In contrast, the charge on the edge state can vary rapidly, because it  is connected directly to the edge states outside the interferometer, and  we consider here a situation where the backscattering probabilities at the constrictions are small. Thus, {\color{black} the edge charge is not quantized, and the area $A_I$, related to  it  by Eq. (\ref{dni}) below, may be considered to be a continuous variable.  }

We now define an energy function $E(N_L, A_I)$, which describes the free energy of the system after all other variables have been integrated out.\cite{coulomb11}  
We assume that the time-average interference current measured in an experiment is proportional to the thermodynamic average of $  {\rm{Re}} \, ( e^ {i \theta } )$, weighted by the factor $e^{- E(N_L,A_I) / T} $.   

It is convenient to introduce {\color{black} another variable $\delta n_L$, so that we can write  }
\be
\label{dni} 
\delta n_I \equiv    {\color{black} - } (\nin - \nout) \,  B (A_I - \bar{A} )/ \Phi_0 ,
\ee
\be
\delta n_L \equiv N_L  \frac {\qin}{e}    {\color{black} -} \nin \frac {B \bar{A} }{\Phi_0} -  \bar Q ,
\ee
where $\bar A$ and $\bar Q$ are quantities chosen such that $\delta n_L$ and $\delta n_I$ would be zero if we were to minimize $E$ without the constraint that $N_L$ be an integer. The values of $\bar A$ and $\bar Q$ should  be smooth monotonic functions of any  applied gate voltages, with perhaps a weak smooth dependence on $B$. The variable $\delta n_I$ describes charge fluctuations on the interfering edge, while ${\color{black}\delta n_L} $ is  determined by fluctuations in the interior.
Then for small fluctuations in the variables, we can expand $E$ in the form
\be
\label{interferometer-Coulomb-energy}
E=\frac{K_L}{2}\delta n_L^2 + \frac{K_I}{2}\delta n_I^2+K_{IL}\delta n_L\delta n_I,
\ee
where the constants $K_L, K_I$ and $K_{IL}$ depend on the geometry and are largely determined by the Coulomb interactions between charges.  
{\color{black} Eq. (\ref{interferometer-Coulomb-energy}) contains only the effect of long-range Coulomb forces and no contribution from the quasiparticle gap since the random potential creates an essentially continuous spectrum for anyons.}

At $T=0$, there will be no thermal fluctuations, and the phase factor $e^{i \theta}$ will exhibit jumps at 
discrete values of the parameters, where $N_L$ increases or decreases by one.  At finite temperatures, fluctuations become important, and one rapidly enters a regime where one or two  Fourier components are  dominant in a plot of the thermal expectation value $\langle e^ {i \theta} \rangle$ as a function of the parameters  $B$ and $V_g$.  The allowed contributions have the form  \cite{coulomb11} 
\be
\label{Dm}
 D_m \exp \left\{ 2 \pi i  \left[ m \left(\frac{B \bar A}{\Phi_0} \right)
- \frac {\bar Q (\qin  -m e) }{ e\nin }  \right]              \right\}, 
\ee 
where $m$  are integers restricted to values of form 
\be
\label{mvalues}
m = - \frac {\nout e}{\qout } + g \frac{\nin e}{\qin} ,
\ee
where $g$ is an integer.  
 The amplitudes $D_m$ fall off exponentially with temperature, $|D_m| \propto \exp ( - 2 \pi^2 T / E_m)$, 
 so typically only  the component with {\color{black} largest}  $E_m$ is visible.  An explicit expression for $E_m$ in terms of the parameters of the model is given by Eqs. (20) and (27) of 
 Ref.~\onlinecite{coulomb11}. 
 {\color{black} 
 According to  those formulas, the largest  value of $E_m$ occurs when $g$ is the closest integer to $- \Delta \theta / 2 \pi$, where 
 \be
 \Delta \theta = - 2 \theta_{\rm{in}} + \frac {2 \pi \qin ^2} {e^2 (\nin-\nout)} \frac {K_{IL}} {K_I} 
 \nn
 \ee
 \be
   =  2 \pi \left( \frac{2s \qin}{e} -1 + \frac{\qin}{\qout} \frac{K_{IL}} {K_I} \right) 
 \label{jump}
 \ee 
 is the jump in interferometer phase that would occur  if $N_L$ is increased by one at $T=0.$  The favored value of $g$ corresponds to the Fourier component of the interference oscillations that is least sensitive to thermal fluctuations in $N_L$ and $A_I$  at higher temperatures. 
 
 The case $g=1$ has been termed the Aharonov-Bohm or AB regime, while the case $g=0$ has been termed the Coulomb-Dominated or  CD regime.  For integer quantum Hall states, where $s=0$, the AB regime occurs  when   $K_{IL} / K_I < 1/2 $, so that the coupling between edge and bulk is relatively weak, while the CD regime occurs for  $1/2 < K_{IL} / K_I < 3/2$, where the coupling is relatively strong.  For a fractional state of the form (\ref{jainstates}),   there will  again be a CD-like  regime with 
$g=0$,  and at least  in principle, an AB  regime with $g=1$.   However,  the value of  $K_{IL} / K_I $  separating the two regimes will be $<1/2$, and the AB regime may be difficult to access.  In fact, for the Laughlin states, with $p=1$ and $s \geq 1$, one is in the CD-like regime, with $g=0$, even for $K_{IL} = 0$, so the traditional CD designation is actually a misnomer in this case. [To reach the AB regime at $\nin=1/3$, one would actually need an  attractive interaction between the edge mode and localized charges, with $-7/2 <   (K_{IL} / K_I ) <  -1/2 $.]
 For  integer states and for Jain states of the form  (\ref{jainstates}), the dominant term in the AB region  has $m=1$, while in the $g=0$ CD-like region, it has     
$m=1- (\nin e / \qin)=1-p $,
 
 }

According to (\ref{Dm}), if the gate voltage is held fixed, and if one can assume that $\bar A$ and $\bar Q$ are insensitive to the magnetic field, then the oscillations in conductance should have a period  in the magnetic field given by $\Delta B = \Phi_0 /  |m|  A$.  In the AB regime, where $m=1$, the flux period is $\Phi_0$ for all the states under consideration. By contrast, in the CD regime, the period depends on the state, and it will be a submultiple of  $\Phi_0$ for states  where there are  two or more fully transmitted edge states, such as $\nin = 3$ or $\nin=3/7$,

If one fixes $B$ and varies $V_g$, one will generally see an oscillating conductance with a period that will depend on  $d \bar A/ dV_g$ and $d \bar Q / d V_g$.   A color plot of the conductance oscillations as a function of $B$ and $V_g$ will lead to a series of parallel stripes, similar to those seen in Fig. 6.   It was argued in Ref. \onlinecite{coulomb11} that at least for integer quantized Hall states,  lines of equal phase should have a negative slope, similar to the stripes in Fig. 6, in the AB regime, but they  should have a positive slope in the CD regime, provided there is at least one fully transmitted edge mode.  Fabry-Perot experiments in the integer quantized regime have  seen both types of behavior, depending on the details of the sample.\cite{Zhang09,AB-heiblum}  Also, in certain samples, the two types of stripes were seen to coexist, leading to a checkerboard pattern of diamond shapes in the color plot.
However, the situation is more complicated in the FQH case.  If we define the measured phase $\tilde {\theta}$ as  ${\rm arg}(\langle e^{i\theta} \rangle) $, then   
following Eq. (\ref{Dm}), $\partial  \tilde {\theta} / \partial B$ will again have the same sign as $m$.  However, for FQH states, the sign of $\partial  \tilde {\theta} / \partial V_g$ can depend on microscopic details.

For the case of $\nin =1/3$ and $\nout=0$, where there are no fully-transmitted edge modes, one has $m=0$ in the {\color{black}CD-like} regime, as noted above.  Then the conductance will not show oscillations as the magnetic field is varied, and stripes of equal phase will be horizontal in the color plot. Behavior of this type was indeed observed in the experiments reported in 
Ref.~[\onlinecite{manfra20}] for magnetic fields on outside of the range shown in Fig. ~\ref{fig-manfra:review}. We remark that this  result differs from the {\color{black} original} prediction of Ref.~\onlinecite{SR-manfra} that there should be a flux period of $\Phi_0$ in this region,  as one would expect in the AB compressible regime; {\color{black} however,  that prediction has now been corrected. }

Note that  in the compressible domain,   FQH states in a higher Landau level will have different flux periods than the corresponding states in the spin-polarized lowest Landau level. For example \cite{coulomb11}, for a state at $\nu=7/3$, which we assume  to consist of a Laughlin liquid at $\nu=1/3$  on top of an integer quantized Hall state with $\nu=2$, we would have $\nin=7/3$ and $\nout=2$, so the allowed values of $m$ in (\ref{Dm}) will be equal to 
{\color{black}  $ -2 +7 g$. As at $\nin=1/3$, we expect that the dominant term should have $g=0$.   
Experimental results at $\nu=7/3$  by Willett and collaborators\cite{willett13} showed } 
a flux period $\Phi_0/2$,  consistent with predictions for the compressible regime with $m = -2$. However, the dependence on gate voltage was not reported in this reference.     

In another experiment at $\nu=7/3$, An and collaborators \cite{an11}  reported a gate period with phase jumps, appearing in the form of telegraph noise, which was consistent, at least qualitatively, with what one might expect for a state in the incompressible regime.  However, the flux period was not reported at this filling fraction, nor was there a reported calibration of the amount of charge entering the interferometer in one gate period.

Motivated by the experiments of Ref.~\onlinecite{marcus-antidot},  Schreier {\it{et al.}}\cite{schreier16} have analyzed interference effects to be expected in a geometry where there is tunneling through an antidot inside a constriction.  In particular, they considered a situation where there are two edge states around the antidot, and they found that the system was likely to be in an AB regime for an FQH state for the same geometry where one would observe CD behavior in the integer case. They advanced this as an explanation for the different behaviors observed in Ref.~\onlinecite{marcus-antidot} between bulk filling factors  $\nu=2/3$ and $\nu=2$.  
    
 It should be cautioned that our discussion of the Fabry-Perot interferometer  ignored the possible effects of tunneling between different edge modes along the perimeter of the interferometer.  While this has been justified by experiments in the integer regime in many cases, it may be more questionable for FQH states, particularly when there are edge modes propagating in two directions.  Inter-mode scattering may contribute to decoherence effects, which may be a reason why interference oscillations have proved much more difficult to observe for FQH states than for integer states.

The analyses which led to the results described above, for both the AB  and CD regimes in the FQH case with a compressible bulk, certainly made use of the property of fractional statistics. More generally, if one accepts that the interfering particles have fractional charge, then one needs to invoke fractional statistics to avoid flux periods which are integer multiples of  $\Phi_0$. However, the flux periods predicted above  for the Jain states are identical to the ones predicted for integer states in both the AB and CD  regimes, where the tunneling particles are electrons.  Consequently, it might be hard to rule out the possibility that the interfering particles in an experiment{\color{black}\cite{GLZ}} are electrons rather than fractionally charged quasiparticles. For this reason, observations of the predicted flux period in either regime might not be accepted as a convincing direct  observation of fractional statistics.

\subsubsection{Quasiparticle charges from Fabry-Perot experiments}  \label{fpicharge} 

Measurements using the Fabry-Perot geometry can be used to measure the charges of quasiparticles in various quantized Hall states in {\color{black} either the CD or AB regime.}  Using expression (\ref{Dm}), if the values of  $d \bar A/ dV_g$ and $d \bar Q / d V_g$ are known, one can predict the  oscillation period $\Delta V_g$ when $B$ is held fixed.  
In the CD regime, this period corresponds to the addition of  a charge equal to $\qout$ 
to the interior of the interferometer.  

Importantly, although (\ref{Dm}) was derived in the regime of weak backscattering, the same result obtains, for a given partially-transmitted edge state,  in the regime of strong backscattering, where the partially transmitted edge state is almost totally reflected at the constrictions, and there is only weak forward scattering.   (See Fig. 8a.)
In  this limit, the area enclosed by the interfering edge forms an isolated droplet of material in a quantum Hall state  with quantum number $\nin$, embedded in a region with quantum number $\nout$.  Charge can then enter or leave the droplet only in units of $\qout$, and the total charge in the droplet must be an integral multiple of this unit. If $V_g$ is varied, periodic oscillations will  occur in the amplitude for forward tunneling through the constriction, as the quantized charges enter or leave the droplet.

\begin{figure}[!htb]
\bigskip
\centering\scalebox{0.3}[0.3]{\includegraphics{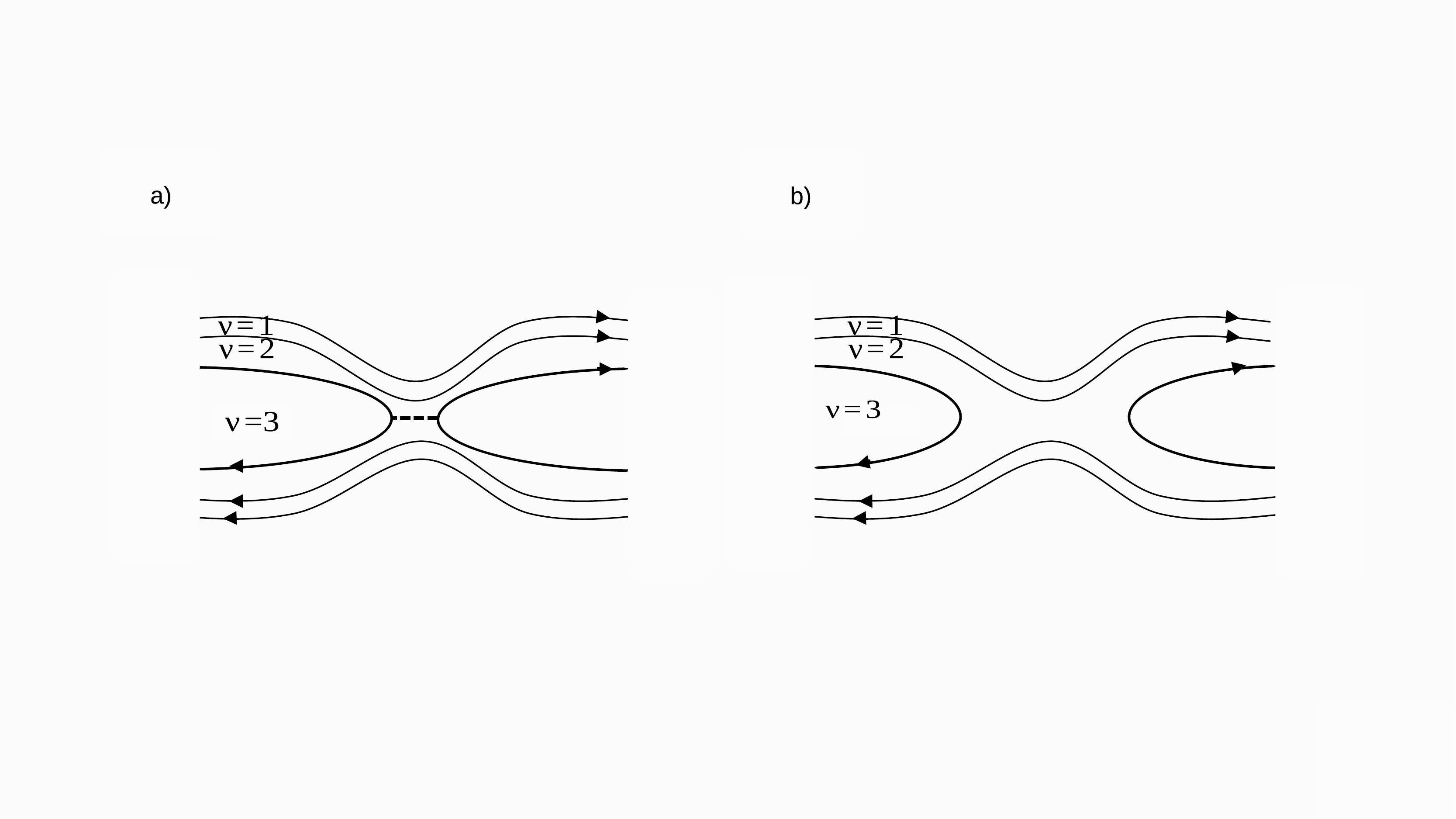}}
\caption{{Schematic of a constriction with (a) weak forward scattering and (b) no scattering. Dashed line shows electron tunneling.
}}
\label{fig-3-regimes:review}
\end{figure}

{\color{black}
In typical experiments, the filling factor $f_c$  in the constrictions and the filling factor in the  bulk are varied simultaneously by changing the magnetic field, while  the overall electron density and gate voltages are held constant. A region of weak forward scattering should occur when $f_c$ is slightly above a rational value $\nu_c$ that corresponds to a well-established quantized Hall state.  (This will obtain  when the magnetic field is slightly lower than the value at which $f_c=\nu_c$.)  In this case, we have $\nout = \nu_c$, so the charges measured in the CD regime will be that of the elementary quasiparticles in the region of the constriction.  Figure 8a shows a case  where $\nu_c = 2$ and $f_c \approx 2.2$.   By contrast, the constrictions shown in Fig.~\ref {fig-interferometer:review}  are in a regime of weak backscattering, where  $f_c$ is slightly below  $\nu_c$, meaning that the magnetic field is slightly higher than the value where  $f_c =\nu_c$.  (In Fig.~\ref {fig-interferometer:review}, we have $\nu_c = 2$ and $f_c < 2$.) In such cases, we have $\nu_c =  \nin$, so if $\nu_c$ is a fraction, the charge $\qout$ measured in the experiment will differ from the elementary charge in the constriction. Note that the partially reflected edge states are not the same in Figs.~\ref   {fig-interferometer:review}       and 8a.
 
In a well-made constriction, as  parameters are varied, there will be intervals where $f_c$  is sufficiently close to some  quantized value $\nu_c$ that there is  neither appreciable forward nor back scattering. (See  Fig. 8b.)  In this regime,  the measured Hall resistance of the device and  the two-terminal conductance will sit on a quantized Hall plateau, where no interference oscillations will be seen.  

}

Quasiparticle charge measurements in the CD regime  obtained in
Ref.~\onlinecite{AB-heiblum} were consistent with the expected anyon charges $e/3$ and $e/5$ at $\nu=1/3$ and $\nu=2/5$ respectively.
Charge $\approx e/3$ was reported  for $\nu=1/3,~2/3,~4/3,$ and $5/3$
in Ref. \onlinecite{marcus-inter}.  A recent Coulomb blockade experiment \cite{ensslin-cb} reveals charges $e/3$ at $\nu=1/3$ and $2/3$.  

As mentioned above, the color plots of conductance as a function of $B$ and $V_g$ presented in Ref. \onlinecite{manfra20} showed a series of horizontal stripes, for fields outside the range of incompressible bulk, which is what one predicts for  the CD regime when  the 
bulk is in a compressible state on the $\nu=1/3$ plateau and the filling in the constrictions is less than that of the bulk. Moreover, the  observed gate period $\Delta V_g$ is consistent with the  predicted period  in the CD regime, since $\qout = e$.

Quasiparticle charge can also be obtained from Aharonov-Bohm oscillations  in the incompressible bulk regime  shown in Fig. 6,  as was done in Ref. \onlinecite{manfra19}.  Assuming the interference area $A$ is known, {\color{black} if one can neglect Coulomb coupling between the edge and the bulk, }  the charge of the interfering particle can be extracted from the magnetic-field period in  an interval where  no localized quasiparticles enter or leave, by use of  the equation $ q_m A \Delta B  = e \Phi_0$.  Alternatively, if the dependence of $A$ on gate voltage is known, the charge may be extracted from the gate period using $q_m B  \Delta V_g   = e \Phi_0 / (\partial A/ \partial V_g)$. The authors of Ref. \onlinecite{manfra19}  used the second method to extract the value of $q_m$ at $\nu=1/3$,   assuming that  value of $ (\partial A/ \partial V_g)$ was unchanged from the value at $\nu=1$, and they obtained the value $q_m = 0.29 e$, in good agreement with the expected value $e/3$.  
On the other hand, measurements of the same type at $\nu = 2/3$ obtained a result of $0.93e$, suggesting that the tunneling charges in that case  might be electrons rather than fractionally charged quasiparticles.

{\color{black}
Using the model defined by Eq.~(\ref{interferometer-Coulomb-energy}), we can  address the effects of the Coulomb interactions, omitted above and  in Section IV A, on interferometer experiments in the incompressible region. }
If we continue to assume that the background parameters $\bar{A}$ and $\bar{Q}$ are insensitive to the magnetic field, then  modifications of the interference area $A_I$ are controlled by the coupling constants  $K_{IL}$ and $K_I$.  In this case one finds that the jump in the interferometer phase on entry of a quasiparticle will be given by  
{\color{black}Eq. (\ref{jump}) 
while the magnetic field period, between jumps,  will be renormalized to 
\be
\label{DBinc} 
\Delta B = \frac {e \Phi_0} { \qin \bar{A}}  \left[ 1 - \frac { K_{IL}} {K_I} \frac {\nin}{(\nin-\nout)} \right]^{-1} .
\ee
The slope of lines of equal phase on the plane of $B$ and $V_g$ should not be affected by a non-zero $K_{IL}$.
The distinct jumps predicted by (\ref{jump}) should be visible in the incompressible regime at temperatures much higher than in the compressible regime, in so far as the energy to create a quasiparticle is typically much higher than the scales of charging energies, $K_I$ and $K_L$. 
}

The claim in Refs.~\onlinecite{manfra19} and \onlinecite{manfra20}  that Coulomb coupling may be neglected in their sample is supported by the fact that the interference stripes they observe at the integer filling $\nu=1$ are consistent with what one would expect  in the incompressible regime on  neglecting the correction proportional to $K_{IL} / K_{I}$ in (\ref{DBinc}), or in the Aharonov-Bohm regime if the bulk is compressible.  

{\color{black} Note that in the incompressible region, one finds only a gradual transition between the regimes of weak and strong Coulomb interaction, as the predictions for $\Delta \theta$ and $\Delta B $ vary continuously as a function of $K_{IL}/K_{I}$. This is in contrast to the  compressible region, where the transition between  AB and CD-like regimes is marked by simultaneous manifestation of two distinct periodicities, rather than a single intermediate period. }

We conclude this section by mentioning puzzling  behavior \cite{goldman-puzzle,LinGoldman09} observed in a geometry with a $\nu=1/3$ channel going around  a $\nu=2/5$ island, 
where the transport data were interpreted as showing a magnetic-field period of $5 \Phi_0$  and a period of  in the interferometer charge of $2e$.  The explanation advanced by the experimenters supposed that the enclosed $\nu=2/5$ region was in a compressible state, where $e/5$ quasiparticles could readily enter or leave, so as to keep the electron-density and area fixed as the magnetic field was varied. However, according to the analysis presented above, in a compressible region, regardless of whether one was in the AB regime or the CD regime, any observed flux periods should be $\Phi_0$ or a submultiple of it, not a period larger than $\Phi_0$. (See Refs. \onlinecite{jain-puzzle,kim-puzzle,fisher-puzzle,coulomb11} for further discussions.)

 A possible resolution of the puzzle might be obtained if the quantum dots in these experiments were actually measured in a magnetic field interval where the interior state was essentially incompressible, as in the central magnetic-field region of Refs.~\onlinecite{manfra20} and \onlinecite{manfra19}.  In that case, if the interfering qusaiparticles have charge $e/5$, one would naturally expect to find a flux period of $5 \Phi_0$ and a gate period corresponding to the addition of two electrons. It should be noted, however, that the varying gate employed in these experiments was  not a side gate but rather a back gate, separated from the sample by the thickness of a sapphire substrate, which may complicate the analysis.  In any case, a more detailed analysis, and perhaps further experiments, are needed to resolve these issues.

 \subsection{Anyon collider}

It is known that the scattering of identical fermions differs from the scattering of identical bosons with the same interaction potential. 
This suggests the use of anyon collisions to probe fractional statistics. 

An anyon collider at $\nu=1/3$ was implemented in Ref. \onlinecite{collider-exp} following 
the proposal from Ref. \onlinecite{collider}. The setup is illustrated in Fig.~\ref{fig8:review}.
Charge from  two sources arrives along the edges to two point contacts QPC1 and QPC2, where tunneling gives rise to two dilute beams of 
anyons propagating along the edges towards cQPC. Anyons, arriving from the two sides to that contact, collide. This affects the currents, collected in the two drains, D1 and D2. 
If the anyons were fermions, the Pauli principle  would prohibit the two arriving anyons from ending up on the same side of cQPC. In other words, the two fermions 
would block each other from tunneling through cQPC. Mathematically, this  would result in an absence of correlations between the two drain currents. 
Bosons are known to bunch, and this would result in non-zero correlations.
Laughlin anyons are intermediate in their properties between bosons and fermions. Thus one might expect some intermediate form of current correlations for a Laughlin liquid.  

\setcounter{figure}{8}

\begin{figure}[!htb]
\bigskip
\centering\scalebox{0.25}[0.25]{\includegraphics{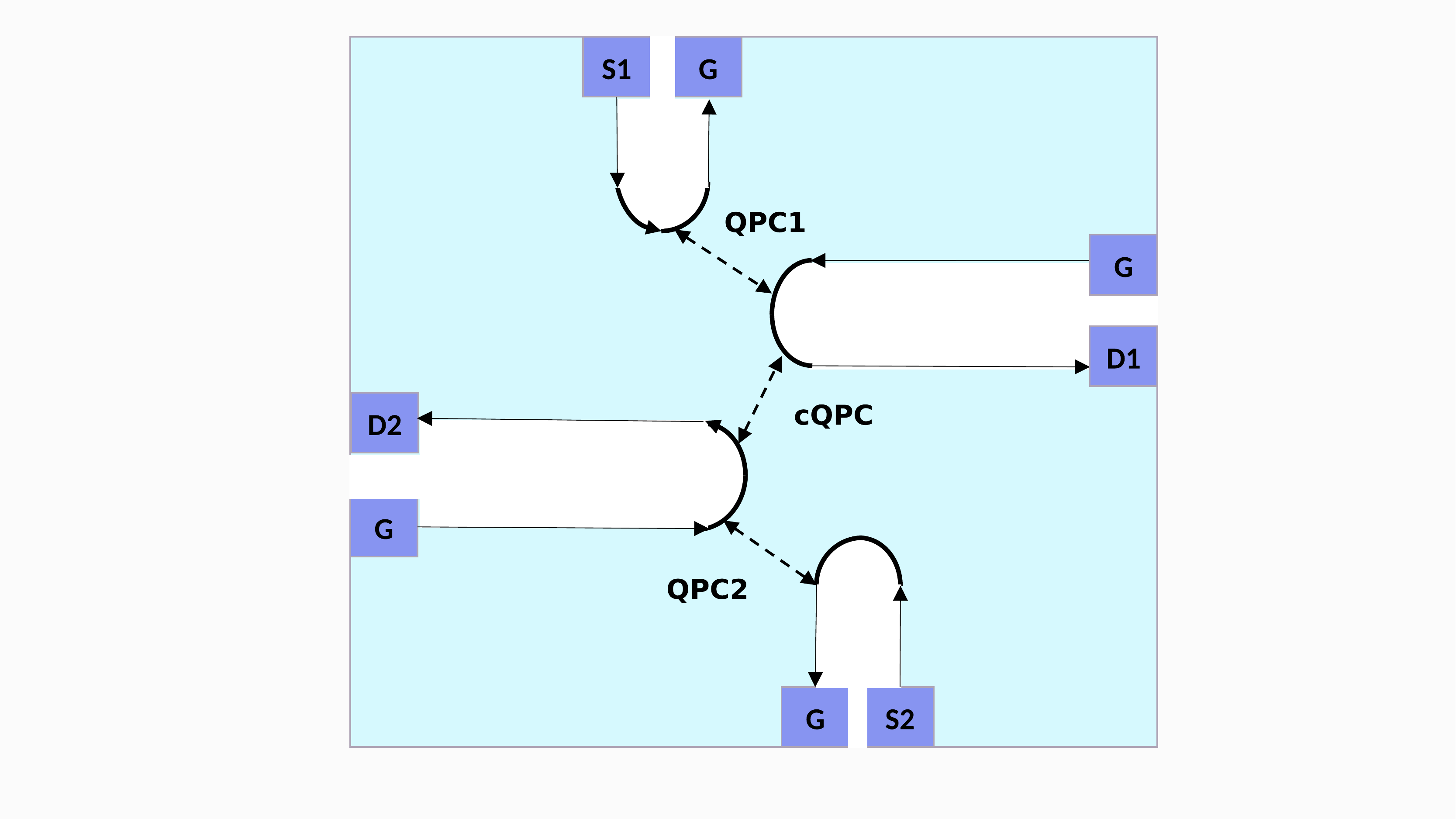}}
\caption{{Anyon Collider. The currents from sources S1 and S2 give rise to dilute beams of quasiparticles from QPC1 and QPC2 to cQPC, where anyons collide. The correlations of the currents in 
drains D1 and D2 are determined experimentally.}}
\label{fig8:review}
\end{figure}

Reference \onlinecite{collider} made specific predictions for various Laughlin liquids, and the experimental results, obtained at $\nu = 1/3$, were found to be in excellent agreement with the theory.  However, there were other ingredients in the theory in addition to the assumption of fractional statistics.  The theory employed a specific model of the 
edge Hamiltonian $\hat H_{\rm edge}$, given by Eq. (\ref{edge-H}). The Hamiltonian is important because a quasiparticle tunneling between two edge states will leave behind  trace in  excitations along the edges, which can affect the amplitude for tunneling of a second quasiparticle.  Indeed, it is expected that the form of the current correlations may be altered if there is reconstruction at the edges of the sample.  The results  of the anyon collider experiment, while very interesting,  would therefore seem to be a less direct measurement of fractional statistics than those obtained from the Fabry-Perot experiments.

{\color{black} Ref. \onlinecite{safi2020} discusses what information can be extracted from finite-frequency noise in an anyon collider.}
Several other setups have been  proposed theoretically to obtain signatures of fractional statistics from other current correlations \cite{safi2001,HBT3,noise1,noise2,noise3,noise4,noise5}, but these have not yet been realized experimentally.   

{\color{black} 
Another type of correlation experiment, which requires the simultaneous presence of two identical particles and which depends on their mutual statistics is the Hanbury-Brown Twiss interferometer. 
A beautiful experiment of this type, demonstrating  the interference between two electrons from independent sources injected into a quantum Hall  edge state at integer filling, was reported in Ref.~\onlinecite{Neder07}.  Results that might be expected for a similar experiment with  FQH edge states have been explored theoretically in  Refs. \onlinecite{Campagnano-1,Campagnano-2,HBT3}.

}

{\color{black}
\section{Non-Abelian statistics} \label{non-abelian} 

In our previous discussion of Abelian anyons, we  focused on the statistical angle  acquired during anyon braiding, that is, in  a process in which anyons exchange their positions or run full circles around other anyons. 
By measuring braiding phases, accumulated by various anyon types on a  circle around a localized anyon,  the localized anyon could be identified. This was the idea behind the interferometry technique in Section~\ref{stat-expts}.A.

Two other important processes  which characterize the topological behavior of anyons are fusion  and splitting. In fusion, two anyons combine into a single excitation. Splitting is the reverse process. These processes will be of particular importance in our discussion of non-Abelian anyons. } 

In the Laughlin states \cite{Laughlin83} at $\nu=1/m$, fusion is trivial. Anyon types are fully determined by anyon charges. Combining two anyons of charges $q_1$ and $q_2$ produces an anyon of charge $q_1+q_2$.
As an example, consider the $\nu=1/2$ liquid of charge $e$ bosons{\color{black}\cite{semion1,semion2}}.    The elementary quasiparticles have charge $e/2$ and are semions, that is they have statistical angle $\theta_m = \pi/2$, half that of a fermion. 
 A semion is a non-local or topologically nontrivial object. This means that there is no way to create an isolated semion. Semions can only be created in pairs. 
Two semions fuse into a boson, which is a topologically trivial object that can be created locally and cannot be detected  with interferometry. We say that it belongs to the vacuum topological sector. Similarly, adding any number of bosons to a semion does not affect the outcome of an interferometry experiment and 
does not change the topological sector of the excitation.
If we label the vacuum sector with 1 and the semion sector with $s$, we get the following fusion rules for the particles from the two sectors:
\be
\label{semion-fusion}
1\times 1=1, ~1\times s=s, ~s\times s=1.
\ee
This is an example of {\it Abelian statistics}. More complicated states with Abelian statistics, such as Halperin's $nnm$ liquids\cite{331}, allow neutral anyonic excitations. Thus, anyons of the same electric charge may belong to different sectors $a_i$. Still, the sectors form an Abelian group with the fusion rules  $a_i\times a_j=a_{k(i,j)}$, 
where $k$ is uniquely determined by $i$ and $j$.

In systems with  {\it non-Abelian statistics}, we encounter situations where  two given anyons can fuse into excitations from more than one  topological sector. 
We shall be particularly interested in systems with the simplest type of non-Abelian statistics, known as {\it{ Ising statistics}}, which  emerges in the exactly solvable Kitaev model\cite{kitaev-rev}
of a magnet on a hexagonal lattice and is 
relevant for vortices in $p$-wave superconductors\cite{ivanov} as well as FQH states at half-integer fillings\cite{willett-rev}. 
Systems with Ising topological order have  three topological sectors: vacuum 1, fermion $\psi$, and Ising anyons $\sigma$. Fusion with the vacuum has no effect on the topological sector. The remaining three fusion rules are
\be
\label{Ising-fusion}
\psi\times\psi=1,~\sigma\times\psi=\sigma,~\sigma\times\sigma=1+\psi.
\ee
The last rule means that the fusion of two Ising anyons may yield a boson or a fermion. {\color{black}If the two $\sigma$ particles are far apart,}  the two fusion channels cannot be distinguished by local measurements and are present at the same local quantum numbers of the two anyons. 
The information about the fusion channel is stored globally. This serves as the foundation for the idea of topological quantum computing \cite{top-comp}.

As an example, consider two vortices in a spinless two-dimensional superconductor with pairing of the form $p_x+ip_y$  \cite{ivanov}. Each vortex binds a Majorana zero mode described by a real fermion $\Psi_{1,2}=\Psi^\dagger_{1,2}$.
 The two real fermions combine into a single complex fermion $\Psi=\Psi_1+i\Psi_2$ that can populate a single energy level. 
The states with the filled and empty level differ by their fermionic parity
but cannot be distinguished locally by looking at a single vortex. 
Thus, we can think of the vortices as $\sigma$-anyons and the filled and empty level as the two fusion channels from Eq. (\ref{Ising-fusion}).

A system with conserved fermionic parity cannot move between the two fusion channels, but the same physics is present in a parity conserving system with four Ising anyons. The trivial total parity can be obtained in two locally indistinguishable ways: anyons 1 and 2 fuse to vacuum and anyons 3 and 4 fuse to vacuum, or alternatively, anyons 1 and 2 fuse to fermion and
anyons 3 and 4 fuse to fermion. A system of $2n$ Ising anyons has $2^{n-1}$ locally indistinguishable states.

\subsection{Basic principles}

The theory of fusion and braiding was dubbed the algebraic theory of anyons in Ref. \onlinecite{kitaev-rev}, which is the approach we follow. In pure mathematics, fractional  statistics
correspond to modular tensor categories \cite{TyuraevBook}. The same mathematical structures emerge in topological field theory\cite{witten} and in conformal field theory\cite{CFT} (CFT). 
This reflects the physics of the problem: topological field theories capture some of the bulk physics in topological liquids; we will see below that CFT captures universal aspects of the edge physics.  
{\color{black}The complete topological classification of a system with non-Abelian statistics goes beyond  the rules stating which sectors can fuse into which others,  
but will depend also on various amplitudes associated with fusions and braidings.  For example, one finds\cite{kitaev-rev} that that there are  eight distinct topological orders for systems obeying the fusion rules (\ref{Ising-fusion})  }

The two-dimensional nature of the problem does not make any difference for fusion. We will follow the convention of placing all anyons on a line. Braiding exchanges anyon positions on that line while fusing and splitting changes the number of the occupied sites. We also ignore all local quantum numbers of the anyons and focus solely on topologically distinct states. 
Thus, we consider just one anyon state in each topological sector. In other words, we introduce a one-dimensional Hilbert space for each anyon type. 

We will use diagrammatic language to speak of fusion and braiding. The key objects are the splitting operators $[\psi_i]_c^{ab}$ and their Hermitian conjugate fusion operators, illustrated below:

\begin{tikzpicture}[baseline=0, thick,scale=.5, shift={(0,-4.8)}]
		\draw (2,2) node[left] {\small $c$};
		\draw (1,4) node[left] {\small $a$};
		\draw (3,4) node[right] {\small $b$};				
		\draw (2,2) -- (2,3); \draw (2,3) -- (1,4); \draw(2,3) -- (3,4);
		\draw (2,3) node[left] {\small $[\psi_i]_c^{ab}$};		
		\draw (4+2,4) node[left] {\small $c$};
		\draw (4+1,2) node[left] {\small $a$};
		\draw (4+3,2) node[right] {\small $b$};				
		\draw (4+2,4) -- (4+2,3); \draw (4+2,3) -- (4+1,2); \draw(4+2,3) -- (4+3,2);
		\draw (4+2,3) node[left] {\small $[\psi^\dagger_i]_c^{ab}$};
		\draw (8,3) node[left]{\small .};
\end{tikzpicture}\\
We use the convention that the time axis runs up.
The right diagram suggests moving two anyons 
$a$ and $b$ into the same point, where they fuse into $c$, but a different way of thinking is often useful. We can assume that particles do not move  and the fusion and splitting operators are just linear maps between
the Hilbert space of the combined system of the two anyons and a one-dimensional space.

The most general fusion rule is

\be
\label{fusion}
a\times b=\sum_c N^c_{ab}c,
\ee
where the fusion multiplicities $N^c_{ab}$ show the number of independent ways to fuse anyons $a$ and $b$ into anyon $c$; in other words,  $N^c_{ab}$ is the dimension of the Hilbert space $V^{ab}_c$ of the states of the two anyons with the total topological charge $c$. 
All fusion multiplicities equal 1 for the  Ising statistics and for any Abelian statistics. Assuming the normalization

\be
\label{normalization}
\begin{tikzpicture}[baseline=0, thick,scale=.5, shift={(0,-4)}]
		\draw (2,2) node[left] {\small $c$};
		\draw (1,4) node[left] {\small $a$};
		\draw (3,4) node[right] {\small $b$};				
		\draw (2,2) -- (2,3); \draw (2,3) -- (1,4); \draw(2,3) -- (3,4);
		\draw (2,3) node[right] {\small $\psi_j$};		
		\draw (2,6) node[left] {\small $c$};		
		\draw (1,4) -- (2,5); \draw (3,4) -- (2,5); \draw(2,5) -- (2,6);
		\draw (2,5) node[right] {\small $\psi^\dagger_k$};

\draw (4,4) node[right]{\small $= ~\delta_{jk} $};

              \draw (6.5,2) -- (6.5,6);
              \draw (6.5,4) node[right]{\small $c$~,};
\end{tikzpicture}
\ee
where $j,k=1,\dots, N^c_{ab}$,
we decompose the identity operator as 

\be
\label{ind-res}
\begin{tikzpicture}[baseline=0, thick,scale=.5, shift={(0,-4)}]
		\draw (2,4) node[left] {\small $a$};
		\draw (4,4) node[left] {\small $b$};			
	        \draw (2,2) -- (2,6); \draw (4,2) -- (4,6); 
	        
	       \draw (5,4) node[right]{\small $=~~\sum_{c,j}$};
	        
		\draw (7+2,5) node[right] {\small $\psi_j$};			
		\draw (6+2,2) -- (7+2,3); \draw (8+2,2) -- (7+2,3); \draw(7+2,3) -- (7+2,5);
		\draw (7+2,5) -- (8+2,6); \draw (7+2,5) -- (6+2,6);
		\draw (7+2,3) node[right] {\small $\psi^\dagger_j$};
		\draw (6+2,2) node[left]{\small $a$};
		\draw (8+2,2) node[right]{\small $b$};
		\draw (6+2,6) node[left]{\small $a$};
		\draw (8+2,6) node[right]{\small $b$};
		\draw (7+2,4) node[left]{\small $c$};
		\draw (8+2.5,4) node[right]{\small .};
\end{tikzpicture}
\ee
One of the anyon sectors is vacuum 1, and the fusion multiplicity with vacuum is always $N_{a1}^a=N_{1a}^a=1$. Also,
$N^1_{ab}$ can only be 0 or 1. It is always possible to add a vacuum line to any diagram. 

Calculations with diagrams often involve $F$-moves

\be
\label{F-moves}
\begin{tikzpicture}[baseline=0, thick,scale=.5, shift={(0,-3.4)}]

		\draw (2,2) node[left] {\small $u$};
		\draw (1.5,3.25) node[left] {\small $x$};
		\draw (0,5) node[left] {\small $a$};
		\draw (2,5) node[right] {\small $b$};	
		\draw (4,5) node[right] {\small $c$};				
	        \draw (2,2) -- (2,3); \draw (2,3) -- (0,5); \draw (1,4) -- (2,5); \draw (2,3)--(4,5);
	        
	       \draw (5,3.5) node[right]{\small $=~~F_u^{abc}$};
	        
		\draw (9,2) node[left] {\small $u$};
		\draw (9.5,3.25) node[right] {\small $y$};
		\draw (7,5) node[left] {\small $a$};
		\draw (9,5) node[right] {\small $b$};	
		\draw (11,5) node[right] {\small $c$};				
	        \draw (9,2) -- (9,3); \draw (9,3) -- (7,5); \draw (10,4) -- (9,5); \draw (9,3)--(11,5);

		\draw (11.5,3.5) node[right]{\small ,};
\end{tikzpicture}
\ee

\noindent
where $F_u^{abc}$ are matrices with the indices $x$ and $y$ and additional numerical indices, if the fusion multiplicity exceeds one in any node of the diagram.
{\color{black} The diagrams on the right and on the left represent two compositions of splitting operators.}
A gauge freedom exists in the choice of the $F$-symbols and other topological data of an order. 
See Ref. \onlinecite{RSW} for numerous examples of such data.
For Abelian statistics the $F$-matrices are $1\times 1$, {\it i. e.}, just numbers. 
For example, for semions, $F_s^{sss}=-1$ and all the other $F$-symbols are 1 in the gauge \cite{RSW} we use.
For the Ising statistics with the braiding rules (\ref{R-Ising}), the following $F$-symbols are non-trivial:

\be
\label{ising-F-matrix}
\left[F^{\sigma\sigma\sigma}_{\sigma}\right]_{rs}
=
\begin{pmatrix}
1/\sqrt{2}  & 1/\sqrt{2} \\
1/\sqrt{2}  & -1/\sqrt{2}
\end{pmatrix},
\ee
\be
\label{Ising-F-number}
\left[F^{\sigma\psi\sigma}_{\psi}\right]_{\sigma\sigma}
=\left[F^{\psi\sigma\psi}_{\sigma}\right]_{\sigma\sigma}
=-1,
\ee
where $r,s=1,\psi$ with $r=s=1$ in the upper left corner of the matrix.

Thinking of fusion operators as linear maps naturally leads to an infinite number of associativity relations such as the pentagon equation (Fig. 10),
which tells that the two upper $F$-moves in the diagram are equivalent to the three lower ones.
It can be proven that any other `obvious' relation follows from the pentagon equation and the hexagon equation (Fig. 11).

Braiding is described by the unitary operators called $R$-symbols:

\be
\label{R-symbol}
\begin{tikzpicture}[baseline=0, thick,scale=.5, shift={(0,-2)}]
\draw (-2,2) node{\small $R_{ab}~~=$};

\draw (0,0) -- (0,1) -- (2,3) -- (2,4);
\draw  (0,-0.5) node{\small $a$};
\draw  (0,4.5) node{\small $b$};

\draw (2,0) -- (2,1) -- (1.15,1.85);
\draw (0.85,2.15) -- (0,3) -- (0,4);

\draw  (2,-0.5) node{\small $b$};
\draw  (2,4.5) node{\small $a$};

\draw (-2+8,2) node{\small $R^{-1}_{ab}~~=$};

\draw (2+8,0) -- (2+8,1) -- (0+8,3) -- (0+8,4);
\draw  (0+8,-0.5) node{\small $b$};
\draw  (0+8,4.5) node{\small $a$};

\draw (0+8,0) -- (0+8,1) -- (0.85+8,1.85);
\draw (1.15+8,2.15) -- (2+8,3) -- (2+8,4);

\draw  (2+8,-0.5) node{\small $a$};
\draw  (2+8,4.5) node{\small $b$};

\end{tikzpicture}
\ee

\begin{widetext}
\begin{tikzpicture}[baseline=0, thick,scale=.5, shift={(0,-4.8)}]

\draw (3,0) -- (3,1); \draw (3,1) -- (0,4); \draw (1,3) -- (2,4); \draw (2,2) -- (4,4); \draw (3,1)-- (6,4);

\draw[->] (6,1.5) -- (14,4.5);

\draw (10,3.5) node[left]{\small $F$};

\draw (16+1,3) -- (16+1,4); \draw (16+1,4) -- (13+1,7); \draw (14+1,6) -- (15+1,7); \draw (16+1,4) -- (19+1,7); \draw (18+1,6)-- (17+1,7);

\draw[->] (20,4.5) -- (28,1.5);

\draw (24,3.5) node[right]{\small $F$};
 
\draw (31,0) -- (31,1); \draw (31,1) -- (28,4); \draw (31,1) -- (34,4); \draw (32,2) -- (30,4); \draw (33,3)-- (32,4);

\draw[->] (3,-1) -- (8,-3.5);

\draw (5.5,-2.75) node[left]{\small $F$};

\draw (10+1,-6) -- (10+1,-5); \draw (10+1,-5) -- (7+1,-2); \draw (10+1,-5) -- (13+1,-2); \draw (9+1,-4) -- (11+1,-2); \draw (10+1,-3)-- (9+1,-2);

\draw[->] (14,-4.5) -- (20,-4.5);

\draw (17.5,-5.25) node[left]{\small $F$};

\draw (24-1,-6) -- (24-1,-5); \draw (24-1,-5) -- (21-1,-2); \draw (24-1,-5) -- (27-1,-2); \draw (25-1,-4) -- (23-1,-2); \draw (24-1,-3)-- (25-1,-2);

\draw[->] (26,-3.5) -- (31,-1);

\draw (28.5,-2.75) node[right]{\small $F$};

\draw (17,-7) node{\small { FIG. 10:} Pentagon equation.};

\end{tikzpicture}
\\
\vskip .4in

%\end{widetext}
%\begin{widetext}

\begin{tikzpicture}[baseline=0, thick,scale=.5, shift={(0,-4.8)}]

\draw (0,0) -- (0,1) -- (-2,3) -- (1,6);
\draw (-1,2) -- (0,3) -- (-0.85, 3.85);
\draw (-1.15,4.15) -- (-2,5) -- (-2,6);
\draw (0,1) -- (1,2) -- (1,4) -- (0.15,4.85);
\draw (-0.15, 5.15) -- (-1,6);

\draw[->] (2,5) -- (8,8);
\draw (5,7.5) node[left]{\small $R$};

\draw (0+8+3,0+5) -- (0+8+3,1+5) -- (-2+8+3,3+5) -- (-2+8+3,4+5) -- (-2+8+3,6+5);
\draw (-1+8+3,2+5) -- (1+8+3,4+5) -- (1+8+3,6+5);
\draw (0+8+3,1+5) -- (1+8+3,2+5)  -- (0.15+8+3,2.85+5);
\draw (-0.15+8+3, 3.15+5) -- (-1+8+3,4+5) -- (-1+8+3,6+5);

\draw[->] (13,8) -- (19,8);
\draw (16,9) node{\small $F$};

\draw (0+15+6,0+5) -- (0+15+6,1+5) -- (-1+15+6,2+5) -- (-1+15+6,4+5) -- (-1+15+6,6+5);
\draw (0+15+6,1+5) -- (2+15+6,3+5) -- (1.15+15+6,3.85+5);
\draw (0.85+15+6,4.15+5) -- (0+15+6,5+5) -- (0+15+6,6+5);
\draw (1+15+6, 2+5) -- (0+15+6,3+5) -- (2+15+6,5+5) -- (2+15+6,6+5);

\draw[->] (24,8) -- (29,5);
\draw (26.5,7.5) node[right]{\small $R$};

\draw (0+22+9,0) -- (0+22+9,1) -- (-1+22+9,2) -- (-1+22+9,4) -- (-1+22+9,6);
\draw (0+22+9,1) -- (2+22+9,3) -- (2+22+9,6);
\draw (1+22+9,2) -- (0+22+9,3) -- (0+22+9,6);

\draw[->] (2,1) -- (8,-2);
\draw (5,-1.5) node[left]{\small $F$};

\draw (0+8-1+3,0-5) -- (0+8-1+3,1-5) -- (-1+8-1+3,2-5) -- (2+8-1+3,5-5) -- (2+8-1+3,6-5);
\draw (0+8-1+3,1-5) -- (2+8-1+3,3-5) -- (1.15+8-1+3,3.85-5);
\draw (0.85+8-1+3,4.15-5) -- (0+8-1+3,5-5) -- (0+8-1+3,6-5);
\draw (1+8-1+3, 2-5) -- (0.15+8-1+3,2.85-5); 
\draw (-0.15+8-1+3,3.15-5) -- (-1+8-1+3,4-5) -- (-1+8-1+3,6-5);

\draw[->] (13,-2) -- (19,-2);
\draw (16,-3) node{\small $R$};

\draw (0+15+1+6,0-5) -- (0+15+1+6,1-5) -- (-2+15+1+6,3-5) -- (-2+15+1+6,6-5);
\draw (0+15+1+6,1-5) -- (1+15+1+6,2-5) -- (1+15+1+6,6-5);
\draw (-1+15+1+6,2-5) -- (0+15+1+6,3-5) -- (0+15+1+6,6-5);

\draw[->] (24,-2) -- (29,1);
\draw (26.5,-1.5) node[right]{\small $F$};

\draw (15.5,-7) node{\small { FIG. 11:} Hexagon equation.};

\end{tikzpicture}

\end{widetext}

\setcounter{figure}{11}

The statistical phase, accumulated at the exchange of $a$ and $b$, is unaffected by local operators acting on each of the anyons. As a consequence, lines can be moved over crossing points of other lines. For example, splitting $b$ into two anyons below or above the crossing point in Eq. (\ref{R-symbol}) produces equivalent diagrams. 
For the semion topological order, the only nontrivial $R$-symbol describes the exchange of two semions: $R_{ss}=i$. For Ising anyons 
{\color{black} with the fusion rules (\ref{Ising-fusion}), eight topological orders are known with different braiding rules \cite{kitaev-rev}.
  In this subsection we consider one example, where the} non-trivial $R$-symbols depend on the fusion channel of the excitations in the following way:

\begin{eqnarray}
\label{R-Ising}
R^{\sigma\sigma}_1&=e^{-i\pi/8}, \quad R^{\sigma\sigma}_{\psi}=e^{3i\pi/8}, \nonumber
\\[3mm]
R^{\sigma\psi}_\sigma&=R^{\psi\sigma}_\sigma=e^{-i\pi/2}, \quad
R^{\psi\psi}_1=-1.
\end{eqnarray}

A combination of $F$-moves and $R$-moves generates the hexagon equation (Fig. 11): the composition of the the two $R$-moves and one $F$-move
in the upper part of the diagram is equivalent to the composition of the two $F$-moves and one $R$-move in the lower part of the diagram. A similar equation holds for
$R^{-1}$-moves.
 The $R$- and $F$-symbols satisfy the equations 
in Figs. 10 and 11 and form a key part of the data that defines a topological order.
Each particle $a$ has a unique antiparticle $\bar a$ {\color{black} for which the vacuum 1 is a possible outcome of the product 
$a\times \bar a$.}  The antiparticle of a Laughlin anyon of charge $q_m$ carries the opposite electric charge $-q_m$.
In the Ising and semion orders, each particle is its own antiparticle. Since the fusion multiplicity with an antiparticle to the vacuum has to be 1, the following diagram is defined up to an arbitrary phase factor $\kappa_a$:

\be
\label{quantum-d}
\begin{tikzpicture}[baseline=0, thick,scale=.5, shift={(0,-1)}]

\draw (0,2.5) node{\small $a$};
\draw (4,-0.5) node{\small $a$};

\draw (0,2) -- (0,1) -- (1,0) -- (3,2) -- (4,1) -- (4,0);

\draw (2,1.25) node[left]{\small $\bar a$};

\draw (6,1) node{ $=~~\frac{\kappa_a}{d_a}$};

\draw (7.5,0) -- (7.5,2);

\draw (7.5,1) node[right]{\small $a~~~,$};

\end{tikzpicture}
\ee
where the quantum dimension $d_a$ describes the scaling of the number of states $~\sim {d_a}^N$ of $N\gg 1$ anyons $a$. All quantum dimensions are $1$ for Abelian statistics.
For the Ising order, one quantum dimension is nontrivial: $d_\sigma=\sqrt{2}$ in agreement with $2^{n-1}$ states for $2n$ anyons. The quantum dimensions are the same for a particle and its antiparticle. {\color{black} A useful identity relates quantum dimensions with fusion multiplicities:

\be
\label{multiplicities-eigenvalues}
\sum_c N^c_{ab} d_c=d_a d_b.
\ee}

If $a=\bar a$, the phase factor $\kappa_a$ in Eq. (\ref{quantum-d}) is no longer arbitrary and is known as the Frobenius-Schur indicator. This invariant equals $\pm 1$
and  indicates the breaking of the spin-statistics correspondence at $\kappa_a=-1$. [See Eq. (\ref{spin-stat}) below.]. The indicator is $1$ for all excitations of the Ising liquid
with the braiding rules (\ref{R-Ising}), while $\kappa_s=-1$ in the semion order.

It proves profitable to redefine the normalization of the splitting and fusion operators (\ref{normalization}) in terms of quantum dimensions:

\be
\label{d-normalization}
\begin{tikzpicture}[baseline=0, thick,scale=.5, shift={(0,-3.8)}]

		\draw (-1,4) node{\small $\sqrt{\frac{d_c}{d_a d_b}}$};
		
		\draw (2,2) node[left] {\small $c$};
		\draw (1,4) node[left] {\small $a$};
		\draw (3,4) node[right] {\small $b$};				
		\draw (2,2) -- (2,3); \draw (2,3) -- (1,4); \draw(2,3) -- (3,4);
		\draw (2,3) node[right] {\small $\psi_j$};		
		\draw (2,6) node[left] {\small $c$};		
		\draw (1,4) -- (2,5); \draw (3,4) -- (2,5); \draw(2,5) -- (2,6);
		\draw (2,5) node[right] {\small $\psi^\dagger_k$};

\draw (4,4) node[right]{\small $= ~\delta_{jk} $};

              \draw (6.5,2) -- (6.5,6);
              \draw (6.5,4) node[right]{\small $c$~.};
\end{tikzpicture}
\ee
The resulting diagrammatic technique has a nice feature that topologically equivalent diagrams are equal. For example, for $\kappa_a=1$,

\be
\label{top-eq}
\begin{tikzpicture}[baseline=0, thick,scale=.5, shift={(0,-1)}]

\draw (0,2.5) node{\small $a$};
\draw (4,-0.5) node{\small $a$};

\draw (0,2) -- (0,1) -- (1,0) -- (3,2) -- (4,1) -- (4,0);

\draw (2,1.25) node[left]{\small $\bar a$};

\draw (5.25,1) node{\small $=$};

\draw (6.5,0) -- (6.5,2);

\draw (6.5,1) node[right]{\small $a~~~.$};

\end{tikzpicture}
\ee
Negative Frobenius-Schur indicators are accounted for by decorating lines with arrows. We will ignore this complication since the Ising order with the $R$-symbols (\ref{R-Ising}), 
which is of primary interest for this review, has trivial $\kappa_a$. With the new normalization, a closed non-self-intersecting loop from a particle line and an antiparticle line equals the quantum dimension of the particle.

Braiding properties can largely be deduced from a single number called the topological spin for each anyon type: 

\be
\label{topspin}
\begin{tikzpicture}[baseline=0, thick,scale=.5, shift={(0,-0)}]

\draw (-4,0) node{\small $\theta_a~=~~{d_a}^{-1}$};

\draw (-0.15,0.15) -- (-1,1) -- (-2,0) -- (-1,-1) -- (1,1) -- (2,0) -- (1,-1) -- (0.15,-0.15);

\draw (0,-0.75) node{\small $a$};

\draw (3,0) node{\small .};

\end{tikzpicture}
\ee
The topological spin is a root of unity \cite{vafa}. For vacuum, $\theta_1=1$. For semions, $\theta_s=i$. For the anyons in Ising liquids, $\theta_\psi=-1$ and 
$\theta_\sigma=\exp(i\pi/8)$.

\begin{figure}[!htb]
\bigskip
\centering\scalebox{0.25}[0.25]{\includegraphics{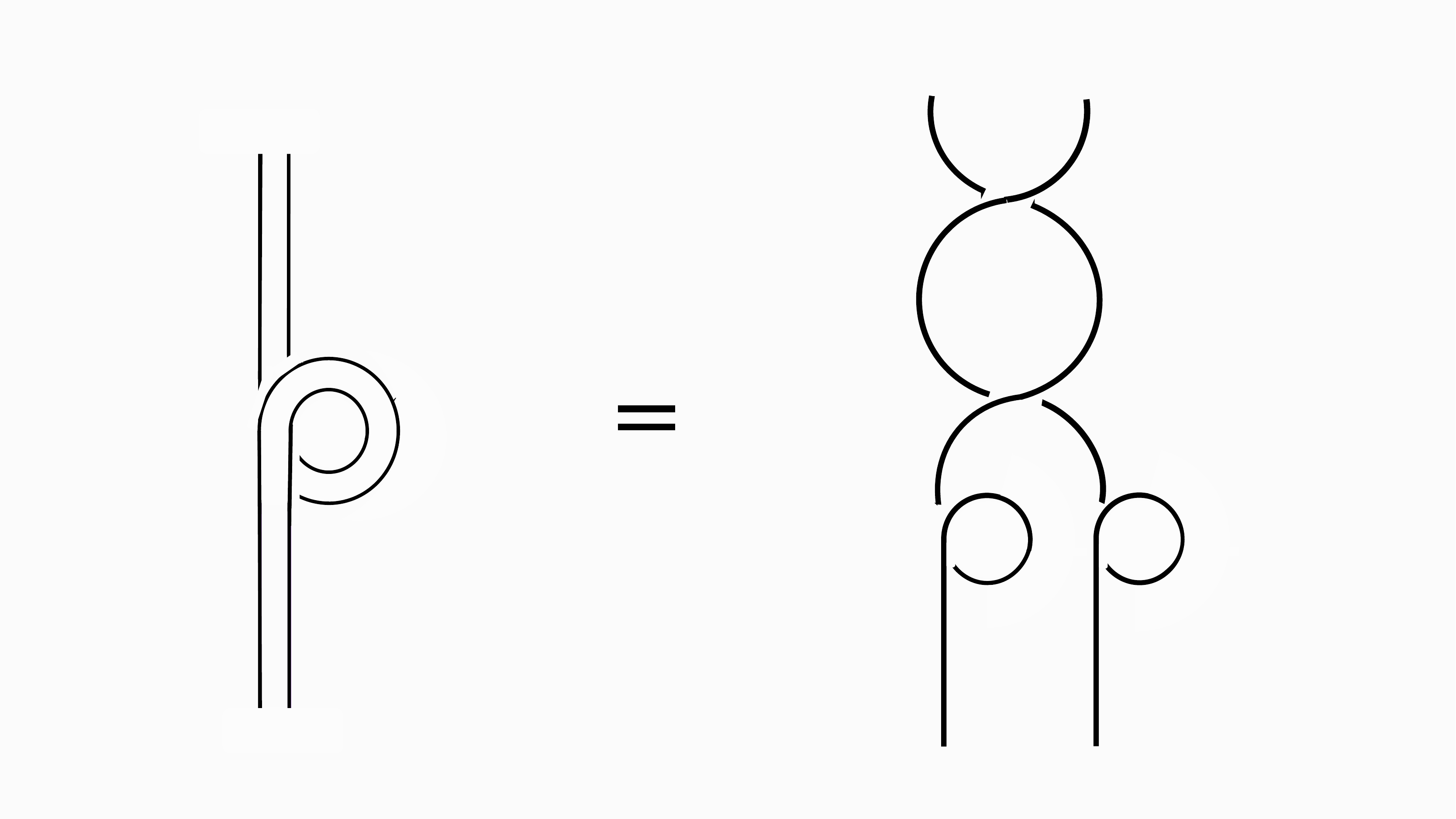}}
\caption{{The two lines of each diagram represent anyons $a$ and $b$. The left diagram can be interpreted as a double line, representing anyon $c$.}}
\label{fig-diagram:review}

\end{figure}

Naively, the topological spin defines the statistical phase at the exchange of a particle with its antiparticle, but this only holds in some cases. First of all, the statistical phase of non-Abelian anyons depends on their fusion channel. Second, even in the vacuum fusion channel, the standard spin-statistics relation may not hold. In particular, for
$a=\bar a$,

\be
\label{spin-stat}
R_1^{aa}=\theta^*_a\kappa_a.
\ee
On the other hand, interferometry involves the phase $\phi_c^{ab}$ accumulated by an anyon $a$ on a full circle around $b$ assuming that $a$ and $b$ fuse to $c$.
This phase depends only on the topological spins:

\be
\label{inter-phase}
\exp(i\phi^{ab}_c)=\frac{\theta_c}{\theta_a\theta_b}.
\ee
{\color{black} The proof of this expression illustrates the power of the diagrammatic approach and immediately follows from the diagrammatic identity in 
Fig. ~\ref{fig-diagram:review}.}

\subsection{Edge modes}

As previously mentioned, a boundary between a gapped FQH liquid and the vacuum necessarily  carries gapless modes. \cite{Halperin82}
The simplest example is the filling factor $\nu=1$ for non-interacting spinless electrons.
Far from the boundary, all electrons occupy degenerate states of the lowest Landau level at the energy $\hbar\omega_C/2$, where $\omega_C$ is the cyclotron frequency.
Assume that the confining potential near the edge changes slowly on the scale of the magnetic length $\sqrt{\hbar c/eB}$. Then the energy of a state localized at the distance $x$ from the boundary
is $E(x)=\hbar\omega_C/2+V(x)$, where $V(x)$ is the confining potential. The boundary $x_0$  of the occupied electron states  corresponds to $E(x=x_0)=E_F$, where $E_F$ is the Fermi energy. Gapless excitations are localized at $x\approx x_0$. The excitations are chiral, that is, they propagate only clockwise or counterclockwise, depending on the direction of the magnetic field. That direction is called downstream. 

In the simplest free-fermion model, the Lagrangian density of the low-energy mode is

\be
\label{f=1-action}
L_1=i\psi^\dagger(\partial_t+v\partial_x)\psi,
\ee
where $v$ is the mode velocity, and $\psi$ is a fermionic Grassmann field.  It is often convenient to bosonize \cite{GiamarchiBook}
the above Lagrangian density, substituting $\psi\sim\exp(i\phi)$, where ${\color{black}-}e\partial_x\phi/2\pi$ is the linear charge density:

\be
\label{f=1-bosonize}
L_B=-\frac{1}{4\pi}\partial_x\phi(\partial_t+v\partial_x)\phi.
\ee
A closely related chiral Luttinger liquid model \cite{WenBook} is often used to describe Laughlin states at $\nu=1/(2n+1)$:

\be
\label{cll}
L_{2n+1}=-\frac{1}{4\nu\pi}\partial_x\phi(\partial_t+v\partial_x)\phi,
\ee
where the electron operator $\psi\sim\exp(i\phi/\nu)$. Generalizations of this model are broadly applied to describe edges of Abelian FQH liquids. Besides a downstream charge mode, 
additional modes are generally present, whose directions can be both downstream and upstream \cite{WenBook}. {\color{black}The additional modes are typically charge-neutral,  due to effects of impurity scattering and/or long-range Coulomb interactions.}

The chiral Luttinger liquid model (\ref{cll}) misses complicated physics due to the long-range Coulomb interaction and often fails to
quantitatively describe the data \cite{tun-1/3,tun-rec,tun-c,sassetti_5/2}. One effect overlooked by the model is edge reconstruction \cite{recon}. It was shown theoretically 
\cite{CSG} and confirmed experimentally \cite{Pascher} that a realistic FQH edge in GaAs heterostructures is formed by a sequence of compressible and incompressible stripes. Their widths depend on the depletion length, where the electron density drops to zero near the sample boundary \cite{CSG}. The latter is set by the gate voltage for gate-defined edges and is determined by the physics of the localized surface states for the edges defined by chemical etching \cite{GelfHalp}. 

Narrow incompressible stripes are fixed-density regions with the filing factor between 0 and the bulk filling factor. They carry current proportional to the voltage difference between their edges. Incompressible stripes are separated by compressible stripes of fixed electrostatic potential and coordinate-dependent charge density. Naively, this picture suggests several co-propagating modes on the edge. Yet, general arguments based on thermal conductance (Section~\ref{other}) show that each downstream mode missed by the chiral Luttinger liquid model must be accompanied by an upstream mode. Inevitable disorder localizes pairs of contra-propagating modes on large lengths. {\color{black}On the longest length scales, the only surviving topologically-protected neutral modes are the ones present even on a sharp edge without reconstruction.}

Despite their limitations, chiral Luttinger liquid models produce deep insights about the FQH effect. One such insight is bulk-edge correspondence \cite{Pf}. 
It turns out that the bulk wave-function of a Laughlin  FQH liquid at $\nu=1/(2n+1)$ can be extracted from the correlation function of the electron operators $\psi(x,t)$ in the chiral CFT (\ref{cll}), where the imaginary time plays the role of the second spatial coordinate $y$. Moore and Read conjectured  \cite{Pf} that this represents a more general relation between conformal field theories of the edges and the ground-state wave functions in topological matter. This led them to a proposal for a non-Abelian state at half-integer filling factors dubbed the Pfaffian state. The Pfaffian edge theory contains two modes: 

\be
\label{L-Pf}
L_{Pf}=-\frac{2}{4\pi}\partial_x\phi_c(\partial_t+v\partial_x)\phi_c+i\psi(\partial_t+v_n\partial_x)\psi,
\ee
where the Bose-mode $\phi_c$ defines the charge density ${\color{black}-}e\partial_x\phi_c/2\pi$, and $\psi=\psi^\dagger$ is a neutral Majorana fermion. The electron operator 
$\Psi=\psi\exp(2i\phi_c)$. Each Bose mode has the central charge of 1 but the central charge \cite{CFT} of the Majorana fermion is $1/2$ (roughly speaking, the central charge counts the degrees of freedom, and a Majorana fermion can be seen as a half of a complex fermion). 
This is an example of a general rule: the chiral central charge of the edge theory in non-Abelian liquids is usually non-integer. 
The topological order in the Pfaffian state is closely related to the Ising order from the previous subsection and will be reviewed below.

Exceptions to bulk-edge correspondence are known \cite{7-3-bulk-edge}. Nevertheless, it remains a useful heuristic principle.
For example, consider particle-hole conjugation of topological orders \cite{phto}
and its effect on the edge structure. Imagine some topological order at $\nu=n+f$ in a fractionally filled Landau level of filling $f$ on top of $n$ filled spin-resolved Landau levels. The same order can be interpreted as a particle-hole conjugate order of holes at the filling factor $1-f$ on top of $n+1$ filled Landau levels.
One can also define a particle-hole conjugate order for electrons at the filling factor $n+1-f$. The classification of the excitations and the fusion rules are the same as in the original order. All braiding phases change their sign. 
The effect of the particle-hole transformation on the edge structure is the following. 
A boundary between $\nu=n$ and $\nu=n+1-f$, with $f < 1/2$, should be understood as a composition of an outer boundary between $\nu=n$ and $\nu=n+1$ and an inner boundary between $\nu=n+1$ and $\nu=n+1-f$.
The original state at $\nu=n+f$ would have had one or more edge modes between $\nu=n$ and $\nu=n+f$. 
The particle-hole conjugate order corresponds to the opposite propagation direction of each of those modes plus an additional downstream integer mode describing the boundary of $\nu=n$ and $\nu=n+1$.

Electron tunneling between edge channels can modify the description of the modes and sometimes reduces their number. For example, the Laughlin liquid at $\nu=1/3$ has a single downstream mode. According to the above prescription, the $2/3$ edge contains two charge modes \cite{edge-2/3}: an integer downstream mode and an upstream mode.
Electron tunneling due to inevitable disorder is known to reorganize the edge into a single downstream charge mode and an upstream neutral mode \cite{KFP}.
 
 We finish the discussion of edge modes by observing that the central charge in a CFT description  is proportional to the heat conductance of a mode \cite{t1,t2,t3}. It was proven that a chiral mode of the central charge $c$ has the thermal conductance $c \kappa_0 T$, where $\kappa_0T=T\pi^2 k_B^2/3h$ is known as a thermal conductance quantum. 
 The central charge is integer for every edge mode in any Abelian state.
 Hence, fractional quantization of the thermal conductance is a sign of non-Abelian statistics.

\subsection{Examples of non-Abelian statistics} 

It was long suspected that the filling factor\cite{5/2} $5/2$ hosts a non-Abelian FQH liquid \cite{Pf}. Experiment has brought strong evidence in favor of that view
\cite{texp2}.  Predictions were made for non-Abelian orders at other filling factors \cite{non-Ab-8/3,non-Ab-7/3,non-Ab-12/5-RR,non-Ab-12/5,non-Ab-12/5-ZGSH,
BS,BS-12/5,non-Ab-12/5-LLM,non-Ab-12/5-LLM-2,non-Ab-19/8,2+3/8-parton}
of the second Landau level in GaAs. 
Experiment is consistent with Abelian orders on the relatively more robust $\nu=7/3$  and $8/3$ plateaus (for a review, see Ref. \onlinecite{MF19}).
At the same time, the analysis of the gap dependence on the filling factor \cite{gaps2LL} supports different nature for FQH states at $\nu=7/3,~8/3$ on the one hand and at $7/3<\nu<8/3$ on the other hand. It might be that all states in the latter interval are non-Abelian. 
Very little is known about the plateaus \cite{2+3/8,gaps2LL,2+4/9,no-13/5} that presumably exist at $\nu=19/8, ~22/9$ and $32/13$. We will not address them 
and will not dwell on possible non-Abelian states in the first Landau level. Our focus will be on the filling factors\cite{5/2,2+3/8} $5/2$ and $12/5$. A fragile state\cite{7/2} at $\nu=7/2$ is expected to be the same or closely related to the state at $\nu=5/2$.

The existing theoretical pictures at $\nu=5/2$ and $12/5$ were influenced by CFT ideas\cite{Pf,RR}. Thus, we start with a brief summary of the CFT approach to the fractional statistics. A reader who is not familiar with CFT will be able to follow the bulk of the discussion in this section.
The starting point is an edge theory, which is a combination of chiral CFTs of perhaps opposite chiralities (downstream and upstream). Anyons correspond to the products of primary fields from each chiral CFT. One such product is postulated to describe electrons. All other allowed anyons must have single-valued operator product expansions 
with the electron operator. The rationale for this requirement comes from two considerations. First, electrons are in the vacuum topological sector and hence braid trivially with all excitations. Second, wave functions of systems of anyons are identified with conformal blocks of the CFT. Trivial braiding implies 
single-valued conformal blocks. The topological spin of each anyon $a$ is determined \cite{kitaev-rev} by its conformal weights $(h_a,\bar h_a)$ in the CFT: 

\be
\label{top-spin-weight}
\theta_a=\exp(2\pi i[h_a-\bar h_a]),
\ee
where $h_a$ comes from the counterclockwise holomorphic part of the edge theory, and $\bar h_a$ from the clockwise antiholomorphic modes.
Below we identify the holomorphic direction with the downstream direction of the charge mode.

\subsubsection{Possible states at $\nu=5/2$}

In our discussion of the proposed $\nu=5/2$ and $\nu=12/5$ orders we will ignore the two filled spin-resolved Landau levels. We will think of electrons in the partially-filled level in terms of composite fermions that combine an electron and two flux quanta \cite{JainBook}. Composite fermions move in zero effective magnetic field. Thus, one might expect that they form a gapless Fermi-liquid-like state.
Gapless states are indeed observed \cite{JainBook} at $\nu=1/2$ and $\nu=3/2$ in GaAs. The gap at $\nu=5/2$ can be explained by Cooper pairing \cite{t2}
of composite fermions. However, multiple ways exist to build a Cooper pair. 
{\color{black} 
In an isotropic system, one can have pairing in various angular momentum channels $l$. In an anisotropic system, where $l$ is not a good quantum number, we can instead talk about the winding number of  the phase of the order parameter as the fermion momentum moves around the Fermi surface.  In general, for a spinless single-component  Fermi surface, only odd values of $l$ are allowed.  However, in the presence of electron-electron interaction it is possible for a Fermi system to spontaneously divide itself into several components with independent Fermi surfaces, and in that case pairing with even values of  $l$ is allowed.  For example, it has been proposed that for a wide quantum well at total filling $\nu=1/2$, electrons might organize themselves into two parallel sheets with 1/4 filling in each.\cite{1/2-bil1}}
 The bulk-edge correspondence gives a convenient principle for classifying the various states.

For a half-filled Landau level,  the charge mode is described by the Lagrangian density (\ref{cll}) with $\nu=1/2$. Operators that create and annihilate an electron charge are proportional to $\Phi_\pm=\exp(\pm 2i\phi)$. One can check that operators $\Phi_\pm$ are bosonic and must be multiplied by a neutral fermion to produce a legitimate electron operator. This means that the edge theory should contain one or more gapless Fermi modes. Since a complex fermion is a combination of two Majorana fermions, we can assume without loss of generality that all fermion modes  are Majorana. We can also assume that all Majorana modes are co-propagating since contra-propagating modes can be gapped out by electron tunneling between edge modes. The net number $C$ of the Majorana modes is often referred to as a Chern number, because it has the form of a Chern index in the analysis presented in Ref.~\onlinecite{kitaev-rev}.   The Chern number is positive for downstream modes and negative for upstream modes. The Lagrangian density is

\be
\label{action-Chern}
L_{Pf}=-\frac{2}{4\pi}\partial_x\phi_c(\partial_t+v\partial_x)\phi_c+\sum_{k=1}^{|C|} i\psi_k(\partial_t+v_n{\rm sign} C\partial_x)\psi_k.
\ee
There is no  {\it a priori} reason for the velocities of the Majorana fermions to be the same, but edge disorder makes them equal in the long-scale limit \cite{apf1,apf2,YF2013}. 

No Majorana modes are present at $C=0$. In that special case\cite{331,WenZ92A,K8}, known as the $K=8$ state, electrons are gapped on the edge. For subtleties in the 113 state at $C=-2$, see Ref. \onlinecite{113}. Subtleties that emerge at $C=-4$ are addressed in Ref. \onlinecite{YF2013}.

The states with even Chern numbers are Abelian and the states with odd Chern numbers are non-Abelian. The topological order depends only on $C~{\rm mod}~16$. Thus, there are 8 Abelian and 8 non-Abelian possibilities known together as the 16-fold way \cite{16-fold,kitaev-rev}. Orders with a large Chern number are seen as unlikely, and the bulk of research has focused on the orders listed in Table I.
The states with the Chern numbers $C$ and $-(C+2)$ are related by the particle-hole conjugation. The PH-Pfaffian order at $C=-1$ is unique in being its own conjugate 
\cite{ph1,PH-MZ}.

\begin{widetext}
\center
\begin{table} [htb]
\renewcommand{\arraystretch}{1.2}
\begin{tabular} {| c | c | c | c|c|c|c|c|}
\hline 
 ~$C$~ & ~$-3$~ & ~$-2$~  & ~$-1$~ & ~$0$~ & ~$1$~ & ~$2$~ & ~$3$~  \\ \hline
name  & anti-Pfaffian & ~$113$~  & PH-Pfaffian & ~$K=8$~ & Pfaffian & ~$331$~ & ~$SU(2)_2$~
\\ \hline
Ref. & ~\onlinecite{apf1,apf2}~ & ~\onlinecite{113}~  & ~\onlinecite{apf2,ph1,PH-MZ} (see \onlinecite{ph3,ph4} for related surface states)~ & ~\onlinecite{331,WenZ92A,K8}~ &~\onlinecite{Pf} &~\onlinecite{331}~ & ~\onlinecite{su1,su2}~ 
\\ \hline
\end{tabular}
\caption{Proposed  topological orders at half-integer filling factors.  The Chern number $C$ is the difference between  the number of forward-  and backward-propagating  Majorana modes on a sample edge.}
\label{tab:states}
\end{table}
\end{widetext}

It appears that multiple orders of the 16-fold way are realized in nature.
Numerical work has brought a preponderance of evidence in favor of the non-Abelian Pfaffian and anti-Pfaffian liquids at $\nu=5/2$ in GaAs 
without impurities \cite{dis-num-1,dis-num-2,dis-num-3}; see, {\it{e.g.}}, 
Refs. \onlinecite{MorfED,num-5/2-1,num-5/2-2,num-5/2-3}. 
Experiment appears consistent with a 
different non-Abelian PH-Pfaffian liquid \cite{PH-MZ,texp2}. Some data were interpreted in terms of the Abelian 113 and 331 states \cite{tun-5/2-2,tun-5/2-3,tun-5/2-4}. Recent theoretical work suggests a complicated phase diagram in realistic disordered systems in which all topological orders with $-3\le C\le 1$ are present \cite{domain1,domain2,domain3} 
(see also Refs. {\color{black}\onlinecite{ph-num-lc,LLM1,LLM2,LLM3}} for the role of Landau level mixing). 
On the other hand, some theoretical proposals question \cite{nogap1,nogap3,nogap2} the existence of an energy gap at $\nu=5/2$.  

Half-integer FQH plateaus have also been found in several systems beyond single-layer GaAs.
The $SU(2)_2$ order was predicted in graphene \cite{su2gr1,su2gr2}. 
The 331 order\cite{331} is believed\cite{331-num1,331-num2,331-num3} to be present in GaAs bilayer\cite{1/2-bil1,1/2-bil2} at the filling factor $1/2$. Recent experiments on single-layer graphene have demonstrated the existence of gapped QH states in the $N=3$ Landau level, corresponding to $\nu =$  21/2, 23/2, 25/2 and 27/2, which have been attributed to a state with $C=3$ or to its particle-hole conjugate with $C=-5$.\cite{su2gr2}
Besides GaAs and 
graphene\cite{,gr1,gr2,gr3,gr4,gr5,su2gr2}, half-integer plateaus have been observed\cite{ZnO1,ZnO2,WSe} in ZnO and ${\rm WSe}_2$.

We finish the discussion of the 16-fold way by describing the quasiparticle types, fusion rules, and topological spins for each order \cite{16-fold}.   (See Table ~\ref{tab:states}.)
All non-Abelian orders possess excitations with three topological charges: 1, $\psi$, and $\sigma$. 1 and $\psi$ carry half-integer electrical charges $ne/2$. $\sigma$ carriers charge $e/4+ne/2$. The fusion rules for the topological charges are given by Eq. (\ref{Ising-fusion}). Electrical charges of the excitations, of course, add up in fusion. The topological spins of the excitations are determined by their topological charge $t$ and their electrical charge $ne/4$ as 
\be
\label{16-fold-top-spin}
\theta_{(t,n)}=\theta_t\exp(i\pi n^2/8),
\ee
 where 
$\theta_1=1$, $\theta_\psi=-1$, and $\theta_\sigma=\exp(i\pi C/8)$. 

The $K=8$ state is effectively a Laughlin-like liquid of charge $2e$ bosons at Landau-level  filling 1/8,  which gives rise to Abelian anyons labeled by their electrical charges $ne/4$, with the topological spins $\exp(i\pi n^2/8)$.
In the remaining Abelian states, there are four topological charges $1,~\psi,~\sigma$ and $\mu$. The electrical charges of $1$- and $\psi$-excitations are 
$ne/2$, while $\sigma$ and $\mu$ carry electrical charges $e/4+ne/2$. The fusion rules depend on the parity of $C/2$. For odd $C/2$, 
$\sigma\times\sigma=\mu\times\mu=\psi$
and $\sigma\times\mu=1$. For even $C/2$, $\sigma\times\sigma=\mu\times\mu=1$ and $\sigma\times\mu=\psi$. In all cases, $\psi\times\psi=1$, $\sigma\times\psi=\mu$, and 
$\mu\times\psi=\sigma$. The topological spins are given by Eq. (\ref{16-fold-top-spin}) with $\theta_1=1$, $\theta_\psi=-1$, $\theta_\sigma=\theta_\mu=\exp(i\pi C/8)$.

A key difference between Abelian and non-Abelian states is the existence of one charge-$e/4$ particle $\sigma$  for the non-Abelian orders and two charge-$e/4$ particles 
$\sigma$ and $\mu$ for the Abelian orders. This leads to subtleties in the interpretation of experiment since two quasiparticle types in Abelian states may have the same experimental consequences  as two fusion 
channels for non-Abelian anyons \cite{even-odd-331}.

Evidence exists for a $\nu=1/4$ plateau in wide GaAs quantum wells \cite{1/4-1,1/4-2,1/4-3}. The 16-fold way was extended to that filling factor in Ref. \onlinecite{MZ-1/4}.

\subsubsection{Read-Rezayi states}

The thermal conductance of a Majorana mode is determined by its central charge $c= 1/2$.  A more general class of CFTs  is known with $ c=(2k-2)/(k+2)$, where $k$ is an arbitrary positive integer. The case $k=2$ reduces to the Ising CFT, while 
the CFTs  with $k >2$ are known as parafermion theories \cite{CFT}. They were used by Read and Rezayi to generate a family of FQH states\cite{RR}  at the filling factors $k/(k+2)$.
Anyon types \cite{fusion-RR} are distinguished by their electrical charge and their topological charge, which comes from the list of the primary fields in the parafermion CFT. There are $k(k+1)/2$ primary fields $\Phi^j_m$ with $j=0,1/2,\dots,k/2$, 
$(j-m)\in\mathbb{Z}$. Two identifications are made: $(j,m)\equiv(j,m+k)$ and $(j,m)\equiv (\frac{k}{2}-j,m+\frac{k}{2})$. This allows choosing $j>0$ and $-j<m\le j$. The topological spin of $\Phi^j_m$ is

\begin{equation}
\label{dima-new-3}
\theta_m^j=\exp\left(2\pi i\left[\frac{j(j+1)}{k+2}-\frac{m^2}{k}\right]\right).
\end{equation}
The fusion channels are given by

\begin{equation}
\label{dima-new-4}
\Phi^j_m\times\Phi^{j'}_{m'}=\sum_{j''=|j-j'|}^{{\rm min}(j+j',k-j-j')}\Phi^{j''}_{m+m'}.
\end{equation}
Electrons carry the topological charge $\Phi^{k/2}_{1-k/2}$. The topological spin of an anyon of electrical charge $se$ is the product of the neutral contribution (\ref{dima-new-3}) and $\exp(\pi i [k+2] s^2/k)$. 
The allowed combinations of the topological and electrical charges make the braiding phase (\ref{inter-phase}) with an electron $\phi^{ae}_{a\times e}$ 
trivial. The lowest quasiparticle charge is $e/(k+2)$.

The state at $k=4$ corresponds to the observed filling factor $8/3=2+2/3$. There is some numerical evidence for a
 Read-Rezayi state at that filling factor \cite{non-Ab-8/3}, but experiment suggests that it hosts a Laughlin-like state (see Ref. \onlinecite{MF19} for a review).
No plateau has been seen\cite{no-13/5} at $\nu=13/5=2+3/5$, which would correspond to $k=3$. A plateau is known \cite{2+3/8} at the particle-hole conjugate filling factor $12/5$. Apparently, Landau level mixing effects\cite{non-Ab-12/5-LLM,non-Ab-12/5-LLM-2} are responsible for the difference between $\nu=12/5$ and $\nu=13/5$.
Numerics suggests a non-Abelian state\cite{non-Ab-12/5-RR,non-Ab-12/5,non-Ab-12/5-ZGSH} at $\nu=12/5$ that is the particle-hole conjugate of the $k=3$ Read-Rezayi state. Such state is interesting from the point of view of quantum computing since it allows universal topological computation\cite{NayakSSFD08}, 
impossible with the topological orders of the 16-fold way.

Note that a generalization of the Read-Rezayi states applies\cite{RR} to the filling factors $\nu=k/(Mk+2)$ with an odd $M$. A negative-flux version \cite{Jolicoeur} of the states was proposed at 
$\nu=k/(3k-2)$, $k>2$.

Another non-Abelian candidate at $\nu=12/5$ is a Bonderson-Slingerland state\cite{BS,BS-12/5}, whose fractional statistics is closely related with that in the Ising topological order.

 {\color{black} 
 
 \section {Fabry-Perot Interferometry with non-Abelian quasiparticles}  \label{fpi-2}
 
\subsection{The even-odd effect}
}

The theory of Fabry-Perot interferometry for non-Abelian anyons has attracted much attention. \cite{an-inf1,an-inf2,eo1,eo2,nAi1,nAi2,nAi3,nAi4,nAi6,nAi5}.
Fabry-Perot interferometry exhibits particularly interesting behavior for non-Abelian states, because the interference picture can depend on the fusion channel of the anyons traveling 
through the interferometer and the anyons trapped inside the device. 
{\color{black}
Let the trapped topological charge be $b$, the topological charge of the tunneling anyon be $a$, and assume that $a$ and $b$ fuse to $c$. Then the tunneling current through the interferometer can be computed from Eq. (\ref{int-I}) with the statistical phase (\ref{inter-phase}) in the cosine. When multiple fusion channels $c$ exist for given $a$ and $b$,
the contributions of each fusion channel should be added with the weight \cite{MZ-noise}

\be
\label{weights-int}
p^c_{ab}=N^c_{ab}\frac{d_c}{d_ad_b}, 
\ee
where $N^c_{ab}$ and $d_x$ are fusion multiplicities and quantum dimensions. 
The weights in Eq. (\ref{weights-int}) reflect the fact  that there is no correlation between the incoming quasiparticle and the particles inside the interferometer, and we may simply add probabilities, because different fusion outcomes are always orthogonal. One can check using Eq. (\ref{multiplicities-eigenvalues}) that the weights add up to one.}

For Ising anyons, multiple fusion channels lead to an  {\it even-odd effect}\cite{eo1,eo2}. Suppose that the leading contribution to the current through the interferometer comes from $e/4$ quasiparticles with the topological charge $\sigma$ (\ref{Ising-fusion}). Consider two possibilities for the trapped topological charge $t$:
(i) $t=1$ or $\psi$; or (ii) $t=\sigma$.
Since the traveling anyon  has topological charge $\sigma$, there will be a  unique fusion channel for it and  the trapped topological charge in case (i). The theory from Section IV applies with the statistical phase determined by the topological and electrical charges inside the interferometer. In case (ii) two equally likely fusion channels exist according to Eq. (\ref{Ising-fusion}). They correspond to the statistical phases (\ref{inter-phase}) that differ by $\pi$. Hence, the two fusion channels interfere destructively with each other and no dependence of the current through the interferometer on the magnetic field can be seen\cite{eo1,eo2}. 

After a sufficiently strong change in the magnetic field, a new anyon of topological charge $\sigma$ is expected to enter the bulk of the interferometer. This leads to the switching between regimes (i) and (ii). The name ``even-odd effect'' reflects that (i) corresponds to an even number of trapped quasiparticles and (ii) corresponds to an odd number.

The even-odd effect is present in all non-Abelian states of the 16-fold way\cite{16-fold} and its absence in an experiment would prove Abelian statistics. The opposite is not necessarily true \cite{even-odd-331}.
Indeed, the existence of two charge-$e/4$ anyons in Abelian states of the 16-fold way may mimic the two fusion channels 
of the $e/4$-particles in non-Abelian topological orders\cite{even-odd-331}.

The above physical picture assumes that the trapped topological charge does not fluctuate randomly  on the laboratory time scale. 
{\color{black} If there are a non-zero even  number of $e/4$ particles inside the interferometer, one should distinguish between the cases where they exist in the topological sector 1 or $\psi$. The interferometer will exhibit the same field periodicity in either case, but  the phase of the signal will differ by $\pi$ between the two cases. Over the  laboratory time scale necessary to accumulate data in an experiment, it  is possible
 that neutral fermions $\psi$ can  tunnel between the edges of the device and one or more localized states in the bulk, and thereby change the  topological sector of the interior. In this case, the mean occupations during the measurement should be a thermal equilibrium distribution, determined by the energy differences between states in the different  sectors.  If one or more of the trapped $e/4$ particles is far from the others and far from the boundaries, then the energy difference between the 1 and $\psi$ sectors will be smaller than $k_B T$, and the two sectors will have equal probability in equilibrium. In this case, the $e/4$ signal would be lost for even occupation numbers as well as for odd. 
}
Effects of frequent $\psi$-tunneling were investigated in Refs. \onlinecite{be-eo1,be-eo2,be-eo3,be-eo4,clarke-shtengel}.

The discussion above also ignores the tunneling of anyons of charge $e/2$ between the edges at the  constrictions in  the interferometer. Such anyons carry topological charges 1 and $\psi$ and do not exhibit an even-odd effect.  A contribution to the tunneling current from $e/2$ particles should exhibit a periodicity  with respect to  the magnetic field that is two times shorter than for charges $e/4$, and it should be present regardless of the number of enclosed $e/4$ particles.  In addition, if there is a significant contribution to the signal from  $e/4$ particles that  wind twice around the interferometer, that contribution should behave similarly to that of $e/2$ particles. 
\vskip .15in

{\color{black}

\subsection{Experimental investigations} 

{\color{black} 

In a series of papers dating back to 2007, Willett and collaborators have reported observations of even-odd alternation in carefully prepared GaAs samples at filling factors 5/2 and 7/2, in a Fabry-Perot geometry. (See Ref.~\onlinecite{willett2019} and references therein.) In the most recent of these papers, they reported extensive measurements on eleven different samples, including  analyses of the oscillatory dependences on magnetic field and gate voltages.

The interpretation of these experiments assumes that the interferometer is  in a compressible Aharonov-Bohm regime, where the periods are strongly affected by underlying filled Landau levels.  It assumes, further,  that 
when there is an even number of $e/4$ particles enclosed by the interferometer path, the system is consistently in one of the two possible topological sectors, 1 or $\psi$, and that it returns to the same sector when  two more quasiparticles are added. At $\nu=5/2$,  ten $e/4$ quasiparticles will leave the interferometer as the flux $\Phi=B\bar{A}$ is increased by $\Phi_0$, so that the parity will switch from even to odd and back five times in this interval. As it turns out, if one takes into account the Abelian phase acquired when an $e/4$ particle encircles an even number of $e/4$ particles in a fixed topological sector but ignores the even-odd switching that turns the interference on and off, one would predict an interference with a flux  period of $\Phi_0$.  When this signal is modulated by the rapid switching with period $\Phi_0 /5$, the power spectrum is predicted to have prominent peaks at frequencies $1/\Phi_0, \, 4/ \Phi_0$ and $6/ \Phi_0$.  A similar analysis at $\nu= 7/2$ predicts that interference peaks due to  the circulation of $e/4$ quasiparticles should occur there  at frequencies $1.5/ \Phi_0, \, 5.5/\Phi_0$ and  $8.5 / \Phi_0$.  

In addition to the signal from $e/4$ particles, one should expect contributions from other processes with different flux periods, as well as aperiodic features due to disorder, etc. These contributions tend to obscure the underlying  periodicities in the raw data and lead to complicating features in the Fourier transform.  Nevertheless, it appears that strong peaks were observed at the predicted positions at $\nu=5/2$, and to a lesser extent at $\nu=7/2$. 
These results give support for the occurrence of even-odd alternations, consistent with the existence of non-Abelian Ising anyons. (The observations do not distinguish between the Pfaffian, Anti-Pfaffian, or PH-Pfaffian states.). 
}

There are, however, some aspects of the experiments which are not well understood.  The interference areas needed to fit the data were very small, typically of order 0.25 $\mu$m$^2$, while the lithographic areas were squares ranging from 2.5 to 5.7 $\mu$m on a side. Moreover, since the interference area should presumably connect the openings  in the defining gates on two ends of the interferometer, the width in the perpendicular direction must be less than 0.1 $\mu$m. It is not clear what are the physical mechanisms that give rise to this unusual geometry. There may also be questions about the extent to which it is appropriate to talk about the existence of a quantized Hall state in a region of these dimensions.

Nevertheless, the small interferometer area appears to be reproducible, as it is seen in many samples, and persists at a variety of filling factors. The samples used in these experiments include a number of special features, including carefully designed screening layers, which may be important for understanding the resulting geometry. It should also be noted that in contrast to the procedures most commonly employed in quantum Hall interference measurements,  the samples in these experiments were illuminated before measurement.  }

{\color{black}
Experiments by An et al. \cite{an11} have measured phase slips and telegraph noise at  $\nu=5/2$ analogous to those  they found at $\nu=7/3$, which were discussed in Section IV.A.2,  above.  They find a distribution of phase-slip sizes with several peaks, including a prominent broad peak centered at $\Delta \theta \approx 5 \pi / 4$, which they attribute to simultaneous entry of an $e/4$ quasiparticle into the interferometer region and tunneling of a Majorana fermion ($\psi$-particle) between the edge of the system and the location of an $e/4$ particle in the interior.  However, their data is less extensive than that of Willett et al., and it seems difficult to rule out alternative explanations for their results. }

{\color{black}
\section{Mach-Zehnder interferometry}    \label{MZ}    }

\begin{figure}[!htb]
\bigskip
\centering\scalebox{0.25}[0.25]{\includegraphics{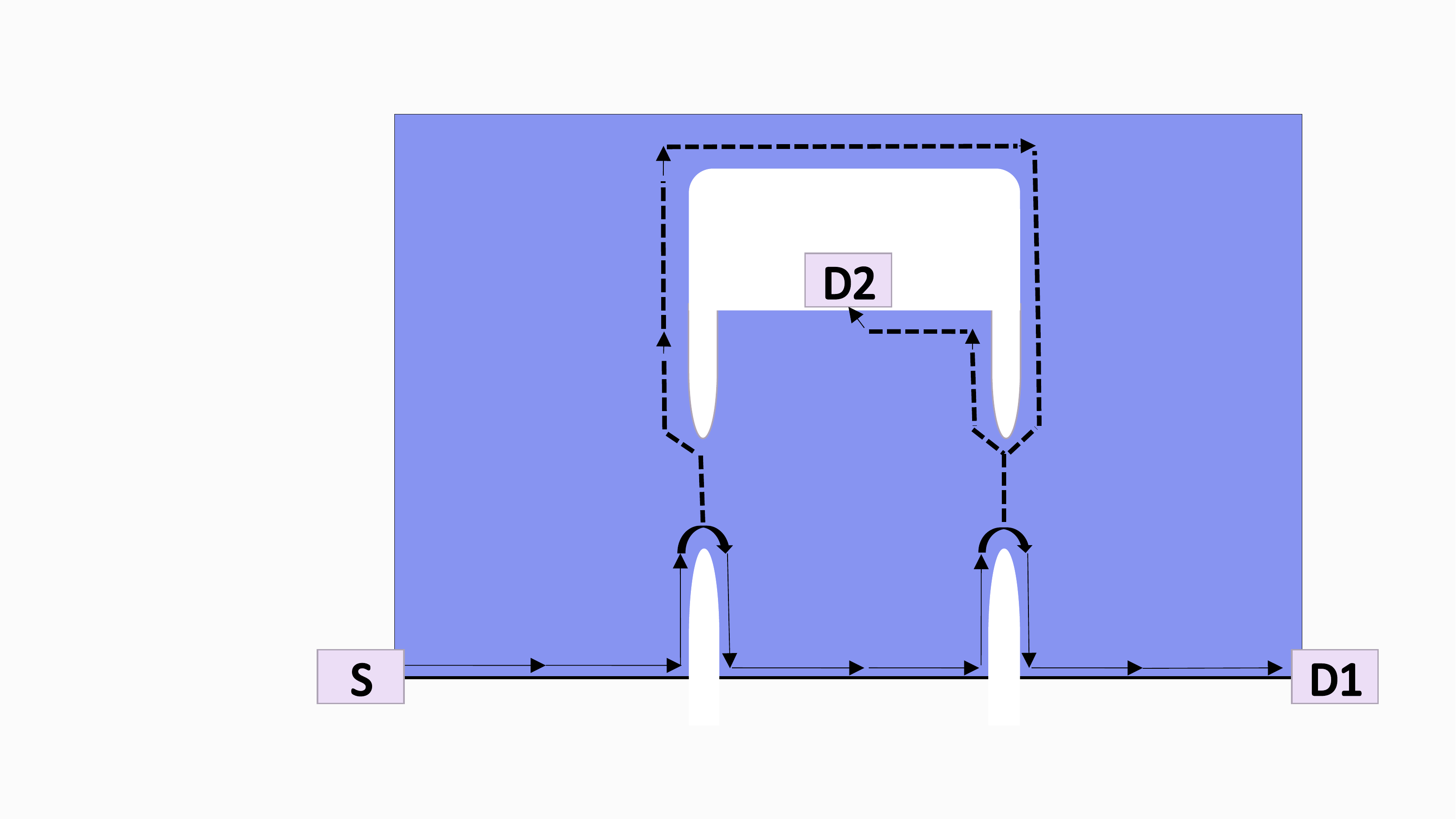}}
\caption{{The current from source S splits between two drains D1 and D2. Drain D2 is inside the interference loop. FQH liquid is shaded.}}
\label{fig7:review}
\end{figure}

A different type of  interference geometry, which has also been realized in quantum Hall states, is
Mach-Zehnder geometry \cite{MZ1,MZ2}, Fig.~\ref{fig7:review}. It is natural to ask whether this geometry can lead to a demonstration of fractional statistics. We shall see that the  geometry is also of theoretical interest since it appears in a general explanation why fractional charge entails fractional statistics \cite{MZ1}.

At first sight, nothing changes compared to the Fabry-Perot case, at least, in the {\color{black}incompressible} limit with {\color{black}weak bulk-edge interactions}. The tunneling rate is still given by Eq. (\ref{int-p}) and the current seems to be the same as above. But, the magnetic-flux periodicity $e\Phi_0/q_m>\Phi_0$ of the so-computed current would conflict with the rigorous Byers-Yang theorem \cite{BY}.
Indeed, the Mach-Zehnder interferometer has a hole. A change of the flux through the hole by one quantum should be invisible to the electrons from which the system is made. Hence, the magnetic flux period cannot possibly exceed $\Phi_0.$ 

The explanation of the paradox lies in that the drain inside the interferometer absorbs a quasiparticle after each tunneling event.
The topology of the device implies that the drain is inside the interference loop.
 Hence, the statistical phase     $\phi_s$ contributing to $\theta =\phi_1-\phi_2$ in Eq. (\ref{int-p}) changes  after each tunneling event, and the {\color{black}tunneling rate} changes accordingly. (The contradiction with the Byers-Yang theorem would be unavoidable if fractional charges could have Bose or Fermi statistics.)

\begin{tikzpicture}[baseline=0, thick,scale=.5, shift={(10,-0)}]

\draw[->] (0-20,0) -- (3-20,5.2); 
\draw (-20,0) node[left]{\small $0$};

\draw[->] (3-20,5.2) -- (9-20,5.2); 
\draw (-17,5.2) node[left]{\small $\nu e$}; 

\draw[->](9-20,5.2) -- (12-20,0); 
\draw (-11,5.2) node[right]{\small $2\nu e$};

\draw (12-20,0) -- (11.5-20,-0.87);
\draw (11-20,-1.73) -- (10.2-20,-3.12);
\draw[->](9.7-20,-3.99) -- (9-20,-5.2); 
\draw (-8,0) node[right]{\small $3\nu e$};

\draw[->](9-20,-5.2) -- (3-20, -5.2); 
\draw (-11,-5.2) node[right]{\small $e-2\nu e$};

\draw[->](3-20,-5.2) -- (0-20,0);
\draw (-17,-5.2) node[left]{\small $e-\nu e$};

\draw (0-15,-8) node{\small { FIG. 14} Transitions between the states of a Mach-Zehnder};
\draw (0-15,-9) node{\small interferometer in a Laughlin liquid at $T=0$. };
\end{tikzpicture}

\vskip .1in

The precise expression for the current depends on the details of statistics 
\cite{MZ1,MZ2,PA1,MZ-12/5,PA2,MZ-331,MZ-113,PH-MZ,16-fold,MZ-1/4}
and simplifies greatly for the Laughlin states at zero temperature \cite{MZ1}.
Fig. 14 illustrates possible transitions between topological charges of the drain at the Laughlin filling factors $\nu=1/m$. At zero temperature, charge only goes from the higher chemical potential to the lower chemical potential and hence all transitions are only possible in the direction of the arrows. The transition rate $p_n$ along the arrow connecting 
the trapped charges $n\nu e ~{\rm mod} ~e$ and $(n+1)\nu e~{\rm mod}~e$ is given by Eq. (\ref{int-p}) with the statistical phase that depends on $n$. The average time a transition takes is $t_n=1/p_n$. The total time for one full circle in the diagram Fig. 14 is thus $\bar t=\sum_{k=0}^{p-1}1/p_n$. Since a charge $me/m=e$ is transmitted in that sequence of tunneling events, the total current $I=e/\bar t$ is the harmonic average of the currents (\ref{int-I}) at all possible values of the statistical phase $\phi_s$. If the 
 flux through the hole is increased by $\Phi_0$, this only has the effect of shifting the transition times $t_n$ to $t_{n+1}$, so the net current is unchanged.

\begin{tikzpicture}[baseline=0, thick,scale=.5, shift={(10,-0)}]

\draw[->] (0-20,0) -- (-16,3); 
\draw (-20,0) node[left]{\small $(-e/4,\sigma)$};
\draw (-16,3.5) node{\small $(0,\psi)$};
\draw(-18.5,1) node[right]{\small $p(\pi)/2$};

\draw[->] (0-20,0) -- (-16,-3); 
\draw (-16,-3.5) node{\small $(0,1)$};
\draw(-18.5,-1) node[right]{\small $p(0)/2$};

\draw[->] (-16,3) -- (-13,0.75); \draw(-13,0.75) -- (-12,0);
\draw (-11.5,0) node[right]{\small $(e/4,\sigma)$}; 
\draw(-15,2.5) node[right]{\small $p(\pi)$};

\draw[->] (-16,-3) -- (-13,-0.75); \draw(-13,-0.75) -- (-12,0);
\draw(-15,-2.5) node[right]{\small $p(0)$};

\draw[->](-12,0) -- (-8,3); 
\draw (-8,3) node[right]{\small $(e/2,1)$};
\draw(-9,2.5) node[left]{\small $p(-\pi/2)/2$};

\draw[->](-12,0) -- (-8,-3); 
\draw (-8,-3) node[right]{\small $(e/2,\psi)$};
\draw(-9,-2.5) node[left]{\small $p(\pi/2)/2$};

\draw (-8,3) -- (-8,5) -- (-20,5);
\draw(-14,5.5) node{\small $p(-\pi/2)$};

\draw (-8,-3) -- (-8,-5) -- (-20,-5);
\draw(-14,-5.5) node{\small $p(\pi/2)$};

\draw[->](-20,5) -- (-20, 1); \draw(-20,1) -- (-20,0);

\draw[->](-20,-5) -- (-20,-1); \draw(-20,-1) -- (-20,0);

\draw (0-15,-8) node{\small { FIG. 15:} Transitions between the states of a Mach-Zehnder};
\draw (0-15,-9) node{\small interferometer in a PH-Pfaffian liquid at $T=0$.};
\end{tikzpicture}

\vskip .1in

Non-Abelian statistics results in more complicated behavior due to multiple fusion channels for non-Abelian anyons. Fig.~15 illustrates possible transitions among drain states for a PH-Pfaffian liquid, in which charge-$e/4$ quasiparticles tunnel at the QPCs. 
Each vertex is labeled by the trapped electric charge ${\rm mod}~e$ and the trapped topological charge in the drain.
The transition rates are defined in terms of

\be
\label{PH-Pf-p}
p(\theta)=[\Gamma_1^2+\Gamma_2^2]r_0(V,T)+2\Gamma_1\Gamma_2\cos(\phi_{AB}+\theta+\alpha)r_1(V,T),
\ee
where the statistical phase $\theta$ can be $0,~\pi,$ or $\pm \pi/2$, $\Gamma_1$ and $\Gamma_2$ are the tunneling amplitudes at the two QPCs, $\phi_{AB}$ is the Aharonov-Bohm phase, and $r_{1,2}$ and $\alpha$ have the same origin as in Eq. (\ref{int-p}). The transition rates equal $p(\theta)/2$ whenever two fusion channels are available.
It is apparent from the figure that the system can return to the initial state in multiple ways. One finds, as in the Abelian case, that the current is unchanged if the flux through the hole is increased by $\Phi_0$.

Shot noise in the weak tunneling limit yields the most striking signature of statistics in Mach-Zehnder interferometry  \cite{MZ-noise}. 
According to Eq. (\ref{Schottky}) one would naively think that the noise does not contain any new information compared to the current. This is indeed the case in the Fabry-Perot geometry. In Mach-Zehnder interferometry, however,  Eq. (\ref{Schottky})
does not hold since tunneling events are not independent, as their probability is affected by the topological charge, accumulated in the drain. 
The noise exhibits particularly interesting behavior in non-Abelian states \cite{16-fold}. For example, it can even diverge at some values of the magnetic flux \cite{PH-MZ}.
Indeed, consider Fig. 15 and suppose that $\Gamma_1\approx\Gamma_2$, $r_0\approx r_1$ in Eq. (\ref{PH-Pf-p}). Changing the magnetic field allows tuning $\phi_{AB}$ so that $p(0)\approx 0$. Consider an interferometer in the initial state $(-e/4,\sigma)$. 
The rate $p(\pi)/2$ of the process $(-e/4,\sigma)\rightarrow (0,\psi)$ is much faster than the rate 
$p(0)/2$ of the process $(-e/4,\sigma)\rightarrow (0,1)$. Hence, before the interferometer enters the $(0,1)$ state, it will evolve through multiple loops $(-e/4,\sigma)\rightarrow (0,\psi)\rightarrow (e/4,\sigma)\rightarrow (e/2,t)\rightarrow(-e/4,\sigma)$, where $t=1$ or $\psi$.
The average charge $q_t$, transmitted during those loops, is large: $q_t\gg e$. Eventually, the drain reaches the $(0,1)$ state. The transition rate $p(0)$ out of that state is small, and the interferometer will be stuck in the $(0,1)$ state for a long time $t\sim 1/p(0)$. At some point, a quasiparticle will tunnel through the device, and it will rapidly reach the $(-e/4,\sigma)$
state again. One observes the alternation of periods of high current and periods of no transport, when the drain is stuck in the $(0,1)$ state. This implies high noise.

{\color{black} On the other hand, possible tunneling between interferometer edges and localized states in the bulk does not have much effect on the current and hence does not lead to telegraph noise as long as the tunneling events are separated by longer time intervals than the tunneling events at the QPCs.}

Calculations tend to be rather involved in the theory of Mach-Zehnder interferometers even in the  lowest order perturbation theory. Yet, curiously, in some cases, the Bethe ansatz  allows an exact solution for the current and noise in a model of a Mach-Zehnder interferometer \cite{PA1,PA2,PA3,PA4}.

It is instructive to reconsider Fabry-Perot interferometry in light of the Byers-Yang theorem in a geometry with a hole in the center of the interferometer \cite{GT}. If the flux through the hole is increased by $\Phi_0$ on a sufficiently  short time scale, the charge in the hole will increase by $\nu e$. For FQH states, this will alter the statistical phase, which, in addition to possible effects of the Coulomb interaction, will generally change the transmission of   the interferometer.  This does not contradict the Byers-Yang theorem, however, because the theorem only applies in equilibrium.   
Equilibrium is established on a long time scale by relatively rare tunneling events between the edge of the interferometer and the inner edge around the hole. 
Such tunneling also leads to telegraphic noise \cite{kane-tel}. A related phenomenon of switching noise in a Fabry-Perot 
interferometer with a fluctuating number of trapped anyons was proposed \cite{switching} as a probe of statistics.

Mach-Zehnder interferometry has not yet been implemented for FQH states,  despite its success in the integer quantum Hall regime. A recent reference  \cite{melting}
sheds light on that challenge by measuring the dependence of the interference visibility on the filling factor. 
The visibility of the interference in the outer $\nu=1$ channel diminishes as the bulk filling factor decreases towards 1.
This is accompanied by signatures of edge reconstruction, 
{\it i.e.}, the emergence of topologically unprotected pairs of contra-propagating edge modes. It has long been recognized that quantum Hall edges exhibit complicated spacial structure.
Progress in interferometry will likely depend on a deeper understanding of edge states.

\setcounter{figure}{15}

Various other geometries have been considered in the literature. Ref. \onlinecite{wormhole} considers a ``wormhole'' geometry in which a tunneling contact creates a shortcut along a chiral FQH edge. Long tunneling contacts were proposed as probes of neutral modes in Ref. \onlinecite{long}.
Ref. \onlinecite{bubbles} introduces a modification of the Fabry-Perot setup that reveals an effect of topological vacuum bubbles.
{\color{black} Ref. \onlinecite{safi2002} addresses setups with a large number of edges.}
Ref. \onlinecite{deviatov} reports an experimental realization of a version of a Mach-Zehnder interferometer in which a single edge is split into two conducting channels that provide two interfering paths.

\vskip .25in

\section{Other techniques} \label{other}

Several other approaches can give  information about topological order. While that evidence may be indirect, it has importance because of the challenges faced by interferometry. In this section we focus on four methods: thermal conductance experiments \cite{texp1,texp2,texp3}, 
detecting upstream neutral modes \cite{ups1}, thermoelectric transport \cite{termoel,termoel-exp2}, and tunneling into the edge \cite{chang03,tun-1/3}. 
Thermal conductance is particularly useful as a probe of non-Abelian statistics. 
 Tunneling seems an enticingly straightforward 
probe of topological order. The actual information it gives turns out rather limited due to the complex physics of real edges. The complications, uncovered in tunneling experiments, are likely relevant for other probes, including interferometry.

\subsection{Thermal transport} 

The quantization of thermal conductance has long been recognized in non-interacting 1D systems \cite{pendry}. Quantum Hall liquids are unique in that their thermal conductance remains quantized even for strong interactions \cite{t1,t2,t3}. 
Consider first an Abelian FQH system with chiral edges such that all edge modes propagate in the same downstream direction, clockwise or counterclockwise, depending on the direction of the magnetic field. 
Since the bulk is gapped, heat is only carried by the edge at the lowest temperatures. A chiral edge, emanating from a source at the temperature $T$,  remains in thermal equilibrium at that temperature. The local thermal current along the edge in any point $x$ depends on the temperature and the details of the Hamiltonian of a local subsystem
around point $x$. At the same time, the heat current must be the same in all points of the edge since energy cannot accumulate on any portion of the edge in a steady state. This implies that the heat current $J_h(T)$ depends only on the temperature and is not sensitive to microscopic details such as the mode velocities and intermode interactions. 
As a consequence, the thermal current on an edge with $n$ chiral modes reduces to the sum of $n$ thermal currents in the simplest chiral systems with harmonic Lagrangians of the form, equivalent to (\ref{f=1-bosonize}):

\be
L=\frac{\hbar}{4\pi}\int dx [\partial_t\phi\partial\phi_x-v(\partial_x\phi)^2]
\ee
with an arbitrary edge velocity $v$. An easy calculation yields the quantized thermal conductance for an FQH bar with two edges emanating from two terminals at the temperatures
$T$ and $T+\Delta T$: 

\be
\kappa= \lim_{\Delta T \to 0} \frac{J_h(T+\Delta T)-J_h(T)}{\Delta T}=n\kappa_0T,
\ee
where $\kappa_0=\pi^2 k_B^2/3h$.

\begin{figure}[!htb]
\bigskip
\centering\scalebox{0.25}[0.25]{\includegraphics{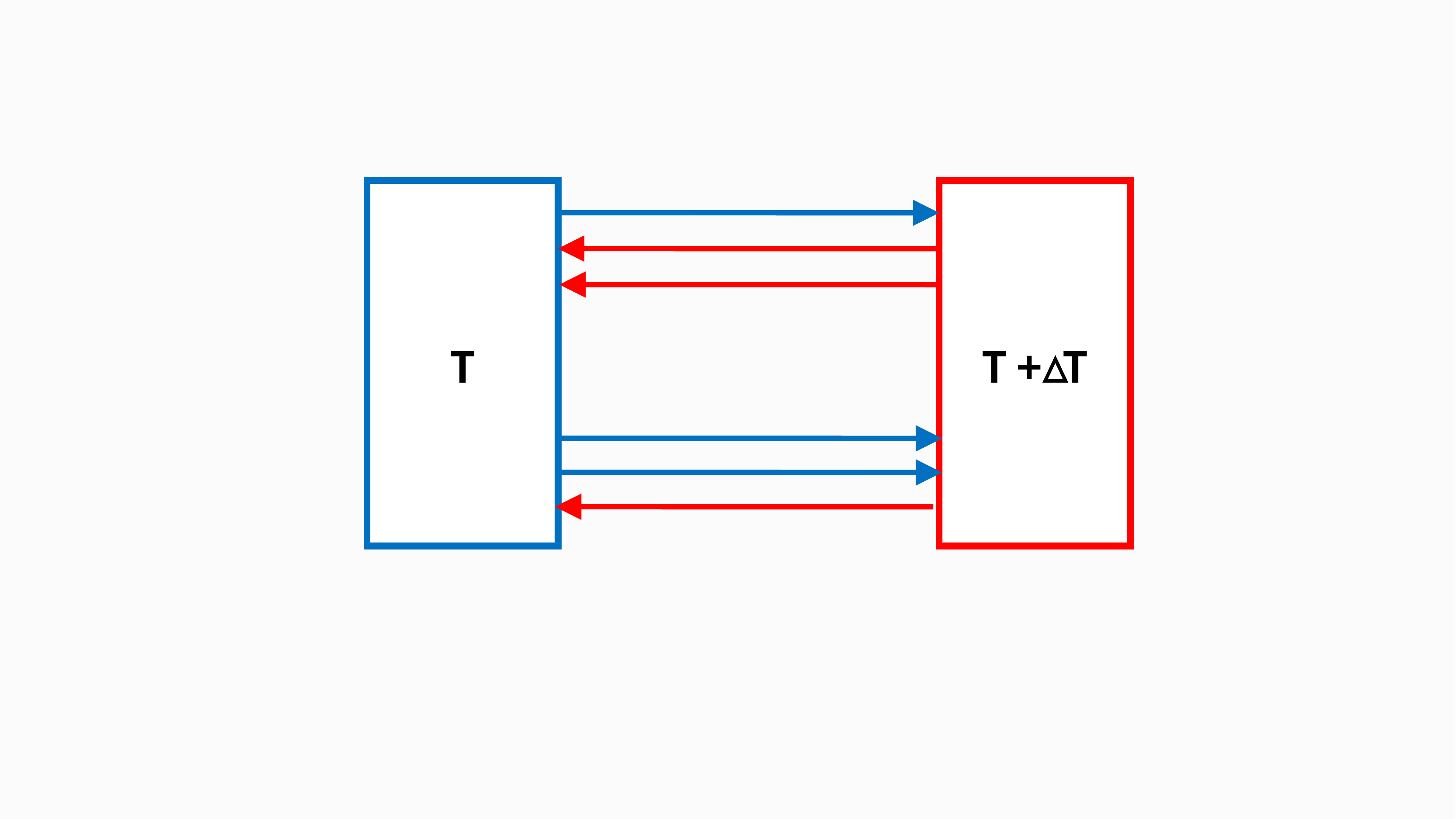}}
\caption{{One downstream and two upstream modes are shown on each edge. The modes, emanating from the left terminal, have the temperature $T$. The modes, emanating from the right terminal, have the temperature $T+\Delta T$.}}
\label{fig9:review}
\end{figure}

Many quantum Hall states possess topologically protected upstream neutral modes that travel in the direction, opposite to that of the charge mode. In particular, 
Jain's states at $\nu=[p+1]/[2p+1]$ have one downstream mode and $p$ upstream modes \cite{WenBook}.
The effect of the upstream modes on the thermal conductance depends on the edge length $L$ in comparison with the equilibration length $\xi$ on which the energy exchange between the upstream and downstream modes is significant \cite{texp1,MF19,MF20}. 
If $L\ll \xi$, the thermal conductances of the $n_u$ upstream modes and the $n_d$ downstream modes add up: $\kappa=(n_u+n_d)\kappa_0 T$. This can be understood by observing that $(n_u+n_d)$ noninteracting modes emanate from each of the two terminals, maintained at different temperatures (Fig.~\ref{fig9:review}).
A long edge reaches thermal equilibrium so that $\kappa=|n_u-n_d|\kappa_0 T$. The absolute value sign arises because heat can only flow from the hotter terminal to the colder terminal. For $n_u\ne n_d$, thermal equilibrium at the temperature of the majority modes is established on the length scale $\sim\xi$.
At $n_u=n_d$ the approach to the equilibrium is slow \cite{texp1}  and the thermal conductance $\kappa\sim\xi/L$. At low $T$, the equilibration length is predicted to diverge as a power of the temperature \cite{MF20}.

Thermal transport in non-Abelian liquids is qualitatively similar to the Abelian case. The integer numbers $n_u$ and $n_d$ should be substituted with the combined central charges of the upstream and downstream modes \cite{t2,t3}. Those central charges are not integer in general. In particular,  a Majorana edge mode contributes $\kappa_0 T/2$
to the thermal conductance. This can be understood by interpreting a real Majorana fermion mode as half of a complex Dirac fermion  mode that can be present on Abelian edges and carries the central charge $c_D=1$. Indeed, a complex fermion $\Psi_D$ can be represented as the combination $\Psi_D=\Psi_1+i\Psi_2$ of two real fermions with  $\Psi_1=\Psi_1^\dagger$ and
$\Psi_2=\Psi_2^\dagger$.

\begin{figure}[!htb] 
\bigskip
\centering\scalebox{0.25}[0.25]{\includegraphics{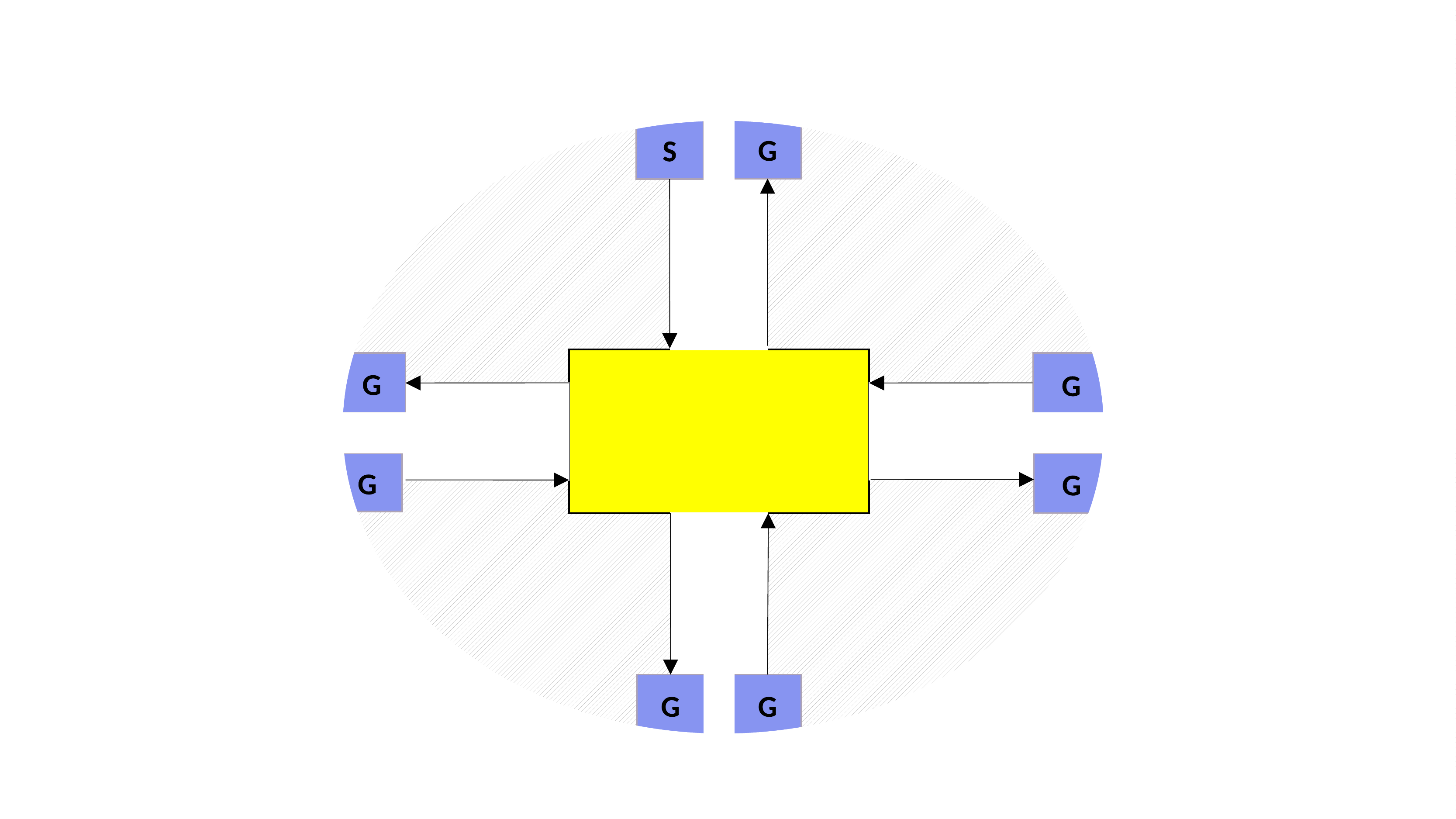}} 
\caption{{The current from source S partitions in the central floating contact into $N=4$ currents along the $N=4$ arms of the device.}}
\label{fig10:review}
\end{figure}

Since the FQH effect is observed at low temperatures, the relevant heat currents are low and challenging to measure. An ingenious approach was introduced in 
Ref.~\onlinecite{Jezoin} in an experiment in the integer quantum Hall effect. The current $I=GV$ enters the central floating contact (Fig.~\ref{fig10:review}) from  a biased source.
The currents $I/N$ leave the contact along $N$ arms. The dissipated Joule heat $Q=[GV^2-NG(V/N)^2]/2$ raises the temperature $T_m$ of the central floating contact and is carried away along the edges of the $n$ arms, so that $Q=N[T_m\kappa(T_m)-T_0\kappa(T_0)]/2$, where $T_0$ is the temperature of the cold contacts. 
$\kappa$ can be found after $T_m$ is determined from the current noise. A possible phonon contribution to the heat escaping the central floating terminal can be eliminated with a subtraction trick\cite{Jezoin}. The success of the experiment depends on how fast charge leaves the central floating contact. For a short dwell time, full equilibration cannot be achieved and the thermal conductance cannot be measured correctly
\cite{Sukhorukov,unc-exp,MF19}. 

Our discussion so far has ignored heat losses from the edge to the bulk by phonons or other possible processes, which can contribute at finite temperatures.  Such processes do not appear to be a major issue in current experiments.  For a theoretical discussion of bulk losses, see Ref. \onlinecite{bulk-loss}.

Using an adaption of the above geometry, Banerjee {\it{et al.}} measured the thermal conductance at several fractionally quantized states in GaAs, finding  the results\cite{texp1}  $\kappa=\kappa_0T$ at $\nu=1/3$ and $3/5$,  and  $\kappa=2\kappa_0T$ at $\nu=4/7$,  consistent with theory. The thermal conductance at $\nu=2/3$ remained relatively  far from the equilibrated value as expected, since there is one upstream mode and one downstream mode a that filling factor.  A recent experiment  on  graphene\cite{texp3}  measured 
 $\kappa\approx 2\kappa_0T$ at $\nu=4/3$, in agreement with theory.

The second Landau-level filling factors $7/3$, $5/2$, and $8/3$ in GaAs were explored in a different sample from the one used for the states of the first Landau level \cite{texp2}. The observed 
$\kappa=2.96\kappa_0 T$ at $\nu=7/3$ is consistent with the Laughlin topological order: two units of thermal conductance come from two integer edge modes and one more unit comes 
from one fractional edge channel. The observed thermal conductance was $2.19\kappa_0 T$ at $\nu=8/3$. The topological order at $\nu=8/3$ is expected to be the same as at the filling factor $2/3$. The predicted equilibrium thermal conductance is $\kappa_{\rm theor}=2\kappa_0 T$ for an infinite edge. 
Indeed, the edge contains two downstream integer edge channels, and one downstream and one upstream fractional channels.
The difference between the theoretical and experimental thermal conductances is similar to the case of $\nu=2/3$. This can be understood by observing that two of the downstream channels interact only weakly with the remaining downstream and upstream channels \cite{MF19}. We first observe that the overall charge mode is much faster than the rest of the modes in the second Landau level \cite{MF19}. Thus, its excitations leave the system before they can exchange energy with the rest of the edge channels
on a realistic finite edge. Besides, the integer spin mode is only weakly coupled with the other modes \cite{MF19}. Thus, the thermal conductance contains three independent contributions: one quantum from the charge mode, one quantum from the spin mode, and the contribution of the remaining downstream and upstream modes. The latter contribution is subject to strong finite-size corrections just like at 
$\nu=2/3$.

The observed thermal conductance at $\nu=5/2$ is $(2.53\pm 0.04)\kappa_0 T$ at higher temperatures and grows rapidly at low temperatures. Both 
properties are consistent with the non-Abelian PH-Pfaffian order \cite{MF19,MF20}, but the interpretation of the data is still 
debated \cite{MF19,int-5/2-1,int-5/2-2,int-5/2-3,int-5/2-4,nogap2}.  

To finish this section, we note that Ref. \onlinecite{T-noise} proposes an experiment with shot noise induced by a temperature gradient in a quantum point contact. 

\subsection{Upstream modes} 

\begin{figure}[!htb]   
\bigskip
\centering\scalebox{0.25}[0.25]
{\includegraphics{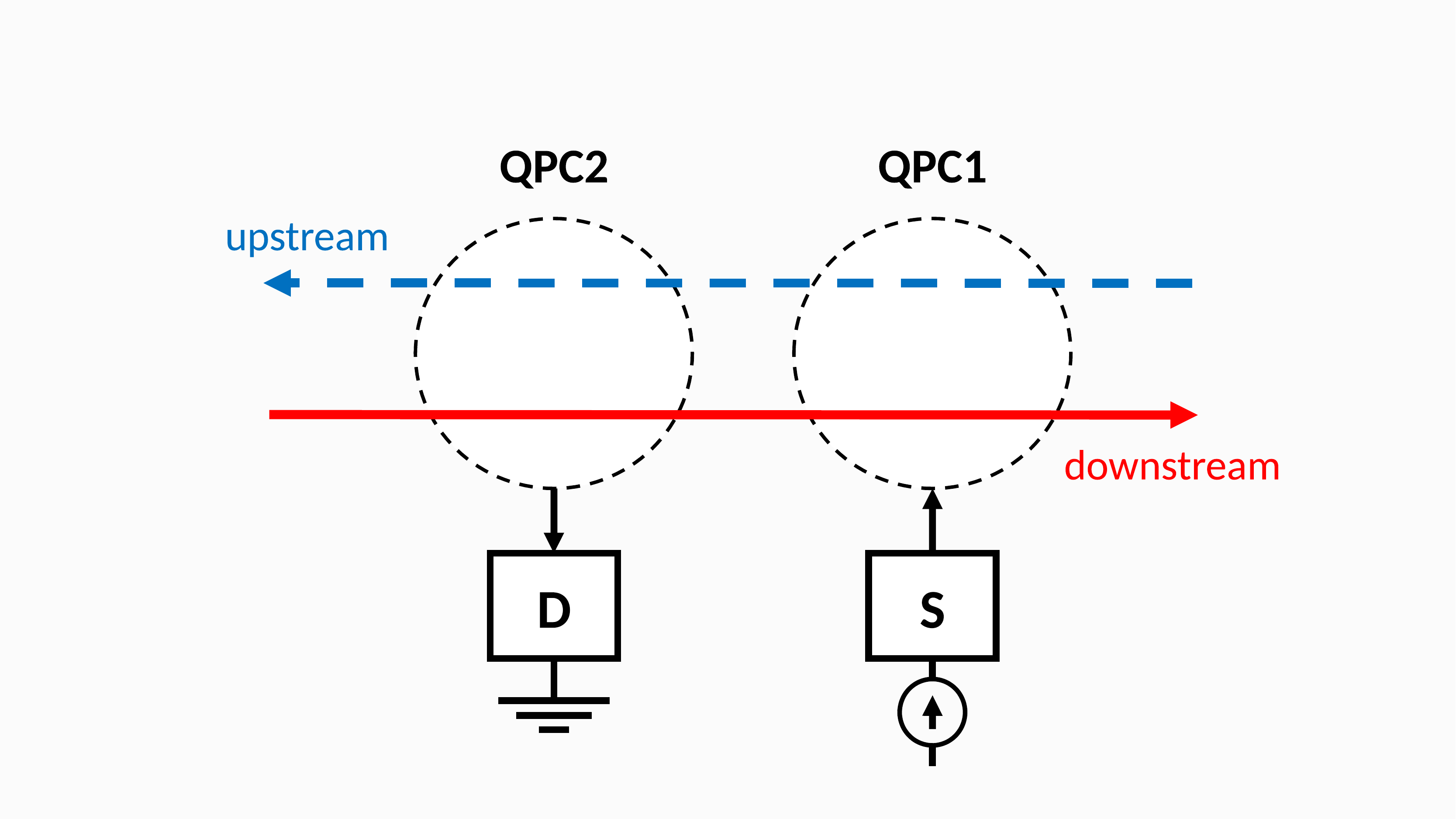}} 
\caption{{Charge tunnels into the edge from source S at QPC1. The upstream neutral mode (dashed line) carriers energy to QPC2. Non-equilibrium noise is generated in drain D at QPC2.}}
\label{fig15:review}
\end{figure}

Thermal conductance experiments cannot distinguish a state with $n$ downstream modes and no upstream modes from a state with $n+m$ downstream modes and
$m$ upstream modes, under conditions where energy is equilibrated between different modes on an edge. Thus, it is helpful to supplement thermal transport experiments with a tool for detecting upstream modes. Several setups have been used for that purpose. 
 Fig.~\ref{fig15:review}  illustrates an early theoretical proposal \cite{FL08}.  Upstream neutral modes carry no current but they can carry energy. Charge tunneling from source S at QPC1 induces Joule heat that is carried upstream to QPC2. 
 A thermoelectric effect generates excess current noise in drain D  and reveals the presence of upstream neutral modes. The role of QPC1 can also be played by a hot spot \cite{hsn} at an ohmic contact. 
 Much of the early theoretical work \cite{FL08} was focused on the states of the 16-fold way at $\nu=5/2$. See Ref. \onlinecite{2QPC-RR} for the application of a version of the setup\cite{FL08} to Read-Rezayi states. 
 
 Experimental probes of upstream neutral modes are well established now.
 Topologically protected upstream modes were observed at the filling factors $2/3$ and  $3/5$ in the first Landau level \cite{ups1,ups3-2/3} in agreement with theory. 
 No evidence of an upstream mode was seen\cite{ups2-5/2} at $\nu=7/3$ in agreement with a Laughlin order at that filling factor. An upstream mode was found at $\nu=8/3$, as it should be for a particle-hole conjugate state of the $7/3$ liquid.
 An upstream mode has also been detected\cite{ups1,ups2-5/2} at $\nu=5/2$ in agreement with the anti-Pfaffian and PH-Pfaffian models.
 
 The above experiments deal with relatively long edges of several tens of microns. At the scale of microns,  evidence of upstream modes was seen\cite{prolif,prolif1} at $\nu=1/3$  and $\nu=4/3$ even though no topologically protected upstream mode is expected at those filling factors.
 This can be understood as an example of edge reconstruction \cite{recon}. The reconstructed upstream modes do not survive on longer edges since inter-channel tunneling localizes them. Topologically protected modes are always delocalized. The dependence of the upstream noise on the length of the edge is addressed theoretically in 
 Refs. \onlinecite{up-noise-length1,up-noise-length2} and experimentally in Ref. \onlinecite{up-noise-length3}. {\color{black} Ref. \onlinecite{up-noise-length4} demonstrates the lack of thermal equilibration between contra-propagating modes in a small sample}.
 
 {\color{black}A very recent experimental development consists in probing upstream modes on interfaces of different filling factors \cite{interface-5/2}.}
Several other approaches have been proposed theoretically for the detection of neutral modes. Much attention has focused on Coulomb blockade 
physics \cite{CB1,CB2,CB3}, which
is closely related to the interferometry ideas addressed above.
Other proposals include the use of microwave absorption \cite{micro} and a proposed experiment to use a quantum dot transport to distinguish FQH states from their particle-hole conjugates \cite{qd-ly}. Neutral modes could be detected with momentum resolved tunneling into the edges \cite{mrt1,mrt2,mrt3}, 
but this technique  requires very weak disorder. The same limitation applies to a proposal \cite{hv-p} 
to probe topological order
by  measuring the Hall viscosity \cite{hv}. 
 
\subsection{Thermoelectric transport}

In the Seebeck effect, a gradient of the electric potential builds in response to a thermal gradient. The strength of the effect is measured by the Seebeck coefficient $Q=-\nabla Q/\nabla T$.
In a uniform FQH system, {\color{black} under conditions where energy is equilibrated more rapidly than momentum is transferred to impurities,}   the Seebeck coefficient  should reflect \cite{termoel} the entropy per charge carrier:
\be 
\label{Seebeck}
Q=-S/(eN_e), 
\ee
where $S$ and $N_e$ are the entropy density and the electron density. For a non-Abelian FQH state with a small number of well-separated localized quasiparticles, the entropy at low temperatures should be 
 determined \cite{termoel} by the number $K$ of states
at the fixed positions of quasiparticles. The latter number depends on the number of the quasiparticles $N_q$ and their quantum dimension $d$, $K \sim d^{N_q}$.
The number $N_q$ is controlled by the magnetic field. Thus, a measurement of the Seebeck coefficient  $Q\sim N_q\log d$ reveals the quantum dimension of non-Abelian anyons.

The existing experimental data \cite{termoel-exp1,termoel-exp2} are limited. Qualitative agreement with the theory for non-Abelian states of the 16-fold way was reported\cite{termoel-exp2} at $\nu=5/2$, but more work is needed before the data are understood. 

Related theoretical ideas are explored in the papers \onlinecite{termoel-2,termoel-3,termoel-rel}. See Ref. \onlinecite{termoel-rel2} for a proposal of a thermoelectric probe of neutral edge modes.

The thermoelectric technique differs profoundly from all the approaches addressed in the previous sections except the single-electron-transistor probe of anyon charges (Section~\ref{charge-expts}.B).
Indeed, all those proposed and implemented probes of fractional charge and statistics involve edge physics. On the other hand, thermoelectric transport occurs in the bulk. 
Thus, this technique should be insensitive to the complications of edge physics (see the next subsection).
We note that it has also been  suggested to use a scanning tunneling microscope for a bulk probe of anyon statistics \cite{stm}. 
{\color{black}Another proposed bulk probe involves Raman scattering \cite{raman-stat,graviton2021}.
{\color{black}See Refs. \onlinecite{mob-imp-1,mob-imp-2} for a discussion of a probe with mobile impurities.}}

\subsection{Tunneling}

It was predicted long ago that the tunneling conductance through a weak link of two FQH liquids follows a universal power dependence $G_t\sim T^{2g_e-2}$, where $g_e$ depends only on the topological order \cite{WenBook}. A similar behavior, $G_t\sim T^{2g_q-2}$ with a universal $q_q$, was predicted for weak quasiparticle tunneling between two edges of an FQH system \cite{WenBook}.
These predictions  were based on the chiral Luttinger liquid model.

Early results \cite{chang96} on electron tunneling at $\nu=1/3$ were consistent with the theoretical expectations for $g_e$. Yet, at other 
filling factors a puzzling dependence $g_e\sim 1/\nu$ was observed \cite{chang98}. This does not agree with the theory \cite{slh98,lsh01}. 
{\color{black} Note that edge reconstruction was predicted to occur in experimental samples \cite{kun-rec-1,kun-rec-2}.}
See Ref. \onlinecite{chang03} for a review.

Later experiments focused on quasiparticle tunneling. The observed $g_q$ is typically greater than the predictions \cite{tun-1/3}. Three mechanisms beyond 
the chiral Luttinger liquid model were introduced to explain the discrepancy: 
edge reconstruction \cite{tun-rec}, long-range Coulomb forces between segments of the edge \cite{tun-c}, and $1/f$ noise and dissipation \cite{sassetti_5/2}. It is possible that a combination of mechanisms is at play. 
Thus, tunneling experiments only yield an upper bound on $g_q$ and provide limited information about topological order \cite{PH-MZ}. This probably explains the difficulties in the interpretation \cite{tun-5/2-1,tun-5/2-2,tun-5/2-3,tun-5/2-4} of the quasiparticle tunneling experiments at $\nu=5/2$.
Different ideal theoretical $g_q$ are predicted for different states of the 16-fold way. The observed $g_q$ {\color{black} has also differed in different experiments \cite{tun-5/2-1,tun-5/2-2,tun-5/2-3,tun-5/2-4}. }
Data from different samples and even from the same sample at different gate voltages were interpreted in terms of several different states of the 16-fold way. However, the tunneling exponent $g_q$ was found to  change continuously with the gate voltage at the gates that form the tunneling contact.
The observed values were  consistent with an upper bound on the ideal theoretical value for the Pfaffian and PH-Pfaffian orders \cite{PH-MZ}.

Tunneling data were used to extract both the tunneling exponent $g_q$ and the quasiparticle charge \cite{tun-5/2-1,tun-5/2-2,tun-5/2-3,tun-5/2-4} at $\nu=5/2$ from a fit to a theoretical $I-V$ curve. 
{\color{black}
The confidence intervals are elongated ovals in the $g_q$-charge plane and hence the uncertainty in both quantities is high. At the same time, the quantized quasiparticle charge is known independently. Fitting for $g_q$ at a fixed charge reduces error bars.}

Note, finally, that tunneling noise was proposed as another probe of non-Abelian statistics \cite{bena}.

 {\color{black} 
 \section{ Concluding Remarks}  \label{conclusions} 
 
 Quantum mechanics textbooks usually state that only two types of quantum  statistics are possible: Fermi and Bose. The argument goes as follows. For two indistinguishable particles, there is no way to tell the configuration with particle 1 in point ${\bf r}_1$ and particle 2 in point ${\bf r}_2$  from the configuration with particle 1 in point ${\bf r}_2$
 and particle 2 in point ${\bf r}_1$. Thus, the probabilities of the two configurations $P({\bf r}_1,{\bf r}_2)=|\psi({\bf r}_1,{\bf r}_2)|^2$ and 
 $P({\bf r}_2,{\bf r}_1)=|\psi({\bf r}_2,{\bf r}_1)|^2$ must equal. Hence, the particle exchange generates a phase change in the wave-function: $\psi({\bf r}_1,{\bf r}_2)=\theta\psi({\bf r}_2,{\bf r}_1)$,
 where $|\theta|=1$. After two particle exchanges, one finds
 
 \be
 \label{exchanges}
 \psi({\bf r}_1,{\bf r}_2)=\theta^2\psi({\bf r}_1,{\bf r}_2) 
 \ee
 so one must have  $\theta=\pm 1$. The plus sign describes bosons and the minus sign describes fermions. 
 
 The argument might look convincing but it contains multiple loopholes. First, it may not be necessary   for the wave-function to be single-valued, as is implicitly assumed in Eq. (\ref{exchanges}).
 Alternatively, the wave function does not have to depend just on the positions of the particles but may depend on how the system reached a particular configuration. In other words, a single-valued wave function may be defined not on the configuration space but on the Riemann surface whose points are equivalency classes of trajectories in the configuration space. Besides, $\theta$ does not have to be a number but may be a unitary operator, if the Hilbert space associated with a fixed  set  of particle positions is multidimensional. This  last loophole opens the particularly interesting possibility of non-Abelian statistics.
 
 The loopholes have some surprising consequences\cite{3d-f-s-1,3d-f-s-2} in 3D, but it is in 2D where things become truly exciting, as systems with  anyons, particles with fractional statistics or non-Abelian statistics,  are mathematically possible.

But, physics is an experimental science, and the theory of anyons is only relevant, if anyons exist in nature. Fortunately, observation of the fractional quantized Hall effect makes their existence an almost mathematical certainty. Indeed,
fractional quantization of the Hall conductance in appropriate systems is well established experimentally. As explained in Section~\ref{meaning}, such fractional quantization of the Hall conductance in an insulator necessarily entails the existence of  fractional charges, and  fractional charges entail fractional statistics.

Yet, general arguments do not tell us everything we might want to know about the particular anyons that might occur in a given quantum Hall system. The quantum number $\nu$ obtained from a measurement of the Hall conductance sets constraints on the possible charges and statistics of the  elementary quasiparticles  hosted by the FQH state, but it does not completely determine them.  Moreover, general arguments  do not tell us whether individual anyons, or small collections of them, will be manifest in any practical experiment.

For a long time, our knowledge about fractional charge and statistics was derived in a rather unsatisfactory way. First, theoretical predictions were made based on assumptions about the nature of the ground state in an observed FQH state.  Second, numerics on small idealized systems would verify some of the theoretical predictions, most importantly, the form of the ground-state wave function. Third, some experimental data   would show agreement with some aspects of numerics, such as the spin polarization. 
This would be interpreted as a proof of the theoretical picture.
Such evidence is inevitably indirect and not always reliable.  For example, there remain persistent discrepancies between calculated energy gaps and the activation gaps measured in experiments. Although these discrepancies have been attributed to effects of disorder, theoretical attempts to understand the precise manner in which impurities affect the measurements have only been partially successful.\cite{gap-ahm} 

The last decade of the twentieth century saw a breakthrough in the detection of fractional charges. The shot noise technique proved particularly fruitful (Section~\ref{charge-expts}.A). A clear direct  evidence of fractional statistics had to wait until very recently. While promising interferometric results for fractional statistics in FQH states at $\nu=1/3$ and $\nu=2/5$ were published more than a decade ago,\cite{goldman-puzzle,LinGoldman09}, interpretation of those data has proved challenging.  Similarly,   though promising interferometry results\cite{willett10} concerning non-Abelian statistics were published some ten years ago at $\nu=5/2$, there have been  questions about the interpretation of those data, particularly because of the very small interferometer area inferred from the experiments.

In 2020, a clear direct observation of the anyonic statistical phase in interferometry at $\nu=1/3$ has finally arrived\cite{manfra20}. Another achievement of 2020 is the implementation of an anyon collider\cite{collider-exp} at $\nu=1/3$. Although the relation of these experiments to fractional statistics may not be direct, the experiments do probe effects of collisions between pairs of diluted anyons, where fractional statistics is an essential ingredient.   Results presented in 2019 of improved interferometer experiments at $\nu=5/2$ and 7/2, using a large number of samples, have confirmed the previous measurements on this system, and give additional support to the existence of particles with  Ising-type non-Abelian statistics in these states. Our understanding of interferometer experiments has increased  as we have seen that one should distinguish measurements  where the central region is in an incompressible state, with at most a few localized quasiparticles, and the more usual situation, where there are many quasiparticles in the system, which can  enter and leave on a laboratory time scale as parameters such as the magnetic field and  gate voltages are varied. 

 Probing potentially non-Abelian states on fragile plateaus of the second Landau level is certainly challenging. Yet,
the distinction between non-Abelian and Abelian statistics is more dramatic than the distinction of Abelian fractional statistics from the Fermi and Bose statistics. This opens
a way for probes that would demonstrate the existence of  non-Abelian statistics even though they would not allow distinguishing Abelian anyons from fermions. One such probe is thermal conductance (Section~\ref{other}.A).
Remarkable evidence of non-Abelian statistics at  $\nu=5/2$ came from a thermal conductance experiment\cite{texp2} in 2018

The main focus of the experimental work on anyonic statistics has been on the simplest Abelian and non-Abelian filling factors $1/3$ and $5/2$. We eagerly await extension of the recent experimental breakthroughs to other filling factors. As this review shows, there is no lack of theoretical proposals to detect fractional statistics, and the ball is in the experimentalists' court.
Yet, there is much work for theory too, since the interpretation of the data is often challenging. Major puzzles surround key probes, such as Fabry-Perot interferometry. For example, it has been found that Fabry-Perot interferometry exhibits an enigmatic pairing effect at certain  {\em{integer}} filling factors \cite{pairing1,pairing2}. Until that effect is understood, it is hard to be confident in the interpretation of FQH data.

Almost all probes that have been proposed or implemented are based on edge physics. This is not surprising, since edges dominate transport and it is easier to access and manipulate the edges than the bulk. Yet, fractional charge and statistics are defined in the bulk. The success of edge probes hinges on the bulk-edge correspondence 
hypothesis (see Section~\ref{non-abelian}.B).  It is noteworthy that measurements of fractional charge in puddles far from the edge of a sample have been successfully carried out using single electron transistors as charge sensors.\cite{local1,local2}.  
It would be highly desirable to also
implement bulk probes of fractional statistics that would not rely on bulk-edge correspondence. Such probes must be robust to the existence of compressible islands in the bulk.

{\color{black} The focus of this review has been on the FQH effect in solids. At the same time, similar physics involving fractional statistics is possible in other settings, including cold atoms \cite{HalperinJainBook,cold-atoms}.}

Recent experiments, particularly at  $\nu=1/3$,  produce direct support for  a theoretical picture, developed almost four decades ago. Yet, other recent experiments on quantum Hall systems  have produced  major surprises. Based on the history of the field, we may expect to see many new surprises, whose influence will likely extend well beyond the FQH effect.
}

\section*{Acknowledgments} DEF was supported in part by the NSF under grant No. DMR-1902356.  
{\color{black} We acknowledge useful discussions with A. C. Barlam, D. C. Glattli, M. Heiblum, M. J. Manfra, T. Martin, B. Rosenow, I. Safi, K. Shtengel, A. Stern, and K. Yang.
We thank D. C. Glattli for sharing unpublished data with us and M. J. Manfra for kindly supplying us with Fig. 6.}

\bibliographystyle{unsrt}
%\bibliography{Fract-Bib}

\end{document}